\documentclass[11pt,final,onecolumn]{IEEEtran}

\usepackage{amsmath,amssymb,float,multirow} 
\usepackage[final]{graphicx} 
\usepackage{wrapfig} 
\usepackage{verbatim} 
\usepackage{epsfig} 
\usepackage{cite}
\usepackage{bibspacing}

\def\RR{\mathbb{R}}

\def\uniqueTheta{\theta}

\def\obsTheta{\theta'}

\newcommand{\vecc}[1]{\mbox{vec}({#1})}
\newcommand{\BF}[1]{\mbox{\boldmath$#1$}}

\newcommand{\DP}[2]{\ensuremath{\mathrm{DP}\!\left({#1},{#2}\right)}}
\newcommand{\bigk}[2]{\mathbf{#1}^{(#2)}}

\newcommand{\bigkT}[2]{\mathbf{#1}^{(#2)^T}}
\newcommand{\bigksub}[3]{\mathbf{#1}_{#3}^{(#2)}}
\newcommand{\bigminksub}[3]{\mathbf{#1}_{#3}^{-(#2)}}
\newcommand{\bigkTsub}[3]{\mathbf{#1}_{#3}^{(#2)^T}}
\newcommand{\symk}[2]{#1^{(#2)}}
\newcommand{\symmink}[2]{#1^{-(#2)}}
\newcommand{\symminkT}[2]{#1^{-(#2)^T}}
\newcommand{\symkT}[2]{#1^{(#2)^T}}
\newcommand{\symkB}[2]{\BF{#1}^{(#2)}}

\newcommand{\Ylag}[1]{\bar{#1}}

\newcommand{\MN}[4]{\mathcal{M}\mathcal{N}\left(#1;#2,#3,#4\right)}
\newcommand{\symsubsup}[3]{{#1}_{#2}^{(#3)}}

\newcommand{\figdir}{figs} 

\newtheorem{crit}{Criterion}[section]

\floatstyle{boxed}
\newfloat{algorithm}{tbh}{loa}
\floatname{algorithm}{Algorithm}
\newcommand{\algtop}{
  \vspace*{-6pt}
  \setlength{\belowdisplayskip}{2pt plus 2pt}
  \setlength{\abovedisplayskip}{2pt plus 2pt}
  \setlength{\itemsep}{4pt}
}
\newcommand{\algend}{
  \vspace*{-2pt}
}

\begin{document}
\title{Bayesian Nonparametric Inference of Switching Linear Dynamical Systems}

\author{Emily~Fox, Erik~Sudderth, Michael~Jordan, and~Alan~Willsky

% <-this % stops a space
\thanks{E. Fox is with the Department of Statistical Science, Duke University, Durham, NC, 27708 USA e-mail: fox@stat.duke.edu. E. Sudderth is with the Department of Computer Science, Brown University, Providence, RI, 02912 USA e-mail: sudderth@cs.brown.edu. M. Jordan is with the Department of Electrical Engineering and Computer Science, and Department of Statistics, University of California, Berkeley, CA, 94720 USA e-mail: jordan@eecs.berkeley.edu. A. Willsky is with the Department of Electrical Engineering and Computer Science, MIT, Cambridge, MA, 02139 USA e-mail: willsky@mit.edu. This work was supported in part by MURIs funded through AFOSR Grant FA9550-06-1-0324 and ARO Grant W911NF-06-1-0076.  Preliminary versions (without detailed development or analysis) of this work have been presented at two conferences~\cite{Fox:NIPS08,Fox:SYSID09}.}}

% <-this % stops a space
\markboth{MIT LIDS Technical Report \#2830}{Fox \MakeLowercase{\textit{et al.}}:Bayesian Nonparametric Inference of Switching Linear Dynamical Systems}

\maketitle 
\vspace{-0.25in}
\begin{abstract}
\vspace{-0.1in}	
	%\boldmath
	Many complex dynamical phenomena can be	effectively modeled by a system that switches among a set of conditionally linear dynamical modes.  We consider two such models: the switching linear dynamical system (SLDS) and the switching vector autoregressive (VAR) process. Our Bayesian nonparametric approach utilizes a hierarchical Dirichlet process prior to learn an unknown number of persistent, smooth dynamical modes.  We additionally employ automatic relevance determination to infer a sparse set of dynamic dependencies allowing us to learn SLDS with varying state dimension or switching VAR processes with varying autoregressive order. We develop a sampling algorithm that combines a truncated approximation to the Dirichlet process with efficient joint sampling of the mode and state sequences. The utility and flexibility of our model are demonstrated on synthetic data, sequences of dancing honey bees, the IBOVESPA stock index, and a maneuvering target tracking application.
\end{abstract}

%\vspace{-0.1in}
\begin{IEEEkeywords}
\vspace{-0.1in}
	Bayesian nonparametric methods, hidden Markov model, Markov jump linear system, time series. 
\end{IEEEkeywords}

\section{Introduction} \label{sec:Intro}
\IEEEPARstart{L}{inear} dynamical systems (LDSs) are useful in describing dynamical phenomena as diverse as human motion \cite{Pavlovic:01,Ren:05}, financial time-series \cite{Kim:94,So:98,Carvalho:06}, maneuvering targets \cite{RongLi:MM,Fox:Fusion07}, and the dance of honey bees \cite{Oh:08}. However, such phenomena often exhibit structural changes over time, and the LDS models which describe them must also change. For example, a ballistic missile makes an evasive maneuver; a country experiences a recession, a central bank intervention, or some national or global event; a honey bee changes from a \emph{waggle} to a \emph{turn right} dance. Some of these changes will appear frequently, while others are only rarely observed. In addition, there is always the possibility of a new, previously unseen dynamical behavior. These considerations motivate us to develop a Bayesian nonparametric approach for learning \emph{switching} LDS (SLDS) models. We also consider a special case of the SLDS---the switching vector autoregressive (VAR) model---in which direct observations of the underlying dynamical process are assumed available. % Although a special case of the general linear systems framework, autoregressive models have simplifying properties that often make them a practical choice in applications.

One can view the SLDS, and the simpler switching VAR process, as an extension of hidden Markov models (HMMs) in which each HMM state, or \emph{mode}, is associated with a linear dynamical process. While the HMM makes a strong Markovian assumption that observations are conditionally independent given the mode, the SLDS and switching VAR processes are able to capture more complex temporal dependencies often present in real data. Most existing methods for learning SLDS and switching VAR processes rely on either fixing the number of HMM modes, such as in \cite{Oh:08}, or considering a change-point detection formulation where each inferred change is to a new, previously unseen dynamical mode, such as in \cite{Xuan:07}. In this paper we show how one can remain agnostic about the number of dynamical modes while still allowing for returns to previously exhibited dynamical behaviors.

Hierarchical Dirichlet processes (HDP) can be used as a prior on the parameters of HMMs with unknown mode space cardinality~\cite{Teh:06,Beal:02}.  In this paper we use a variant of the HDP-HMM---the \emph{sticky HDP-HMM} of~\cite{Fox:ICML08}---that provides improved control over the number of modes inferred; such control is crucial for the problems we examine.
%Although the HDP-HMM and its sticky extension are very flexible time
%series models, they do make a strong Markovian assumption that
%observations are conditionally independent given the state. This
%assumption is often insufficient for capturing the temporal
%dependencies of the observations in real data.
%Our nonparametric Bayesian approach for learning switching dynamical
%processes extends the sticky HDP-HMM formulation to learn an unknown
%number of persistent, smooth dynamical modes and thereby capture a
%wider range of temporal dependencies. We then present a method for
Our Bayesian nonparametric approach for learning switching dynamical processes extends the sticky HDP-HMM formulation to learn an unknown number of persistent dynamical modes and thereby capture a wider range of temporal dependencies. We then explore a method for learning which components of the underlying state vector contribute to the dynamics of each mode by employing \emph{automatic relevance determination} (ARD)~\cite{MacKay:94,Neal:96,Beal:03}. The resulting model allows for learning realizations of SLDS that switch between an unknown number of dynamical modes with possibly varying state dimensions, or switching VAR processes with varying autoregressive orders.
\subsection{Previous System Identification Techniques}
Paoletti et. al.~\cite{Paoletti:07} provide a survey of recent approaches to identification of switching dynamical models. The most general formulation of the problem involves learning: (i) the number of dynamical modes, (ii) the model order, and (iii) the associated dynamic parameters. %Most approaches assume that the model order is the same for each dynamical mode. 
For noiseless switching VAR processes, Vidal et. al.~\cite{Vidal:03} present an exact algebraic approach, though relying on fixing a maximal mode space cardinality and autoregressive order. %However, the method relies on fixing the maximal mode space cardinality and autoregressive order, which is assumed shared among modes. Additionally, extensions to the noisy case rely on heuristics. 
Psaradakis and Spagnolog~\cite{Psaradakis:06} alternatively consider a penalized likelihood approach to identification of stochastic switching VAR processes.

For SLDS, identification is significantly more challenging, and methods 
%such as those in~\cite{Huang:04,Vidal:07}
typically rely on simplifying assumptions such as deterministic dynamics or knowledge of the mode space. %If one knew the mode sequence, then one could partition the data according to the underlying mode sequence and then employ standard techniques for identification of single LDS~\cite{Lindquist:94}. However, when addressing the issue of stochastic realization or system identification of SLDS, we assume that the mode sequence is a latent variable of our model. 
Huang et. al.~\cite{Huang:04} present an approach that assumes deterministic dynamics and embeds the input/output data in a higher-dimensional space and finds the switching times by segmenting the data into distinct subspaces~\cite{Vidal:TR}. %This algebraic approach assumes deterministic dynamics and claims robustness to moderate amounts of noise. 
Kotsalis et. al.\cite{Kotsalis:06} develop a balanced truncation algorithm for SLDS assuming the mode switches are i.i.d. within a fixed, finite set; the authors also present a method for model-order reduction of HMMs\footnote{The problem of identification of HMMs is thoroughly analyzed in~\cite{Anderson:99}.}. In \cite{Vidal:07}, a realization theory is presented for \emph{generalized jump-Markov linear systems} (GJMLS) in which the dynamic matrix depends both on the previous mode and current mode. %The authors mention that it is unclear whether a similar theory can be developed for the standard SLDS we consider in this chapter. 
Finally, when the number of dynamical modes is assumed known, Ghahramani and Hinton~\cite{Ghahramani:00} present a variational approach to segmenting the data into the linear dynamical regimes and learning the associated dynamic parameters\footnote{This formulation uses a \emph{mixture of experts} SLDS in which $M$ different continuous-valued state sequences evolve independently with linear dynamics and the Markovian dynamical mode selects which state sequence is observed at a given time.}. For questions of observability and identifiability of SLDS in the absence of noise, see~\cite{Vidal:02}.

%Many questions of observability and identifiability of SLDS in the absence of noise are addressed in \cite{Vidal:02}. Specifically, a set of sufficient conditions are provided for the initial continuous state $\BF{x}_0$ and discrete mode sequence $z_{1:T}$ to be observable given fixed model parameters. The authors argue that if both the dimension $d$ of the continuous state and the number of possible modes $K$ are unconstrained, there is an infinite set of systems that realize the same set of observations $\BF{y}_{1:T}$ while differing in $\{\BF{x}_0, z_{1:T}\}$. Even when limiting one of these two degrees of freedom, the problem of realization is ill-posed. In this paper, we circumvent having to limit both $d$ and $K$, and having to check the detailed conditions provided in \cite{Vidal:02}, by taking a Bayesian approach. One can interpret the control literature on non-identifiable systems as arising in classical statistics when multiple models have equivalent likelihood. 
In the Bayesian approach that we adopt, we coherently incorporate noisy dynamics and uncertainty in the mode space cardinality.  %it is the specific prior we place on the parameters that allows us to distinguish between the set of these equivalent models. 
Our choice of prior penalizes more complicated models, both in terms of the number of modes and the state dimension describing each mode, allowing us to distinguish between the set of equivalent models described in~\cite{Vidal:02}. Thus, instead of placing hard constraints on the model, 
%number of modes we allow,
we simply increase the posterior probability of simpler explanations of the data. As opposed to a penalized likelihood approach using \emph{Akaike's information criterion} (AIC)~\cite{Akaike:74} or the \emph{Bayesian information criterion} (BIC)~\cite{Schwarz:78},
%(equivalently, \emph{minimum description length} (MDL)
%criterion~\cite{Rissanen:78}),
our approach provides a model complexity penalty in a purely Bayesian manner.
%
%In summary, previous methods for identification of the SLDS we consider in this thesis rely on assuming either: (i) deterministic dynamics, (ii) a fixed number of dynamical modes, or (iii) non-Bayesian penalties on model complexity. The approach we present herein aims to address identification of mode space cardinality and model order of SLDS and switching VAR processes within a Bayesian nonparametric framework. We also demonstrate that allowing for variable order models provides insight into the structure of the underlying phenomenon.
% TODO: \textcolor{red}{and can improve predictive performance of
% the learned model. ??????}

In Sec.~\ref{sec:background}, we provide background on the switching linear dynamical systems we consider herein, and previous Bayesian nonparametric methods of learning HMMs.  Our Bayesian nonparametric switching linear dynamical systems are described in Sec.~\ref{sec:models}.  We proceed by analyzing a conjugate prior on the dynamic parameters, and a sparsity-inducing prior that allows for variable-order switching processes.  The section concludes by outlining a Gibbs sampler for the proposed models.  In Sec.~\ref{sec:results} we present results on synthetic and real datasets, and in Sec.~\ref{sec:ModelVariants} we analyze a set of alternative formulations that are commonly found in the maneuvering target tracking and econometrics literature.
\section{Background}
\label{sec:background}
\subsection{Switching Linear Dynamic Systems}
\label{sec:backgroundSLDS}
A state space (SS) model %provides a general framework for analyzing many dynamical phenomena. The model 
consists of an underlying state, $\BF{x}_t \in \mathbb{R}^n$, with %linear
dynamics observed via $\BF{y}_t \in \mathbb{R}^d$. A linear time-invariant (LTI) SS model is given by
%,in which the dynamics do not depend on time, is given by
%
\begin{align}
\BF{x}_{t} = A\BF{x}_{t-1} + \BF{e}_t \hspace{0.25in} \BF{y}_t =
C\BF{x}_{t} + \BF{w}_t,
\end{align}
where $\BF{e}_t$ and $\BF{w}_t$ are independent Gaussian noise processes with covariances $\Sigma$ and $R$, respectively. 
%From this formulation we see that $\BF{x}_{1:T}$ forms a continuous, discrete-time Markov process.  Thus, the state $\BF{x}_t$ yields the past, $\BF{x}_{1:t-1}$, and the future, $\BF{x}_{t+1:T}$, conditionally independent.  We do not, however, have direct observations of the state and instead rely on noisy observations $\BF{y}_t$ for our inferences.

An order $r$ VAR process, denoted by VAR($r$), with observations $\BF{y}_t \in \mathbb{R}^d$, can be defined as
\begin{align}
\BF{y}_t = \sum_{i=1}^r A_i\BF{y}_{t-i} + \BF{e}_t \hspace{0.25in}
\BF{e}_t \sim \mathcal{N}(0,\Sigma).
\end{align}
%
%Here, the observations depend linearly on the previous $r$ observation vectors. 
Every VAR($r$) process can be described in SS form, %by, for example, the following transformation: 
though not every SS model may be expressed as a VAR($r$) process for finite $r$ \cite{Aoki:91}.
%%
%\begin{align}
%\BF{x}_{t+1} = \begin{bmatrix} A_1 & \dots & A_r\\ I & \dots & 0\\ &
%\vdots & \\ 0 & \dots & I\end{bmatrix}\BF{x}_t + \begin{bmatrix} 1\\
%0\\ \vdots \\ 0\end{bmatrix} \BF{v}_t \hspace{0.25in} \BF{y}_t =
%\begin{bmatrix} I & 0 & \dots & 0 \end{bmatrix}\BF{x}_{t} +
%\BF{e}_t.
%\end{align}
%%
%
%\begin{align}
%\BF{x}_{t} = \begin{bmatrix} A_1 & A_2 &\dots & A_r\\ I & 0 & \dots & 0\\
%\vdots & \ddots & \vdots &\vdots \\ 0 & \dots & I &
%0\end{bmatrix}\BF{x}_{t-1} + \begin{bmatrix} I\\
%0\\ \vdots \\ 0\end{bmatrix} \BF{e}_t \hspace{0.25in} \BF{y}_t =
%\begin{bmatrix} I & 0 & \dots & 0 \end{bmatrix}\BF{x}_{t}.
%\end{align}
%%
%Note that there are many such equivalent \emph{minimal} SS representations that result in the same input-output relationship,
% MJ this is well known; no need for a reference when there's a
% space crunch.
%\cite{Lutkepohl,Jonga:04}
%where minimality implies that there does not exist a realization with lower state dimension. On the other hand, not every SS model may be expressed as a VAR($r$) process for finite $r$ \cite{Aoki:91}. We can thus conclude that considering a class of SS models with state dimension $r\cdot d$ and arbitrary dynamic matrix $A$ subsumes the class of VAR($r$) processes.

The dynamical phenomena we examine in this paper exhibit behaviors better modeled as switches between a set of linear dynamical models. We define a \emph{switching linear dynamical system} (SLDS) by
\begin{equation}
	\begin{aligned}
		\begin{array}{c}
		z_t\mid z_{t-1} \sim \pi_{z_{t-1}}\\
		\BF{x}_{t} = A^{(z_{t})}\BF{x}_{t-1} + \BF{e}_t(z_{t}) \hspace{0.25in} \BF{y}_t = C\BF{x}_{t} + \BF{w}_t.
		\end{array} 
	\end{aligned}
	\label{eqn:SLDS} 
\end{equation}
The first-order Markov process $z_t$ with transition distributions $\{\pi_j\}$ indexes the mode-specific LDS at time $t$, which is driven by Gaussian noise $\BF{e}_t(z_t) \sim \mathcal{N}(0,\Sigma^{(z_t)})$. One can view the SLDS as an extension of the classical hidden Markov model (HMM)~\cite{Rabiner:89}, which has the same mode evolution, but conditionally \emph{independent} observations:
\begin{equation}
	\begin{aligned}	
		z_t\mid z_{t-1} &\sim \pi_{z_{t-1}}\\
		y_t \mid z_t &\sim F(\uniqueTheta_{z_t})
	\end{aligned} \label{eqn:finiteHMM}
\end{equation}
for an indexed family of distributions $F(\cdot)$ where $\theta_i$ are the \emph{emission parameters} for mode $i$.

We similarly define a \emph{switching} VAR($r$) process by
\begin{equation}
	\begin{aligned}
		z_t\mid z_{t-1} &\sim \pi_{z_{t-1}}\\
		\BF{y}_t &= \sum_{i=1}^r A_i^{(z_t)}\BF{y}_{t-i} + \BF{e}_t(z_t). 
	\end{aligned}
	\label{eqn:SVAR} 
\end{equation}
%
%Note that the underlying state dynamics of the SLDS are equivalent to a switching VAR($1$) process.
%
\subsection{Dirichlet Processes and the Sticky HDP-HMM}
\label{sec:backgroundDP}
To examine a Bayesian nonparametric SLDS, and thus relax the assumption that the number of dynamical modes is known and fixed, it is useful to first analyze such methods for the simpler HMM. % which can be thought of as a special case of an SLDS with $A^{(k)}=0$ for all $k$.  
One can equivalently represent the finite HMM of Eq.~\eqref{eqn:finiteHMM} via a set of \emph{transition probability measures} $G_j = \sum_{k=1}^K \pi_{jk}\delta_{\uniqueTheta_k}$, where $\delta_{\theta}$ is a mass concentrated at $\theta$. %Instead of employing \emph{transition distributions} on the set of integers (i.e., modes) which index into the collection of emission parameters, 
We then operate directly in the parameter space $\Theta$ and transition between emission parameters with probabilities given by $\{G_j\}$.  That is,
\begin{equation}
	\begin{aligned}
		\obsTheta_t \mid \obsTheta_{t-1} &\sim G_{j : \obsTheta_{t-1}=\uniqueTheta_j}\\
		y_t \mid \obsTheta_t &\sim F(\obsTheta_t). 
	\end{aligned} \label{eqn:finiteHMMmeas}
\end{equation}
Here, $\obsTheta_t \in \{\uniqueTheta_1,\dots,\uniqueTheta_K\}$ and is equivalent to $\uniqueTheta_{z_t}$ of Eq.~\eqref{eqn:finiteHMM}.  A Bayesian nonparametric HMM takes $G_j$ to be \emph{random}\footnote{Formally, a random measure on a measurable space $\Theta$ with sigma algebra $\mathcal{A}$ is defined as a stochastic process whose index set is $\mathcal{A}$. That is, $G(A)$ is a random variable for each $A\in\mathcal{A}$.} with an infinite collection of atoms corresponding to the infinite HMM mode space.

The \emph{Dirichlet process} (DP), denoted by $\DP{\gamma}{H}$, provides a distribution over discrete probability measures with an infinite collection of atoms
\begin{align}
	G_0 = \sum_{k=1}^\infty \beta_{k}\delta_{\theta_k} \hspace{0.25in} \theta_k \sim H,
	\label{eqn:stickDP}
\end{align}
on a parameter space $\Theta$. The weights are sampled via a \emph{stick-breaking construction} \cite{Sethuraman:94}:
\begin{align}
\beta_k = \nu_k\prod_{\ell=1}^{k-1}(1-\nu_\ell) \qquad
\nu_k \sim \mbox{Beta}(1,\gamma).
\end{align}
In effect, we have divided a unit-length stick into lengths given by the weights $\beta_k$: the $k^{th}$ weight is a random proportion $\nu_k$ of the remaining stick after the previous $(k-1)$ weights have been defined. We denote this distribution by $\beta \sim \mbox{GEM}(\gamma)$.

The Dirichlet process has proven useful in many applications due to its clustering properties, which are clearly seen by examining the \emph{predictive distribution} of draws $\obsTheta_i \sim G_0$. Because probability measures drawn from a Dirichlet process are discrete, there is a strictly positive probability of multiple observations $\obsTheta_i$ taking identical values within the set $\{\uniqueTheta_k\}$, with $\uniqueTheta_k$ defined as in Eq.~\eqref{eqn:stickDP}. For each value $\obsTheta_i$, let $z_i$ be an indicator random variable that picks out the unique value $\uniqueTheta_k$ such that $\obsTheta_i = \uniqueTheta_{z_i}$. Blackwell and MacQueen~\cite{Blackwell:73} introduced a P\'{o}lya urn representation of the $\obsTheta_i$:
\begin{align}
\obsTheta_{i} \mid \obsTheta_1, \dots, \obsTheta_{i-1} &\sim
\frac{\gamma}{\gamma + i-1}H + \sum_{j=1}^{i-1} \frac{1}{\gamma +
i-1}\delta_{\obsTheta_j} = \frac{\gamma}{\gamma + i-1}H + \sum_{k=1}^{K}
\frac{n_k}{\gamma + i-1}\delta_{\uniqueTheta_k}.
\label{eqn:PolyaUrn}
\end{align}
Here, $n_k$ is the number of observations $\obsTheta_i$ taking the value $\uniqueTheta_k$. From Eq.~\eqref{eqn:PolyaUrn}, and the discrete nature of $G_0$, we see a reinforcement property of the Dirichlet process that induces sparsity in the number of inferred mixture components.

\begin{figure}[t]
  \centering
  \begin{tabular}{cc}
  \includegraphics[height=1.75in]{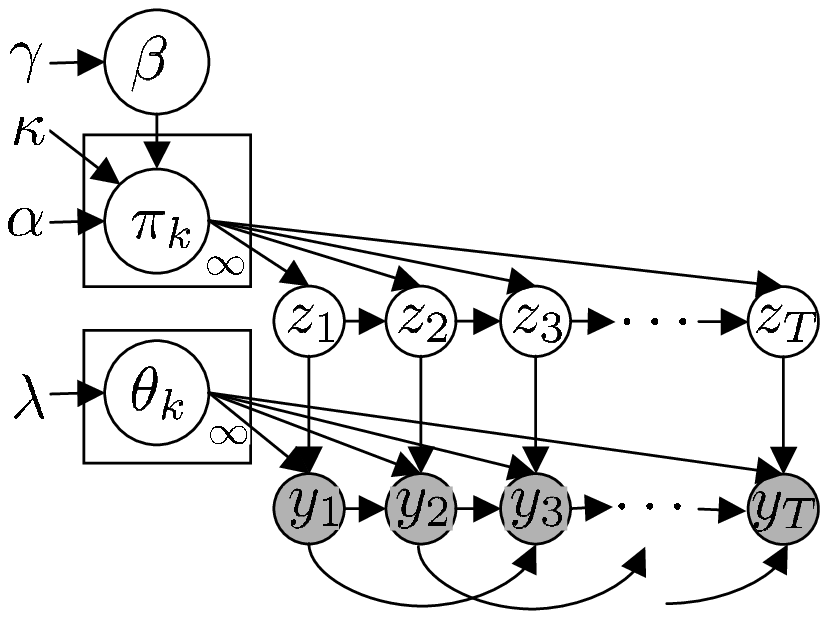} & \includegraphics[height=1.75in]{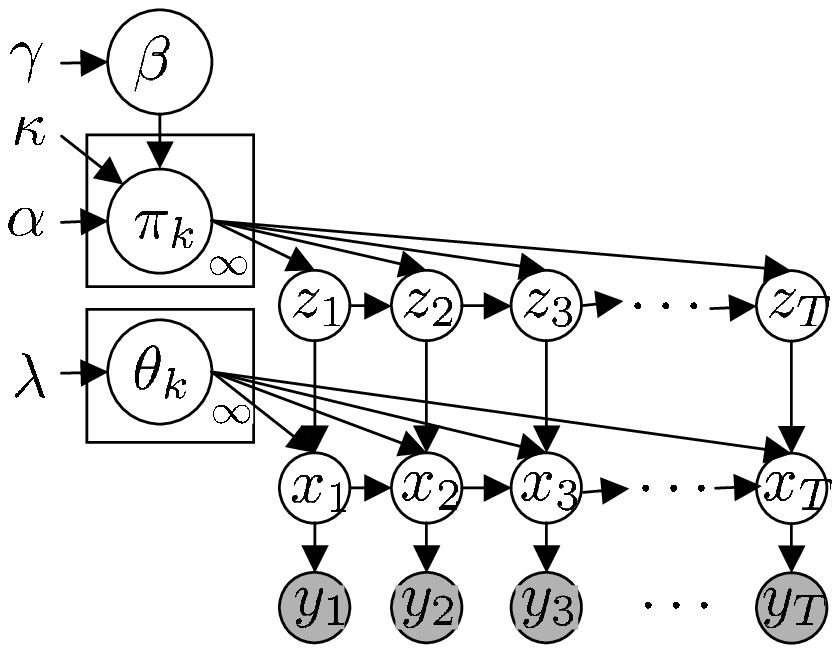} \\%\vspace{-0.1in}\\
  (a) & (b)
  %\vspace{-0.1in}
  \end{tabular}
 \caption{{\small Sticky HDP-HMM prior on (a) switching VAR(2) and (b) SLDS processes with the mode evolving as $z_{t+1}|\{\pi_k\}_{k=1}^{\infty},z_t \sim \pi_{z_t}$ for $\pi_k|\alpha,\kappa,\beta \sim \mbox{DP}(\alpha+\kappa, (\alpha\beta+\kappa\delta_k)/(\alpha+\kappa))$. Here, $\beta\mid \gamma \sim \hbox{GEM}(\gamma)$ and $\theta_k\mid H,\lambda \sim H(\lambda)$. The dynamical processes are as in Table~\ref{table:models}.}} \label{fig:DPMM}
%\vspace{-0.2in}
\end{figure}

A hierarchical extension of the Dirichlet process, the hierarchical Dirichlet process (HDP)~\cite{Teh:06}, has proven useful in defining a prior on the set of HMM transition probability measures $G_j$. The HDP defines a collection of probability measures $\{G_j\}$ on the same support points $\{\uniqueTheta_1,\uniqueTheta_2, \dots\}$ by assuming that each discrete measure $G_j$ is a variation on a global discrete measure $G_0$. Specifically, the Bayesian hierarchical specification takes $G_j\sim \DP{\alpha}{G_0}$, with $G_0$ itself a draw from a Dirichlet process $\DP{\gamma}{H}$. Through this construction, one can show that the probability measures are described as
\begin{align}
	\begin{array}{ll}
			\begin{array}{ll}
				G_0 = \sum_{k=1}^\infty \beta_k \delta_{\uniqueTheta_k}
				\hspace{0.25in} & \beta\mid \gamma \sim \mbox{GEM}(\gamma)\\
				G_j = \sum_{k=1}^\infty \pi_{jk} \delta_{\uniqueTheta_{k}}
				\hspace{0.25in} & \pi_j\mid \alpha,\beta \sim \DP{\alpha}{\beta}
			\end{array}
			& \uniqueTheta_k \mid H \sim H.
		\end{array}
\label{eqn:HDPMM}
\end{align}
Applying the HDP prior to the HMM, we obtain the \emph{HDP-HMM} of Teh et. al.~\cite{Teh:06}. This corresponds to the model in Fig.~\ref{fig:DPMM}(a), but without the edges between the observations.
%See Fig.~\ref{fig:DPMM}(a), ignoring the edges between the nodes representing the observations.
%

By defining $\pi_j \sim \DP{\alpha}{\beta}$, the HDP prior encourages modes to have similar transition distributions. Namely, the mode-specific transition distributions are \emph{identical} in expectation:
%\footnote{In addition, the mean of these distributions, $\beta$, has, in expectation, a monotonically decreasing set of weights due to properties of its stick-breaking construction.}:
%
\begin{align}
\mathbb{E}[\pi_{jk}\mid \beta]=\beta_k. \label{eqn:meanHDPHMM}
\end{align}
However, it does not differentiate self--transitions from moves between modes. When modeling dynamical processes with mode persistence, the flexible
nature of the HDP-HMM prior allows for mode sequences with unrealistically fast dynamics to have large posterior probability. Recently, it has been shown~\cite{Fox:ICML08} that one may mitigate this problem by instead considering a \emph{sticky} HDP-HMM where $\pi_j$ is distributed as follows:
\begin{equation}
%\beta &\sim& \mbox{GEM}(\gamma)\nonumber\\
\pi_j\mid \beta,\alpha,\kappa \sim \mbox{DP}\left(\alpha + \kappa,
 \frac{\alpha\beta + \kappa\delta_j}{\alpha + \kappa}\right).
\label{eqn:stickyHDPHMM}
\end{equation}
Here, $(\alpha\beta + \kappa\delta_j)$ indicates that an amount $\kappa > 0 $ is added to the $j^{th}$ component of $\alpha\beta$. This construction increases the expected probability of self-transition by an amount proportional to $\kappa$. Specifically, the expected set of weights for transition distribution $\pi_j$ is a convex combination of those defined by $\beta$ and mode-specific weight defined by $\kappa$:
\begin{align}
\mathbb{E}[\pi_{jk}\mid \beta,\alpha,\kappa]=\frac{\alpha}{\alpha +
\kappa}\beta_k + \frac{\kappa}{\alpha + \kappa}\delta(j,k).
\end{align}
When $\kappa=0$ the original HDP-HMM of Teh et. al.~\cite{Teh:06} is recovered. We place a prior on $\kappa$ and learn the self-transition bias from the data.
\section{The HDP-SLDS and HDP-AR-HMM} \label{sec:models}
We now consider a significant extension of the sticky HDP-HMM for both SLDS and VAR modeling, capturing dynamic structure underlying the observations by allowing switching among unknown number of unknown dynamics using Bayesian nonparametric methods to capture these uncertainties (and to allow both learning the number of modes and estimating system state).  Fig.~\ref{fig:DPMM}(b) illustrates the \emph{HDP-SLDS} model, while Fig.~\ref{fig:DPMM}(a) illustrates the \emph{HDP-AR-HMM} model (for the case of VAR(2)).  The generative processes for these two models are summarized in Table~\ref{table:models}.
%
%For greater modeling flexibility within the Bayesian framework, we take a nonparametric approach in defining the mode space of our switching dynamical processes. Specifically, we develop extensions of the sticky HDP-HMM for both the SLDS and switching VAR models.
% MJ: this was already noted earlier.
%The HDP-HMM's assumption of
%conditionally independent emissions given the mode sequence is often
%insufficient for capturing how previous emissions influence future
%emissions.
%For the SLDS, we consider conditionally-dependent emissions of which only noisy observations are available (see Fig.~\ref{fig:DPMM}(d).) We refer to this model as the \emph{HDP-SLDS}. The switching VAR($r$) process, with $r$ denoting the autoregressive order, can similarly be posed using an HDP-HMM in which the observations are modeled as conditionally VAR($r$). This model is referred to as the \emph{HDP-AR-HMM} and is depicted in Fig.~\ref{fig:DPMM}(c). The generative processes for these two models are summarized in Table~\ref{table:models}.
%
\begin{table}
	\centering
	\begin{tabular}{|l|l|l|}
		\hline & HDP-AR-HMM & HDP-SLDS\\
		\hline Mode dynamics & $z_t\mid z_{t-1} \sim \pi_{z_{t-1}}$ & $z_t\mid z_{t-1} \sim \pi_{z_{t-1}}$\\
		Observation dynamics & $\BF{y}_t = \sum_{i=1}^r A_i^{(z_t)}\BF{y}_{t-i} + \BF{e}_t(z_t)$ & $\BF{x}_t = A^{(z_t)}\BF{x}_{t-1}+\BF{e}_t(z_t)$\\
		& & $\BF{y}_t = C\BF{x}_t + \BF{w}_t$\\
		\hline 
	\end{tabular}
	\caption{Dynamic equations for the HDP-AR-HMM and HDP-SLDS.  Here, $\pi_j$ is as defined in Eq.~\eqref{eqn:stickyHDPHMM} for the sticky HDP-HMM.  The additive noise processes are distributed as $\BF{e}_t(k) \sim \mathcal{N}(0,\symk{\Sigma}{k})$ and $\BF{w}_t \sim \mathcal{N}(0,R)$.}\label{table:models} \vspace{-0.3in}
\end{table}
%
%
%%
%\begin{figure*}[t] \centering 
%	\begin{tabular}{cc} \hspace{-0.0in}
%		\includegraphics[height=1.7in]{\figdir/HDP-AR2} & \hspace{-0.0in}
%		\includegraphics[height=1.75in]{\figdir/HDP-SLDS}\\
%		\hspace{-0.0in} (a) & \hspace{-0.0in} (b)		
%		%  \vspace{-0.1in}
%	\end{tabular}
%	\caption[Graphical models of the HDP-SLDS and an order two HDP-AR-HMM.] {Graphical models of sticky HDP-HMM prior on switching (a) VAR(2) and (b) SLDS processes with the mode evolving as $z_{t+1}|\{\pi_k\}_{k=1}^{\infty},z_t \sim \pi_{z_t}$ for $\pi_k|\alpha,\kappa,\beta \sim \mbox{DP}(\alpha+\kappa, (\alpha\beta+\kappa\delta_k)/(\alpha+\kappa))$. Here, $\beta\mid \gamma \sim \hbox{GEM}(\gamma)$ and $\theta_k\mid H,\lambda \sim H(\lambda)$. The dynamical processes are as in Eq.~\eqref{eqn:models}.} \label{fig:HDPSLDS} 
%\end{figure*}
%%

For the HDP-SLDS, we place priors on the \emph{dynamic parameters} $\{\symk{A}{k},\symk{\Sigma}{k}\}$ and on measurement noise $R$ and infer their posterior from the data. However, without loss of generality\footnote{This is, in essence, an issue of choosing a similarity transformation for the state of a minimal system, exploiting the fact that the measurement matrix is shared by all modes of the HDP-SLDS so that the same transformation can be used for all modes.}, we fix the measurement matrix to $C=[I_d \,\, 0]$ implying that it is the first $d$ components of the state that are measured. Our choice of the state dimension $n$ is, in essence, a choice of model order, and an issue we address in Sec.~\ref{sec:ARD}. For the HDP-AR-HMM, we similarly place a prior on the dynamic parameters, which in this case consist of $\{A_{1}^{(k)},\dots,A_{r}^{(k)},\symk{\Sigma}{k}\}$.

%
%We do, however, fix the measurement matrix, $C$, for reasons of identifiability. For every SS realization, there exists a set of equivalent systems in terms of the input-output relationship. Let $\tilde{C}\in \mathbb{R}^{d \times n}$, $n\geq d$, be the measurement matrix associated with a dynamical system defined by $\tilde{A}$ and $\tilde{G}$, where $\tilde{G} = \tilde{A}\tilde{P}_x\tilde{C}^T$ with $\tilde{P}_x$ the steady-state covariance matrix. Assume that $\tilde{C}$ has full row rank. Then, without loss of generality, we may consider $C=[I_d \ \ 0]$ since there exists an invertible transformation $T$ such that the triplet $(A = T^{-1}\tilde{A}T, C = \tilde{C}T = [I_d \ \ 0], G = T^{-1}\tilde{G})$ is in the equivalence class defined by $(\tilde{A},\tilde{C},\tilde{G})$. Since the measurement matrix is shared for all modes of the HDP-SLDS, the similarity transformation $T$ is identical for all modes, implying that the change-in-basis of our state vector remains consistent over time. Our choice of the number of columns of zeros in $C$ is, in essence, a choice of model order, and one which we address in Sec.~\ref{sec:ARD}.

In Sec.~\ref{sec:GibbsHDPSLDS} we derive a Gibbs sampling inference scheme for our models.  There is, of course, a difference between the steps required for SLDS-based model (in which there is an unobserved continuous-valued state $\BF{x}_t$) and the AR-based model.  In particular, for the HDP-SLDS the algorithm iterates among the following steps:
% The specific choices of priors we explore, and the resulting posterior distributions conditioned on a set of observations, are described in Sec.~\ref{sec:posterior}. The Gibbs sampling inference scheme derived in Sec.~\ref{sec:GibbsHDPSLDS} iterates between the following steps for the HDP-SLDS:
%
\begin{enumerate}
	\item Sample the state sequence $\BF{x}_{1:T}$ given the mode sequence $z_{1:T}$ and SLDS parameters $\{\symk{A}{k},\symk{\Sigma}{k},R\}$. 
	\item Sample the mode sequence $z_{1:T}$ given the state sequence $\BF{x}_{1:T}$, HMM parameters $\{\pi_k\}$, and dynamic parameters $\{\symk{A}{k},\symk{\Sigma}{k}\}$. 
	\item Sample the HMM parameters $\{\pi_k\}$ and SLDS parameters $\{\symk{A}{k},\symk{\Sigma}{k},R\}$ given the sequences, $z_{1:T}$, $\BF{x}_{1:T}$, and $\BF{y}_{1:T}$. 
\end{enumerate}
For the HDP-AR-HMM, step (1) does not exist. Step (2) then involves sampling the mode sequence $z_{1:T}$ given the observations $\BF{y}_{1:T}$ (rather than $\BF{x}_{1:T}$), and step (3) involves conditioning solely on the sequences $z_{1:T}$ and $\BF{y}_{1:T}$ (not $\BF{x}_{1:T}$).  Also, we note that step (2) involves a fairly straightforward extension of the sampling method developed in~\cite{Fox:ICML08} for the simpler HDP-HMM model; the other steps, however, involve new constructs, as they involve capturing and dealing with the temporal dynamics of the underlying continuous state models.  Sec.~\ref{sec:posterior} provides the necessary priors and structure of the posteriors needed to develop these steps.
%The sampler for the HDP-AR-HMM reuses many steps of the HDP-SLDS sampler, except for directly sampling the mode sequence $z_{1:T}$ and dynamic parameters based on the \emph{observations} $\BF{y}_{1:T}$ (whereas the HDP-SLDS relied on a hidden, sampled state sequence $\BF{x}_{1:T}$ for these steps.) 
%A block diagram depicting the connections between these samplers is shown in Fig.~\ref{fig:samplerBlockDiagram}.
%%
%\begin{figure*}[t] \centering \hspace{-0.0in}
%	\includegraphics[height=2.25in]{\figdir/samplerBlockDiagram} \caption[Block diagram of one iteration of the Gibbs sampler for the HDP-SLDS and HDP-AR-HMM.]{ Block diagram of one iteration of the Gibbs sampler for the HDP-SLDS and HDP-AR-HMM. Assume there exists a previous sample of model parameters. Based on the model type, the appropriate switch closes. If HDP-SLDS, there is an extra stage of sampling the latent, continuous state sequence $\BF{x}_{1:T}$. This sequence then becomes the pseudo-observations. If HDP-AR-HMM, the pseudo-observations are simply the original observations $\BF{y}_{1:T}$. The pseudo-observations are then used to block sample the mode sequence $z_{1:T}$. Subsequently, a new set of model parameters are sampled conditioned on the mode sequence and pseudo-observations.} \label{fig:samplerBlockDiagram} 
%\end{figure*}
%%
%
\subsection{Priors and Posteriors of Dynamic Parameters} \label{sec:posterior}
We begin by developing a prior to regularize the learning of the dynamic parameters (and measurement noise) conditioned on a fixed mode assignment $z_{1:T}$. %The joint learning of the number of modes and resampling the mode sequence $z_{1:T}$ follows as a straightforward extension of the sampling for the sticky HDP-HMM~\cite{Fox:ICML08}.
To make the connections between the samplers for the HDP-SLDS and HDP-AR-HMM explicit, we introduce the concept of \emph{pseudo-observations} $\BF{\psi}_{1:T}$ and rewrite the dynamic equation for both the HDP-SLDS and HDP-AR-HMM generically as
\begin{align}
	\BF{\psi}_t = \bigk{A}{k}\BF{\bar{\psi}}_{t-1} + \BF{e_t}, \label{eqn:psi} 
\end{align}
where we utilize the definitions outlined in Table~\ref{table:rewritten_models}.
\begin{table}
	\centering
	\begin{tabular}{|l|l|l|}		
		\hline & HDP-AR-HMM & HDP-SLDS\\
		\hline Dynamic matrix & $\bigk{A}{k} = [A^{(k)}_1 \dots A^{(k)}_r] \in \RR^{d \times (d*r)}$ & $\bigk{A}{k} = A^{(k)} \in \RR^{n \times n}$\\
		Pseudo-observations & $\BF{\psi_t} = \BF{y}_t$ & $\BF{\psi_t} = \BF{x}_t$\\
		Lag pseudo-observations & $\BF{\bar{\psi}}_t = [\BF{y}^T_{t-1} \dots \BF{y}^T_{t-r}]^T$ & $\BF{\bar{\psi}}_t = \BF{x}_{t-1}$.\\
		\hline 
	\end{tabular}
	\caption{Notational conveniences used in describing the Gibbs sampler for the HDP-AR-HMM and HDP-SLDS.}\label{table:rewritten_models} \vspace{-0.3in}
\end{table}

For the HDP-AR-HMM, we have simply written the dynamic equation in Table~\ref{table:models} in matrix form by concatenating the lag matrices into a single matrix $\bigk{A}{k}$ and forming a \emph{lag observation vector} $\BF{\bar{\psi}}_t$ comprised of a series of previous observation vectors. For this section (for the HDP-SLDS), we assume such a sample of the state sequence $\BF{x}_{1:T}$ (and hence $\{\BF{\psi}_t,\BF{\bar{\psi}}_t\}$) is available so that Eq.~\eqref{eqn:psi} applies equally well to both the HDP-SLDS and the HDP-AR-HMM.  Methods for resampling this state sequence are discussed in Sec.~\ref{sec:GibbsHDPSLDS}. 

Conditioned on the mode sequence, one may partition this dynamic sequence into $K$ different linear regression problems, where $K=|\{z_1,\dots,z_T\}|$. That is, for each mode $k$, we may form a matrix $\bigk{\Psi}{k}$ with $n_k$ columns consisting of the $\BF{\psi}_t$ with $z_t=k$. Then,
\begin{align}
	\bigk{\Psi}{k} &= \bigk{A}{k}\bigk{\Ylag{\Psi}}{k} + \bigk{E}{k}, \label{eqn:Psi} 
\end{align}
where $\bigk{\Ylag{\Psi}}{k}$ is a matrix of the associated $\BF{\bar{\psi}}_{t-1}$, and $\bigk{E}{k}$ the associated noise vectors.
\subsubsection{Conjugate Prior on $\{\bigk{A}{k},\symk{\Sigma}{k}\}$}
The \emph{matrix-normal inverse-Wishart} (MNIW) prior~\cite{West} is conjugate to the likelihood model defined in Eq.~\eqref{eqn:Psi} for the parameter set $\{\bigk{A}{k},\symk{\Sigma}{k}\}$. Although this prior is typically used for inferring the parameters of a single linear regression problem, it is equally applicable to our scenario since the linear regression problems of Eq.~\eqref{eqn:Psi} are independent conditioned on the mode sequence $z_{1:T}$. We note that although the MNIW prior does not enforce stability constraints on each mode, this prior is still a reasonable choice since each mode need not have stable dynamics for the SLDS to be stable~\cite{Costa}, and conditioned on data from a stable mode, the posterior distribution will likely be sharply peaked around stable dynamic matrices.

Let $\bigk{D}{k} = \{\bigk{\Psi}{k},\bigk{\Ylag{\Psi}}{k}\}$. The posterior distribution of the dynamic parameters for the $k^{th}$ mode decomposes as
\vspace{-0.2in}
\begin{align}
	p(\bigk{A}{k},\symk{\Sigma}{k} \mid \bigk{D}{k}) = p(\bigk{A}{k} \mid \symk{\Sigma}{k},\bigk{D}{k})p(\symk{\Sigma}{k} \mid \bigk{D}{k}). 
\end{align}
The resulting posterior of $\bigk{A}{k}$ is straightforwardly derived to be (see~\cite{Fox:PhD}) %in Appendix~\ref{app:conjugatePrior} to be 
\begin{align}
	p(\bigk{A}{k} \mid \symk{\Sigma}{k},\bigk{D}{k}) =\MN{\bigk{A}{k}}{\bigksub{S}{k}{\psi\Ylag{\psi}}\bigminksub{S}{k}{\Ylag{\psi}\Ylag{\psi}}}{\symk{\Sigma}{k}}{\bigksub{S}{k}{\Ylag{\psi}\Ylag{\psi}}}, 
\end{align}
with $\BF{B}^{-(k)}$ denoting $(\BF{B}^{(k)})^{-1}$ for a given matrix $\BF{B}$, $\MN{A}{M}{K}{V}$ denoting a matrix-normal distribution with mean matrix $M$ and left and right covariances $K$ and $V$, and 
\begin{equation}
	\begin{aligned}
		\bigksub{S}{k}{\Ylag{\psi}\Ylag{\psi}} = \bigk{\Ylag{\Psi}}{k}\bigkT{\Ylag{\Psi}}{k} + K \hspace{0.25in} \bigksub{S}{k}{\psi\Ylag{\psi}} = \bigk{\Psi}{k}\bigkT{\Ylag{\Psi}}{k} + MK \hspace{0.25in} \bigksub{S}{k}{\psi\psi} = \bigk{\Psi}{k}\bigkT{\Psi}{k} + MKM^T. 
	\end{aligned}
\end{equation}
The marginal posterior of $\symk{\Sigma}{k}$ is
\begin{align}
	p(\symk{\Sigma}{k} \mid \bigk{D}{k}) = \mbox{IW}\left(n_k + n_0,\bigksub{S}{k}{\psi|\Ylag{\psi}}+S_0\right), \label{eqn:IWposterior_MNIWprior_HDPSLDS} 
\end{align}
where $\mbox{IW}(n_0,S_0)$ denotes an inverse-Wishart prior with $n_0$ degrees of freedom and scale matrix $S_0$, and is updated by data terms $\bigksub{S}{k}{\psi|\Ylag{\psi}} = \bigksub{S}{k}{\psi\psi} - \bigksub{S}{k}{\psi\Ylag{\psi}}\bigminksub{S}{k}{\Ylag{\psi}\Ylag{\psi}}\bigkTsub{S}{k}{\psi\Ylag{\psi}}$ and $n_k = |\{t \mid z_t =k,t=1,\dots,T\}|$.
\subsubsection{Alternative Prior --- Automatic Relevance Determination} \label{sec:ARD}
The MNIW prior leads to full $\bigk{A}{k}$ matrices, which (i) becomes problematic as the model order grows in the presence of limited data; and (ii) does not provide a method for identifying irrelevant model components (i.e. state components in the case of the HDP-SLDS or lag components in the case of the HDP-AR-HMM.) To jointly address these issues, we alternatively consider \emph{automatic relevance determination} (ARD) \cite{MacKay:94,Neal:96,Beal:03}, which encourages driving components of the model parameters to zero if their presence is not supported by the data.

For the HDP-SLDS, we harness the concepts of ARD by placing independent, zero-mean, spherically symmetric Gaussian priors on the columns of the dynamic matrix $\bigk{A}{k}$:
\begin{align}
	p(\bigk{A}{k} | \bigk{\alpha}{k}) 	
	%= \prod_{j=1}^d p(\BF{a}^{(k)}_j |\alpha^{(k)}_j )
	= \prod_{j=1}^n \mathcal{N}\left(\BF{a}^{(k)}_j; 0,\alpha^{-(k)}_jI_n\right). \label{eqn:ARD_SLDS} 
\end{align}
Each precision parameter $\alpha^{(k)}_j$ is given a $\mbox{Gamma}(a,b)$ prior. The zero-mean Gaussian prior penalizes non-zero columns of the dynamic matrix by an amount determined by the precision parameters. Iterative estimation of these hyperparameters $\alpha^{(k)}_j$ and the dynamic matrix $\bigk{A}{k}$ leads to $\alpha^{(k)}_j$ becoming large for columns whose evidence in the data is insufficient for overcoming the penalty induced by the prior. Having $\alpha^{(k)}_j\rightarrow \infty$ drives $\BF{a}^{(k)}_j\rightarrow 0$, implying that the $j^{th}$ state component does not contribute to the dynamics of the $k^{th}$ mode. Thus, examining the set of large $\alpha^{(k)}_j$ provides insight into the order of that mode. Looking at the $k^{th}$ dynamical mode alone, having $\BF{a}^{(k)}_j=0$ implies that the realization of \emph{that mode} is not minimal since the associated Hankel matrix
\begin{align}
\mathcal{H} = \begin{bmatrix} C^T & CA^T & \cdots & (CA^{d-1})^T
\end{bmatrix}^T \begin{bmatrix} G & AG & \cdots & A^{d-1}G
\end{bmatrix}\equiv \mathcal{O}\mathcal{R}
\label{eqn:hankel}
\end{align}
has reduced rank. However, the overall SLDS realization may still be minimal.

For our use of the ARD prior, we restrict attention to models satisfying the property that the state components that are observed are relevant to \emph{all} modes of the dynamics:
%In order to ensure that our choice of $C=[I_d \,\, 0]$ does not interfere with learning a sparse realization if one exists, we must constrain the class of dynamical phenomenon we may analyze. %Imagine a realization of an LDS with
%%
%\begin{align*}
%	\tilde{A} = 
%	\begin{bmatrix}
%		0.8 & 0\\
%		0.2 & 0 
%	\end{bmatrix}
%	, \hspace{0.25in} \tilde{C} = 
%	\begin{bmatrix}
%		1 & 1 
%	\end{bmatrix}. 
%\end{align*}
%%
%Then, the transformation to $C=[1 \,\, 0]$ leads to
%%
%\begin{align*}
%	A = T^{-1}\tilde{A}T = 
%	\begin{bmatrix}
%		0.5 & 1\\
%		0.15 & 0.3
%	\end{bmatrix}, \hspace{0.25in} \mbox{for} \hspace{0.25in} 	T=
%		\begin{bmatrix}
%			0.5 & 1\\
%			0.5 & -1 
%		\end{bmatrix}.
%\end{align*}
%
%So, for this example, fixing $C=[1 \,\, 0]$ would not lead to learning a sparse dynamical matrix $A$. 
%Thus, we restrict ourselves to ARD modeling of dynamical phenomena that satisfy the following criterion.  See Appendix~\ref{app:ARD} for further discussion on this criterion and illustrative examples. If there does not exist a realization $\mathcal{R}$ satisfying Criterion~\ref{crit:nested}, we may instead consider a more general model where the measurement equation is mode-specific and we place a prior on $C^{(k)}$ instead of fixing this matrix. However, this model leads to identifiability issues that are considerably less pronounced in the above case.
%
\begin{crit}
	If for some realization $\mathcal{R}$ a mode $k$ has $\BF{a}_j^{(k)}=0$, then that realization must have $\BF{c}_j = 0$, where $\BF{c}_j$ is the $j^{th}$ column of $C$. Here we assume, without loss of generality, that the observed states are the first components of the state vector. \label{crit:nested} 
\end{crit}
This assumption implies that our choice of $C=[I_d \,\, 0]$ does not interfere with learning a sparse realization\footnote{If there does not exist a realization $\mathcal{R}$ satisfying Criterion~\ref{crit:nested}, we may instead consider a more general model where the measurement equation is mode-specific and we place a prior on $C^{(k)}$ instead of fixing this matrix. However, this model leads to identifiability issues that are considerably less pronounced in the above case.}.
The ARD prior may also be used to learn variable-order switching VAR processes. Here, the goal is to ``turn off'' entire \emph{lag blocks} $A^{(k)}_i$ (whereas in the HDP-SLDS we were interested in eliminating columns of the dynamic matrix.) Instead of placing independent Gaussian priors on each column of $\bigk{A}{k}$ as we did in Eq.~\eqref{eqn:ARD_SLDS}, we decompose the prior over the lag blocks $A^{(k)}_i$:
\begin{align}
	p(\bigk{A}{k}| \bigk{\BF{\alpha}}{k}) 
	%= \prod_{i=1}^r p(A^{(k)}_i | \alpha^{(k)}_i )
	= \prod_{i=1}^r \mathcal{N}\left(\vecc{A^{(k)}_i}; 0,\alpha_i^{-(k)}I_{d^2}\right).\label{eqn:ARD_VAR} 
\end{align}
Since each element of a given lag block $\symsubsup{A}{i}{k}$ is distributed according to the same precision parameter $\symsubsup{\alpha}{i}{k}$, if that parameter becomes large the entire lag block will tend to zero.

In order to examine the posterior distribution on the dynamic matrix $\bigk{A}{k}$, it is useful to consider the Gaussian induced by Eq.~\eqref{eqn:ARD_SLDS} and Eq.~\eqref{eqn:ARD_VAR} on a vectorization of $\bigk{A}{k}$. Our ARD prior on $\bigk{A}{k}$ is equivalent to a $\mathcal{N}(0,\symk{\Sigma_0}{k})$ prior on $\vecc{\bigk{A}{k}}$, where
\begin{align}
	\symk{\Sigma_0}{k} = \mbox{diag}\left(\symk{\alpha_1}{k},\dots,\symk{\alpha_1}{k},\dots, \symk{\alpha_m}{k},\dots,\symk{\alpha_m}{k}\right)^{-1}. \label{eqn:Sigma0} 
\end{align}
%
%Although this
%prior is not conjugate, the sampler of Sec.~\ref{sec:sampler}
%instantiates parameters rather than integrating over them, so
%conjugacy is not necessary.
Here, $m=n$ for the HDP-SLDS with $n$ replicates of each $\symk{\alpha_i}{k}$, and $m=r$ for the HDP-AR-HMM with $d^2$ replicates of $\symk{\alpha_i}{k}$. (Recall that $n$ is the dimension of the HDP-SLDS state vector $\BF{x}_t$, $r$ the autoregressive order of the HDP-AR-HMM, and $d$ the dimension of the observations $\BF{y}_t$.) To examine the posterior distribution of $\bigk{A}{k}$, we note that we may rewrite the state equation as,
\begin{align}
	\BF{\psi}_{t+1} &= \left[
	\begin{array}{cccc}
		\BF{\bar{\psi}}_{t,1}I_\ell & \BF{\bar{\psi}}_{t,2}I_\ell & \cdots & \BF{\bar{\psi}}_{t,\ell*r}I_\ell
	\end{array}
	\right] \vecc{\bigk{A}{k}} + \BF{e}_{t+1}(k) \hspace{0.25in} \forall t|z_t=k\nonumber\\
	&\triangleq \tilde{\Psi}_{t} \vecc{\bigk{A}{k}} + \BF{e}_{t+1}(k), \label{eqn:rewritePsi} 
\end{align}
where $\ell=n$ for the HDP-SLDS and $\ell=d$ for the HDP-AR-HMM. Using Eq.~\eqref{eqn:rewritePsi}, we derive the posterior distribution as
\begin{align}
	p(\vecc{\bigk{A}{k}} \mid \bigk{D}{k},\symk{\Sigma}{k},\bigk{\BF{\alpha}}{k})
	= \mathcal{N}^{-1}\bigg(\sum_{t|z_t=k}\tilde{\Psi}_{t-1}^T\symmink{\Sigma}{k}\BF{\psi_t}, \symmink{\Sigma_0}{k}+\sum_{t|z_t=k}\tilde{\Psi}_{t-1}^T\symmink{\Sigma}{k}\tilde{\Psi}_{t-1}\bigg). 
\end{align}
See~\cite{Fox:PhD} for a detailed derivation. Here, $\mathcal{N}^{-1}(\vartheta,\Lambda)$ represents a Gaussian $\mathcal{N}(\mu,\Sigma)$ with information parameters $\vartheta = \Sigma^{-1}\mu$ and $\Lambda = \Sigma^{-1}$. Given $\bigk{A}{k}$, and recalling that each precision parameter is gamma distributed, the posterior of $\alpha_\ell^{(k)}$ is given by 
\vspace{-0.1in}
\begin{align}
	p(\alpha_\ell^{(k)} \mid \bigk{A}{k}) = \mbox{Gamma}\left(a+\frac{|\mathcal{S}_\ell|}{2},b+\frac{\sum_{(i,j) \in \mathcal{S}_\ell}a_{ij}^{(k)^2}}{2}\right). \label{eqn:alphaPost} 
\end{align}
The set $\mathcal{S}_\ell$ contains the indices for which $a_{ij}^{(k)}$ has prior precision $\alpha_\ell^{(k)}$. Note that in this model, regardless of the number of observations $\BF{y}_t$, the size of $\mathcal{S}_\ell$ (i.e., the number of $a_{ij}^{(k)}$ used to inform the posterior distribution) remains the same. Thus, the gamma prior is an informative prior and the choice of $a$ and $b$ should depend upon the cardinality of $\mathcal{S}_\ell$. For the HDP-SLDS, this cardinality is given by the maximal state dimension $n$, and for the HDP-AR-HMM, by the square of the observation dimensionality $d^2$.

We then place an inverse-Wishart prior $\mbox{IW}(n_0,S_0)$ on $\symk{\Sigma}{k}$ and look at the posterior given $\bigk{A}{k}$: 
\begin{align}
	p(\symk{\Sigma}{k} \mid \bigk{D}{k}, \bigk{A}{k}) = \mbox{IW}\left(n_k + n_0,\bigksub{S}{k}{\psi|\Ylag{\psi}}+S_0\right), 
\end{align}
where here, as opposed to in Eq.~\eqref{eqn:IWposterior_MNIWprior_HDPSLDS}, we define
\begin{align}
	\bigksub{S}{k}{\psi|\Ylag{\psi}} = \sum_{t|z_t = k}(\BF{\psi}_t-\bigk{A}{k}\BF{\bar{\psi}}_{t-1})(\BF{\psi}_t-\bigk{A}{k}\BF{\bar{\psi}}_{t-1})^T. 
\end{align}
\subsubsection{Measurement Noise Posterior}
For the HDP-SLDS, we additionally place an $\mbox{IW}(r_0,R_0)$ prior on the measurement noise covariance $R$. The posterior distribution is given by 
\begin{align}
	p(R \mid \BF{y}_{1:T},\BF{x}_{1:T}) = \mbox{IW}(T + r_0,S_R+R_0), 
\end{align}
where $S_R = \sum_{t=1}^T(\BF{y}_t-C\BF{x}_t)(\BF{y}_t-C\BF{x}_t)^T$. Here, we assume that $R$ is shared between modes. The extension to mode-specific measurement noise is straightforward. %\footnote{If we wished to consider a model with mode-specific measurement noise, we would simply partition the data according to the mode sequence and examine the posterior $p(R^{(k)} \mid \BF{y}_{1:T},\BF{x}_{1:T},z_{1:T}) = \mbox{IW}\left(N_k + r_0,S^{(k)}_R+R_0\right)$ for each $k$, where $S^{(k)}_R = \sum_{t|z_t = k}(\BF{y}_t-C\BF{x}_t)(\BF{y}_t-C\BF{x}_t)^T$.}.
\subsection{Gibbs Sampler} \label{sec:GibbsHDPSLDS}
For inference in the HDP-AR-HMM, we use a Gibbs sampler that iterates between sampling the mode sequence, $z_{1:T}$, and the set of dynamic and sticky HDP-HMM parameters. The sampler for the HDP-SLDS is identical with the additional step of sampling the state sequence, $\BF{x}_{1:T}$, and conditioning on this sequence when resampling dynamic parameters and the mode sequence. Periodically, we interleave a step that sequentially samples the mode sequence $z_{1:T}$ marginalizing over the state sequence $\BF{x}_{1:T}$ in a similar vein to that of Carter and Kohn~\cite{Carter:96}. We describe the sampler in terms of the pseudo-observations $\BF{\psi}_t$, as defined by Eq.~\eqref{eqn:psi}, in order to clearly specify the sections of the sampler shared by both the HDP-AR-HMM and HDP-SLDS. 
\subsubsection{Sampling Dynamic Parameters $\{\bigk{A}{k},\Sigma^{(k)}\}$}
Conditioned on the mode sequence, $z_{1:T}$, and the pseudo-observations, $\BF{\psi}_{1:T}$, we can sample the dynamic parameters $\BF{\uniqueTheta}\hspace{-1pt} =\hspace{-1pt} \{\bigk{A}{k},\symk{\Sigma}{k}\}$ from the posterior densities of Sec.~\ref{sec:posterior}. For the ARD prior, we then sample $\bigk{\BF{\alpha}}{k}$ given $\bigk{A}{k}$. In practice we iterate multiple times between sampling $\bigk{\BF{\alpha}}{k}$ given $\bigk{A}{k}$ and $\bigk{A}{k}$ given $\bigk{\BF{\alpha}}{k}$ before moving to the next sampling stage.
%(which we recall are equal to the observations,
%$\BF{y}_{1:T}$, in the case of the HDP-AR-HMM or the sampled state
%sequence, $\BF{x}_{1:T}$, in the case of the HDP-SLDS)
%
\subsubsection{Sampling Measurement Noise $R$ (HDP-SLDS only)}
For the HDP-SLDS, we additionally sample the measurement noise covariance $R$ conditioned on the sampled state sequence $\BF{x}_{1:T}$.
\subsubsection{Block Sampling $z_{1:T}$} \label{sec:HDPARHMMzsampling}
As shown in~\cite{Fox:ICML08}, the mixing rate of the Gibbs sampler for the HDP-HMM can be dramatically improved by using a \emph{truncated} approximation to the HDP and jointly sampling the mode sequence using a variant of the forward-backward algorithm. In the case of our switching dynamical systems, we must account for the direct correlations in the observations in our likelihood computation. The variant of the forward-backward algorithm we use here then involves computing backward messages $m_{t+1,t}(z_{t})\propto p(\BF{\psi}_{t+1:T}|z_{t},\BF{\bar{\psi}}_t,\BF{\pi},\BF{\uniqueTheta})$ for each $z_t \in \{1,\dots,L\}$ with $L$ the chosen truncation level, followed by recursively sampling each $z_t$ conditioned on $z_{t-1}$ from
\begin{align}
	p(z_t\mid z_{t-1},\BF{\psi}_{1:T},\BF{\pi},\BF{\uniqueTheta}) \propto p(z_t\mid \pi_{z_{t-1}})p(\BF{\psi}_t\mid \BF{\bar{\psi}}_{t-1},\bigk{A}{z_t},\Sigma^{(z_t)})m_{t+1,t}(z_t). 
\end{align}
Joint sampling of the mode sequence is especially important when the observations are directly correlated via a dynamical process since this correlation further slows the mixing rate of the sequential sampler of Teh et. al.~\cite{Teh:06}. Note that using an order $L$ weak limit approximation to the HDP still encourages the use of a sparse subset of the $L$ possible dynamical modes.
\subsubsection{Block Sampling $\BF{x}_{1:T}$ (HDP-SLDS only)}
Conditioned on the mode sequence $z_{1:T}$ and the set of SLDS parameters $\BF{\uniqueTheta} = \{\bigk{A}{k},\Sigma^{(k)},R\}$, our dynamical process simplifies to a time-varying linear dynamical system. We can then block sample $\BF{x}_{1:T}$ by first running a backward Kalman filter to compute $m_{t+1,t}(\BF{x}_{t})\propto p(\BF{y}_{t+1:T}|\BF{x}_{t},z_{t+1:T},\BF{\uniqueTheta})$ and then recursively sampling each $\BF{x}_t$ conditioned on $\BF{x}_{t-1}$ from
\begin{align}
	p(\BF{x}_t\mid \BF{x}_{t-1},\BF{y}_{1:T},z_{1:T},\BF{\uniqueTheta}) \propto p(\BF{x}_t\mid \BF{x}_{t-1},\symk{A}{z_t},\Sigma^{(z_t)})p(\BF{y}_t\mid \BF{x}_t, R)m_{t+1,t}(\BF{x}_t). 
\end{align}
The messages are given in information form by $m_{t,t-1}(\BF{x}_{t-1})$ $\propto \mathcal{N}^{-1}(\BF{x}_{t-1}; \vartheta_{t,t-1},\Lambda_{t,t-1})$, where the information parameters are recursively defined as
\begin{equation}
	\begin{aligned}
		\vartheta_{t,t-1} &= \symkT{A}{z_t}\symmink{\Sigma}{z_t}\tilde{\Lambda}_t(C^TR^{-1}\BF{y}_t + \vartheta_{t+1,t})\\
		\Lambda_{t,t-1} &= \symkT{A}{z_t}\symmink{\Sigma}{z_t}\symk{A}{z_t} - \symkT{A}{z_t}\symmink{\Sigma}{z_t}\tilde{\Lambda}_t\symmink{\Sigma}{z_t}\symk{A}{z_t}, 
	\end{aligned}
	\label{eq:filterrecursion} 
\end{equation}
with $\tilde{\Lambda}_t = (\symmink{\Sigma}{z_t} + C^TR^{-1}C + \Lambda_{t+1,t})^{-1}$. The standard $\vartheta_{t|t}^b$ and $\Lambda_{t|t}^b$ updated information parameters for a backward running Kalman filter are given by
\begin{align}
	\Lambda^b_{t|t} &= C^TR^{-1}C + \Lambda_{t+1,t}\nonumber\\
	\vartheta^b_{t|t} &= C^TR^{-1}y_t + \vartheta_{t+1,t}. 
	\label{eqn:backwardrecursions}
\end{align}
See~\cite{Fox:PhD} for a derivation and for a more numerically stable version of this recursion.
\subsubsection{Sequentially Sampling $z_{1:T}$ (HDP-SLDS only)} 
For the HDP-SLDS, iterating between the previous sampling stages can lead to slow mixing rates since the mode sequence is sampled conditioned on a sample of the state sequence. For high-dimensional state spaces $\RR^n$, this problem is exacerbated. Instead, one can analytically marginalize the state sequence and sequentially sample the mode sequence from $p(z_t\mid z_{\backslash t}, \BF{y}_{1:T},\BF{\pi},\BF{\theta})$. %\footnote{Note that the instantiated parameter values are used in this sequential sampling, and the resampling of these parameters relies on the sampled state sequence so that one cannot simply iterate between sampling the mode sequence and parameters.} 
This marginalization is accomplished by once again harnessing the fact that conditioned on the mode sequence, our model reduces to a time-varying linear dynamical system. When sampling $z_t$ and conditioning on the mode sequence at all \emph{other} time steps, we can run a forward Kalman filter to marginalize the state sequence $\BF{x}_{1:t-2}$ producing $p(\BF{x}_{t-1}\mid y_{1:t-1},z_{1:t-1},\BF{\theta})$, and a backward filter to marginalize $\BF{x}_{t+1:T}$ producing $\mbox{$p(\BF{y}_{t+1:T}\mid x_{t},z_{t+1:T},\BF{\theta})$}$. Then, for each possible value of $z_t$, we combine these forward and backward messages with the local likelihood $p(\BF{y}_t \mid \BF{x}_t)$ and local dynamic $\mbox{$p(\BF{x}_{t} \mid \BF{x}_{t-1},\BF{\theta},z_t=k)$}$ and marginalize over $\BF{x}_{t}$ and $\BF{x}_{t-1}$ resulting in the likelihood of the observation sequence $\BF{y}_{1:T}$ as a function of $\BF{z}_t$. This likelihood is combined with the prior probability of transitioning from $z_{t-1}$ to $z_t=k$ and from $z_t=k$ to $z_{t+1}$. The resulting distribution is given by:
\begin{multline}
	p(z_t = k\mid z_{\backslash t}, \BF{y}_{1:T},\BF{\pi},\BF{\uniqueTheta}) \propto \pi_{z_{t-1}}(k)\pi_k(z_{t+1})\\
	\frac{|\Lambda_t^{(k)}|^{1/2}}{|\Lambda_t^{(k)}+\Lambda_{t|t}^b|^{1/2}}\exp\left(-\frac{1}{2}\vartheta_t^{(k)^T}\Lambda_t^{-(k)}\vartheta_t^{(k)} + \frac{1}{2}(\vartheta_t^{(k)}+\vartheta_{t|t}^b)^T(\Lambda_t^{(k)}+\Lambda_{t|t}^b)^{-1}(\vartheta_t^{(k)}+\vartheta_{t|t}^b)\right) 
\end{multline}
with
\begin{equation}
	\begin{aligned}
		\Lambda_t^{(k)} &= (\symk{\Sigma}{k} + \bigk{A}{z_t}\Lambda_{t-1|t-1}^{-f}\bigkT{A}{z_t})^{-1}\\
		\vartheta_t^{(k)} &= (\symk{\Sigma}{k} + \bigk{A}{z_t}\Lambda_{t-1|t-1}^{-f}\bigkT{A}{z_t})^{-1} \bigk{A}{z_t}\Lambda_{t-1|t-1}^{-f}\vartheta_{t-1|t-1}^f. 
	\end{aligned}
	\label{eqn:Lambda_tk} 
\end{equation}
See~\cite{Fox:PhD} for full derivations. Here, $\vartheta_{t|t}^f$ and $\Lambda_{t|t}^f$ are the updated information parameters for a forward running Kalman filter, defined recursively as
\begin{align}
	\Lambda^f_{t|t} &= C^TR^{-1}C + \symmink{\Sigma}{z_t} - \symmink{\Sigma}{z_t}\bigk{A}{z_t}(\bigkT{A}{z_t}\symmink{\Sigma}{z_t}\bigk{A}{z_t} + \Lambda^f_{t-1|t-1})^{-1}\bigkT{A}{z_t}\symmink{\Sigma}{z_t}\nonumber\\
	\vartheta^f_{t|t} &= C^TR^{-1}y_t + \symmink{\Sigma}{z_t}\bigk{A}{z_t}(\bigkT{A}{z_t}\symmink{\Sigma}{z_t}\bigk{A}{z_t} + \Lambda^f_{t-1|t-1})^{-1}\vartheta^f_{t-1|t-1}. 
	\label{eqn:forwardrecursions}
\end{align}
Note that a sequential node ordering for this sampling step allows for efficient updates to the recursively defined filter parameters. However, this sequential sampling is still computationally intensive, so our Gibbs sampler iterates between blocked sampling of the state and mode sequences many times before interleaving a sequential mode sequence sampling step. 

The resulting Gibbs sampler is outlined in Algorithm~\ref{alg:full}.
\begin{algorithm}
	[htbp] \vspace*{-1pt} \hspace*{-6pt} 
	\begin{flushleft}
		Given a previous set of mode-specific transition probabilities $\BF{\pi}^{(n-1)}$, the global transition distribution $\beta^{(n-1)}$, and dynamic parameters $\BF{\uniqueTheta}^{(n-1)}$: 
		\begin{enumerate}
			%\algtop 
			\item Set $\BF{\pi}=\BF{\pi}^{(n-1)}$, $\BF{\beta}=\BF{\beta}^{(n-1)}$, and $\BF{\uniqueTheta} = \BF{\uniqueTheta}^{(n-1)}$. %\algtop 
			\item If HDP-SLDS, 
			\begin{enumerate}
				%\algtop 
				\item For each $t \in \{1,\ldots,T\}$, compute $\{\vartheta^f_{t|t},\Lambda^f_{t|t}\}$ as in Eq.~\eqref{eqn:forwardrecursions}. %\algtop 
				\item For each $t \in \{T,\ldots,1\}$, 
				\begin{enumerate}
					\item Compute $\{\vartheta^b_{t|t},\Lambda^b_{t|t}\}$ as in Eq.~\eqref{eqn:backwardrecursions}. 
					\item For each $k \in \{1,\dots,L\}$, compute $\{\vartheta_t^{(k)},\Lambda_t^{(k)}\}$ as in Eq.~\eqref{eqn:Lambda_tk} and set \vspace{-0.1in}					
					\begin{multline*}
						f_k(\BF{y}_{1:T}) = |\Lambda_t^{(k)}|^{1/2}|\Lambda_t^{(k)}+\Lambda_{t|t}^b|^{-1/2}\\\exp\left(-\frac{1}{2}\vartheta_t^{(k)^T}\Lambda_t^{-(k)}\vartheta_t^{(k)} + \frac{1}{2}(\vartheta_t^{(k)}+\vartheta_{t|t}^b)^T(\Lambda_t^{(k)}+\Lambda_{t|t}^b)^{-1}(\vartheta_t^{(k)}+\vartheta_{t|t}^b)\right). 
					\end{multline*}					
					\item Sample a mode assignment \vspace{-0.1in}
					\begin{equation*}
						z_t \sim \sum_{k=1}^L \pi_{z_{t-1}}(k)\pi_k(z_{t+1})f_k(\BF{y}_{1:T})\delta(z_t,k). 
					\end{equation*}
				\end{enumerate}
				\algtop 
				\item Working sequentially forward in time sample 
				\begin{align*}
					\BF{x}_t &\sim \mathcal{N}(\BF{x}_t;(\Sigma^{-(z_t)}+\Lambda^b_{t|t})^{-1}(\Sigma^{-(z_t)}A^{(z_t)}\BF{x}_{t-1}+\vartheta^b_{t|t}),(\Sigma^{-(z_t)}+\Lambda^b_{t|t})^{-1}). 
				\end{align*}
				\item Set pseudo-observations $\BF{\psi}_{1:T} = \BF{x}_{1:T}$. 
			\end{enumerate}
			%\algtop 
			\item If HDP-AR-HMM, set pseudo-observations $\BF{\psi}_{1:T} = \BF{y}_{1:T}$. \algtop 
			\item Block sample $z_{1:T}$ given transition distributions $\BF{\pi}$, dynamic parameters $\BF{\theta}$, and pseudo-observations $\BF{\psi}_{1:T}$ as in Algorithm~\ref{alg:blockZsampler}. %\algtop 
			\item Update the global transition distribution $\beta$ (utilizing auxiliary variables $\BF{m}$, $\BF{w}$, and $\BF{\bar{m}}$), mode-specific transition distributions $\pi_k$, and hyperparameters $\alpha$, $\gamma$, and $\kappa$ as in~\cite{Fox:ICML08}. %\algtop			
			%\item Update the global transition distribution by sampling
			%    \begin{equation*}
			%        \beta \sim \mbox{Dir}(\gamma/L + \bar{m}_{.1},\dots,\gamma/L +
			%        \bar{m}_{.\bar{K}}).
			%    \end{equation*}
			%\vspace{-0.25in}
			\item For each $k \in \{1,\dots,L\}$, sample dynamic parameters $(\bigk{A}{k},\Sigma^{(k)})$ given the pseudo-observations $\BF{\psi}_{1:T}$ and mode sequence $z_{1:T}$ as in Algorithm~\ref{alg:MNIW} for the MNIW prior and Algorithm~\ref{alg:ARD} for the ARD prior.			
			%%The ARD algorithm also
			%%relies providing the previous set of dynamic parameters.
			%\begin{enumerate}
			%\algtop
			% \item Compute the transition counts $n_{k\ell} = \sum_{t=1}^{T-1} \delta(z_t,k)\delta(z_{t+1},\ell)$ for each $\ell \in \{1,\dots,K\}$.
			% \algtop
			% \item Sample the mode-specific transition distribution:
			%    \begin{eqnarray*}
			%        \pi_k &\sim& \mbox{Dir}(\alpha\beta_1 + n_{k1},\dots,\alpha\beta_k +
			%        \kappa + n_{kk}, \dots, \alpha\beta_L + n_{kL}),
			%    \end{eqnarray*}
			% \item Sample dynamic parameters $(\bigk{A}{k},\Sigma^{(k)})$ given the
			%pseudo-observations $\BF{\psi}_{1:T}$ and mode sequence $z_{1:T}$ as
			%in Algorithm~\ref{alg:MNIW} for the MNIW prior and
			%Algorithm~\ref{alg:ARD} for the ARD prior.
			%%The ARD algorithm also
			%%relies providing the previous set of dynamic parameters.
			%\end{enumerate}
			%\algtop
			\item If HDP-SLDS, also sample the measurement noise covariance 
			\begin{eqnarray*}
				R &\sim& \mbox{IW}\left(T + r_0,\sum_{t=1}^T(\BF{y}_t-C\BF{x}_t)(\BF{y}_t-C\BF{x}_t)^T+R_0\right). 
			\end{eqnarray*}
			\item Fix $\BF{\pi}^{(n)}=\BF{\pi}$, $\beta^{(n)}=\beta$, and $\BF{\uniqueTheta}^{(n)} = \BF{\uniqueTheta}$. \algend 
		\end{enumerate}
	\end{flushleft}	
	%\begin{singlespace}
	\caption{HDP-SLDS and HDP-AR-HMM Gibbs sampler. \vspace{-0.2in}} \label{alg:full}
	%\end{singlespace}
\end{algorithm}
\begin{algorithm}
	[htbp] \vspace*{-1pt} \hspace*{-6pt} 
	\begin{flushleft}
		Given mode-specific transition probabilities $\BF{\pi}$, dynamic parameters $\BF{\uniqueTheta}$, and pseudo-observations $\BF{\psi}_{1:T}$: 
		\begin{enumerate}
			%\algtop 
			\item Calculate messages $m_{t,t-1}(k)$, initialized to $m_{T+1,T}(k)=1$, and the sample mode sequence $z_{1:T}$: 
			\begin{enumerate}
				%\algtop 
				\item For each $t \in \{T,\ldots,1\}$ and $k\in \{1,\dots,L\}$, compute 
				\begin{equation*}
					m_{t,t-1}(k) = \sum_{j=1}^L \pi_k(j)\mathcal{N}\left(\BF{\psi}_t;\sum_{i=1}^r A^{(j)}_i\BF{\psi}_{t-i},\Sigma^{(j)}\right)m_{t+1,t}(j) 
				\end{equation*}
				\vspace{-4pt} \algend \vspace{0in}
				\item Working sequentially forward in time, starting with transitions counts $n_{jk}=0$: 
				\begin{enumerate}
					\item For each $k\in \{1,\dots,L\}$, compute the probability 
					\begin{equation*}
						f_k(\BF{\psi}_t) = \mathcal{N}\left(\BF{y}_t;\sum_{i=1}^r A^{(k)}_i\BF{\psi}_{t-i},\Sigma^{(k)}\right)m_{t+1,t}(k) 
					\end{equation*}
					\item Sample a mode assignment $z_t$ as follows and increment $n_{z_{t-1}z_t}$: \vspace{-0.1in}
					\begin{equation*}
						z_t \sim \sum_{k=1}^L \pi_{z_{t-1}}(k)f_k(\BF{\psi}_t)\delta(z_t,k) 
					\end{equation*}
					\algend \vspace{-0.1in}
				\end{enumerate}
				Note that the likelihoods can be precomputed for each $k \in \{1,\dots,L\}$. 
			\end{enumerate}
		\end{enumerate}
	\end{flushleft}	
	%\begin{singlespace}
	\caption{Blocked mode-sequence sampler for HDP-AR-HMM or HDP-SLDS. \vspace{-0.2in}} \label{alg:blockZsampler}
	%\end{singlespace}
\end{algorithm}
\begin{algorithm}
	[htbp] \vspace*{-1pt} \hspace*{-6pt} 
	\begin{flushleft}
		Given pseudo-observations $\BF{\psi}_{1:T}$ and mode sequence $z_{1:T}$, for each $k \in \{1,\dots,K\}$: 
		\begin{enumerate}
			%\algtop 
			\item Construct $\bigk{\Psi}{k}$ and $\bigk{\Ylag{\Psi}}{k}$ as in Eq.~\eqref{eqn:Psi}. \algend 
			\item Compute sufficient statistics using pseudo-observations $\BF{\psi}_{t}$ associated with $z_t=k$: 
			\begin{align*}
				\bigksub{S}{k}{\Ylag{\psi}\Ylag{\psi}} &= \bigk{\Ylag{\Psi}}{k}\bigkT{\Ylag{\Psi}}{k} + K \hspace{0.25in}
				\bigksub{S}{k}{\psi\Ylag{\psi}} &= \bigk{\Psi}{k}\bigkT{\Ylag{\Psi}}{k} + MK \hspace{0.25in}
				\bigksub{S}{k}{\psi \psi} &= \bigk{\Psi}{k}\bigkT{\Psi}{k} + MKM^T. 
			\end{align*}
			\item Sample dynamic parameters: 
			\begin{eqnarray*}
				\Sigma^{(k)} \sim \mbox{IW}\left(n_k + n_0,\bigksub{S}{k}{\psi\mid\Ylag{\psi}} +S_0\right) \hspace{0.2in}
				\bigk{A}{k} \mid \Sigma^{(k)} \sim \MN{\bigk{A}{k}}{\bigksub{S}{k}{\psi \Ylag{\psi}}\bigminksub{S}{k}{\Ylag{\psi}\Ylag{\psi}}}{\symk{\Sigma}{k}}{\bigksub{S}{k}{\Ylag{\psi}\Ylag{\psi}}}. 
			\end{eqnarray*}
			\algend \vspace{-0.1in}
		\end{enumerate}
	\end{flushleft}
	%\begin{singlespace}
	\caption{Parameter sampling using MNIW prior.} \label{alg:MNIW}
	%\end{singlespace}
\end{algorithm}
\begin{algorithm}
	[htbp] \vspace*{-1pt} \hspace*{-6pt} 
	\begin{flushleft}
		Given pseudo-observations $\BF{\psi}_{1:T}$, mode sequence $z_{1:T}$, and a previous set of dynamic parameters $(\bigk{A}{k},\Sigma^{(k)},\BF{\alpha}^{(k)})$, for each $k \in \{1,\dots,K\}$: 
		\begin{enumerate}
			%\algtop 
			\item Construct $\tilde{\Psi}_{t}$ as in Eq.~\eqref{eqn:rewritePsi}. 
			\item Iterate multiple times between the following steps: 
			\begin{enumerate}
				\item Construct $\Sigma_0^{(k)}$ given $\BF{\alpha}^{(k)}$ as in Eq.~\eqref{eqn:Sigma0} and sample the dynamic matrix: 
				\begin{align*}
					\vecc{\bigk{A}{k}} \mid \symk{\Sigma}{k},\bigk{\BF{\alpha}}{k}
					\sim \mathcal{N}^{-1}\bigg(\sum_{t|z_t=k}\tilde{\Psi}_{t-1}^T\symmink{\Sigma}{k}\BF{\psi_t}, \symmink{\Sigma_0}{k}+\sum_{t|z_t=k}\tilde{\Psi}_{t-1}^T\symmink{\Sigma}{k}\tilde{\Psi}_{t-1}\bigg). 
				\end{align*}
				\item For each $\ell \in \{1,\dots,m\}$, with $m=n$ for the SLDS and $m=r$ for the switching VAR, sample ARD precision parameters: 
				\begin{eqnarray*}
					\alpha_\ell^{(k)} \mid \bigk{A}{k} \sim \mbox{Gamma}\left(a+\frac{|\mathcal{S}_\ell|}{2},b+\hspace{-0.1in}\frac{\sum_{(i,j) \in \mathcal{S}_\ell} a_{ij}^{(k)^2}}{2}\right). 
				\end{eqnarray*}
				\item Compute sufficient statistic: 
				\begin{eqnarray*}
					\bigksub{S}{k}{\psi|\Ylag{\psi}} = \sum_{t|z_t = k}(\BF{\psi}_t-\bigk{A}{k}\BF{\bar{\psi}}_{t-1})(\BF{\psi}_t-\bigk{A}{k}\BF{\bar{\psi}}_{t-1})^T 
				\end{eqnarray*}
				and sample process noise covariance: 
				\begin{eqnarray*}
					\Sigma^{(k)}\mid \bigk{A}{k} &\sim& \mbox{IW}\left(n_k + n_0, \bigksub{S}{k}{\psi\mid\Ylag{\psi}} +S_0\right). 
				\end{eqnarray*}
			\end{enumerate}
			\algend \vspace{-0.1in}
		\end{enumerate}
	\end{flushleft}
	%\begin{singlespace}
	\caption{Parameter sampling using ARD prior.} \label{alg:ARD}	
	%\end{singlespace}
\end{algorithm}
\section{Results} \label{sec:results}
%
%
%%
%\begin{figure*}[t!]
%\vspace{-0.1in}
%  \centering
%  \begin{tabular}{cc}
%  \hspace{-0.2in}\includegraphics[height = 1.25in]{\figdir/obs_VAR2} & \hspace{-0.1in}\includegraphics[height =
%  1.25in]{\figdir/hamming_dist_VAR2}\vspace{-0.1in}\\
%  \hspace{-0.2in}{\small (a)} & \hspace{-0.1in}{\small (b)}
%  \end{tabular}
%  \vspace{-0.2in}\caption{(a) Observation sequence and associated mode sequence
%  (magenta) for a 3-mode switching VAR($2$) process. The components of
%  the observation vector have been shifted for clarity. (b) Plot
%  of 10th, 50th, and 90th Hamming distance quantiles over 100 trials using
% an HDP-AR($2$)-HMM model.} \label{fig:VAR}
%\end{figure*}
%%
%
\subsection{MNIW prior} \label{sec:MNIWresults}
\begin{figure*}
	[t!] \centering 
	\begin{tabular}
		{ccccc} \hspace{-0.15in}
		\includegraphics[height = 1.1in]{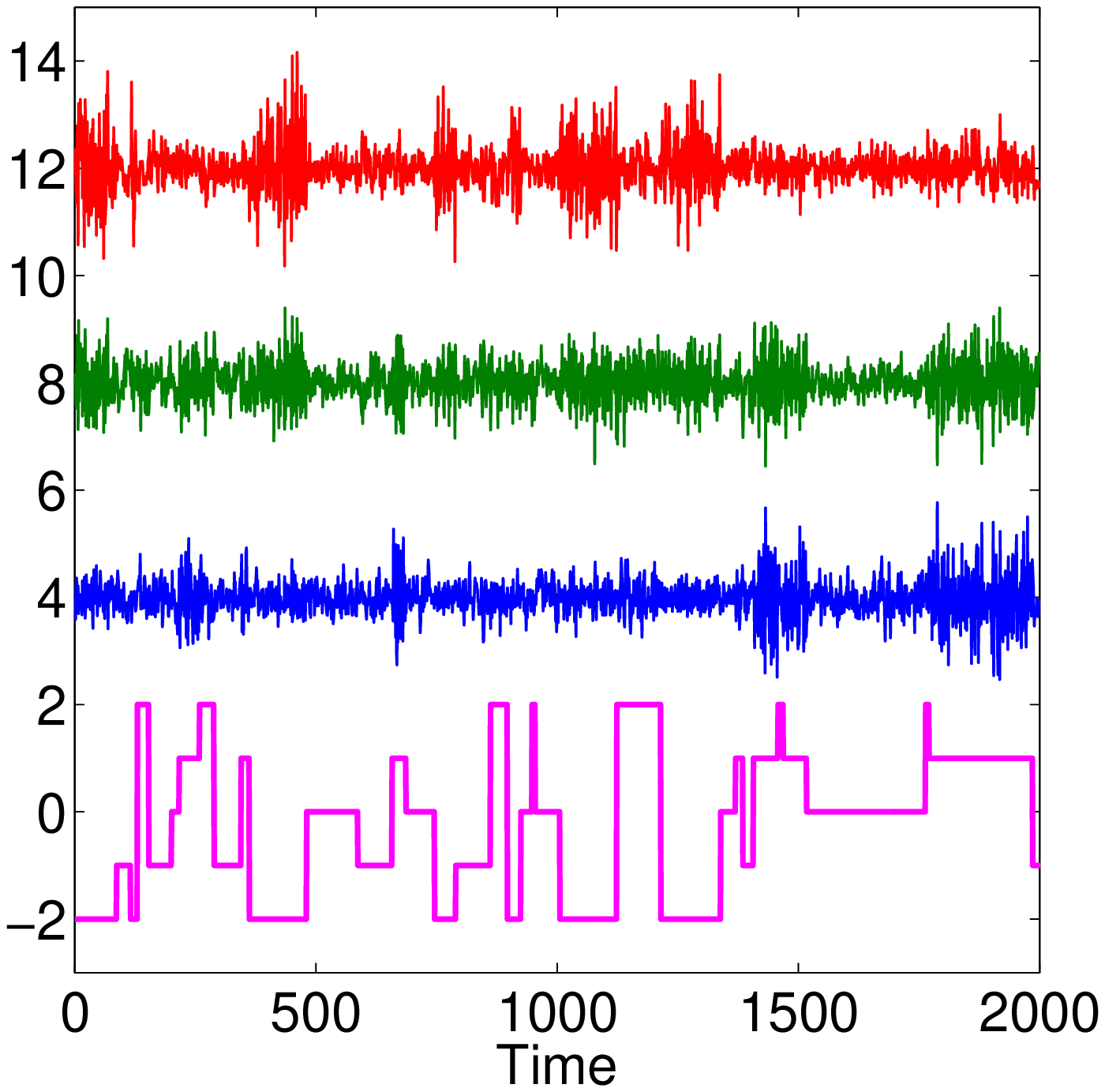} & \hspace{-0.15in}
		\includegraphics[height = 1.1in]{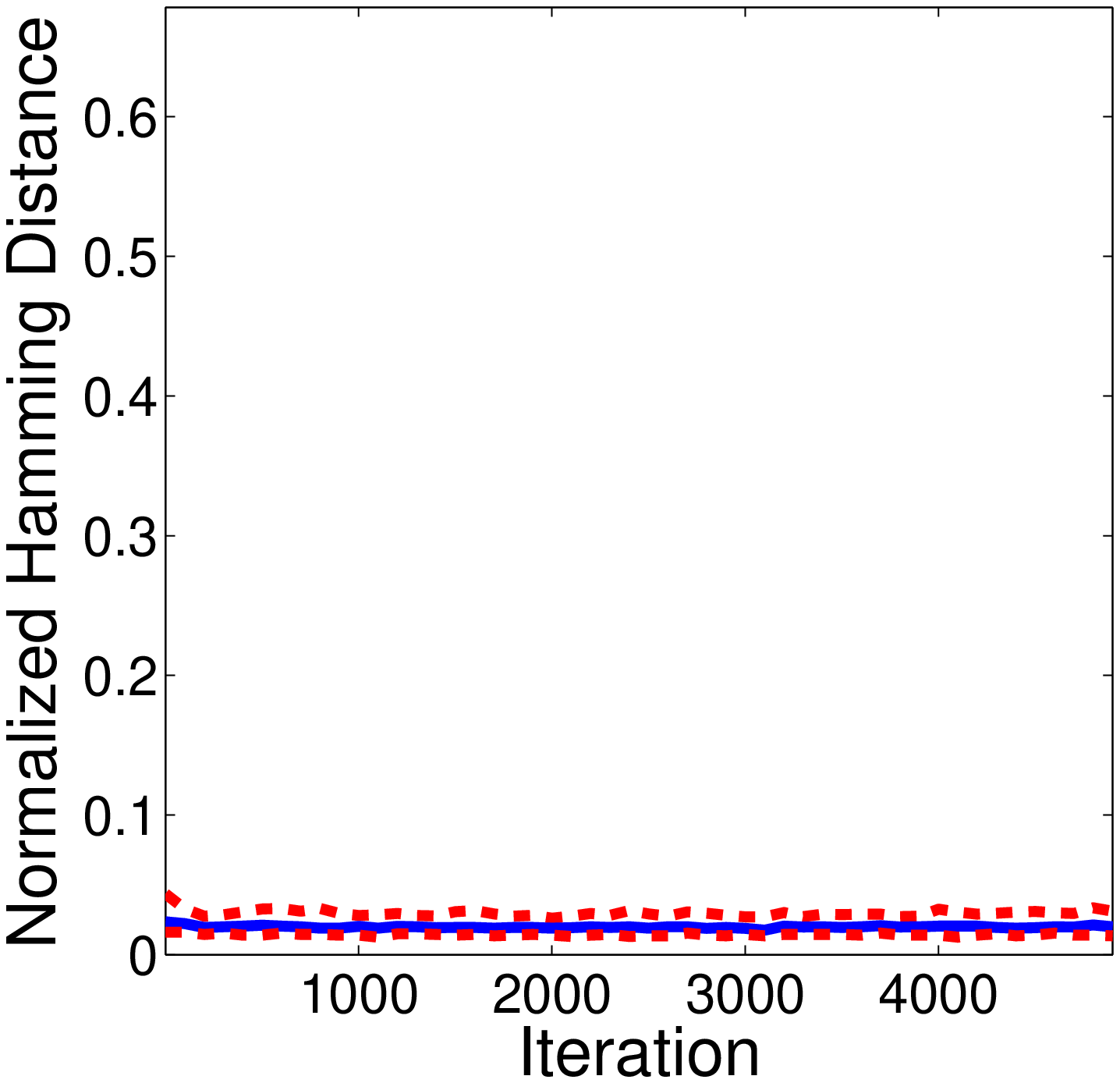} & \hspace{-0.15in}
		\includegraphics[height = 1.1in]{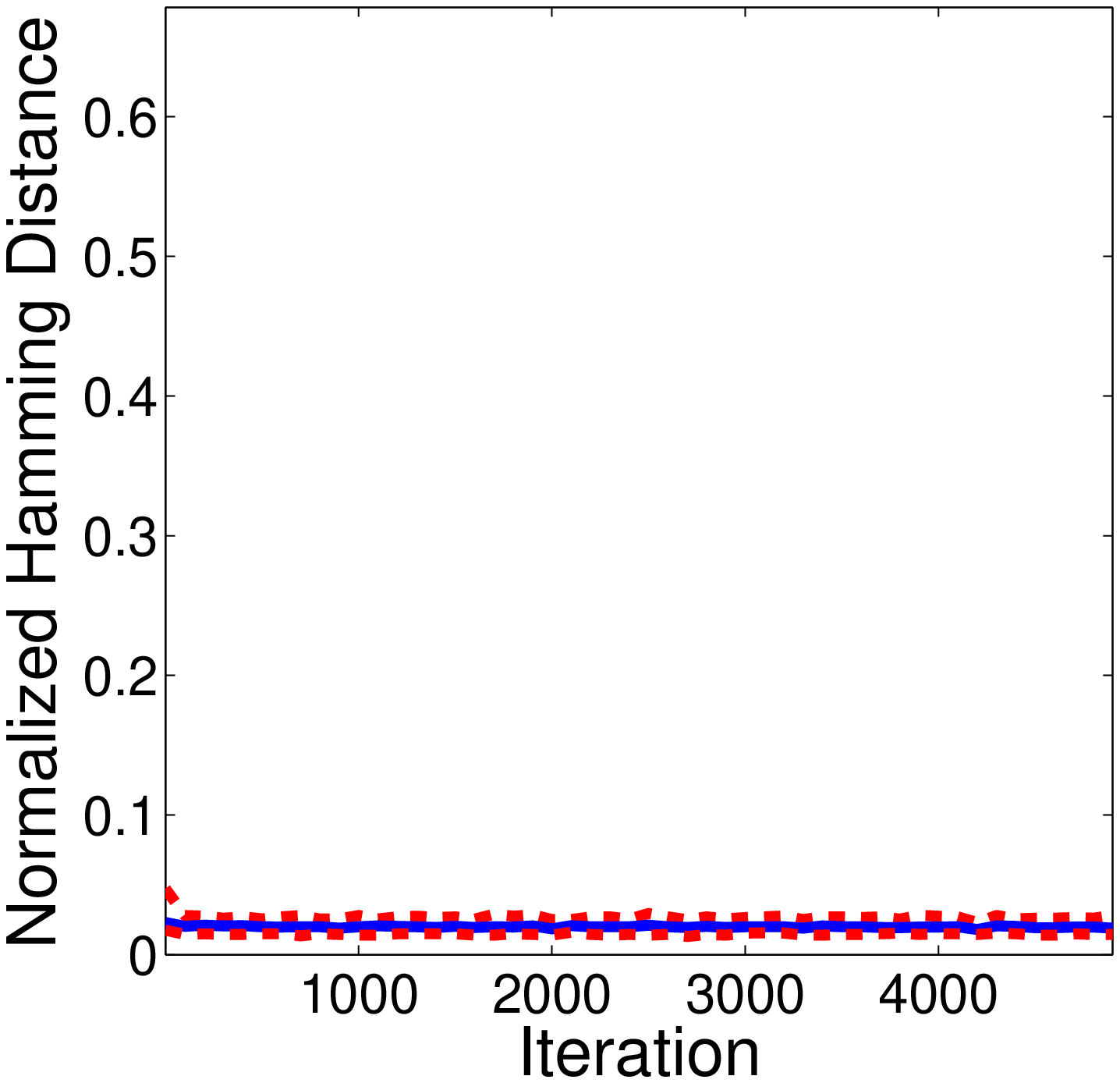} & \hspace{-0.15in}
		\includegraphics[height = 1.1in]{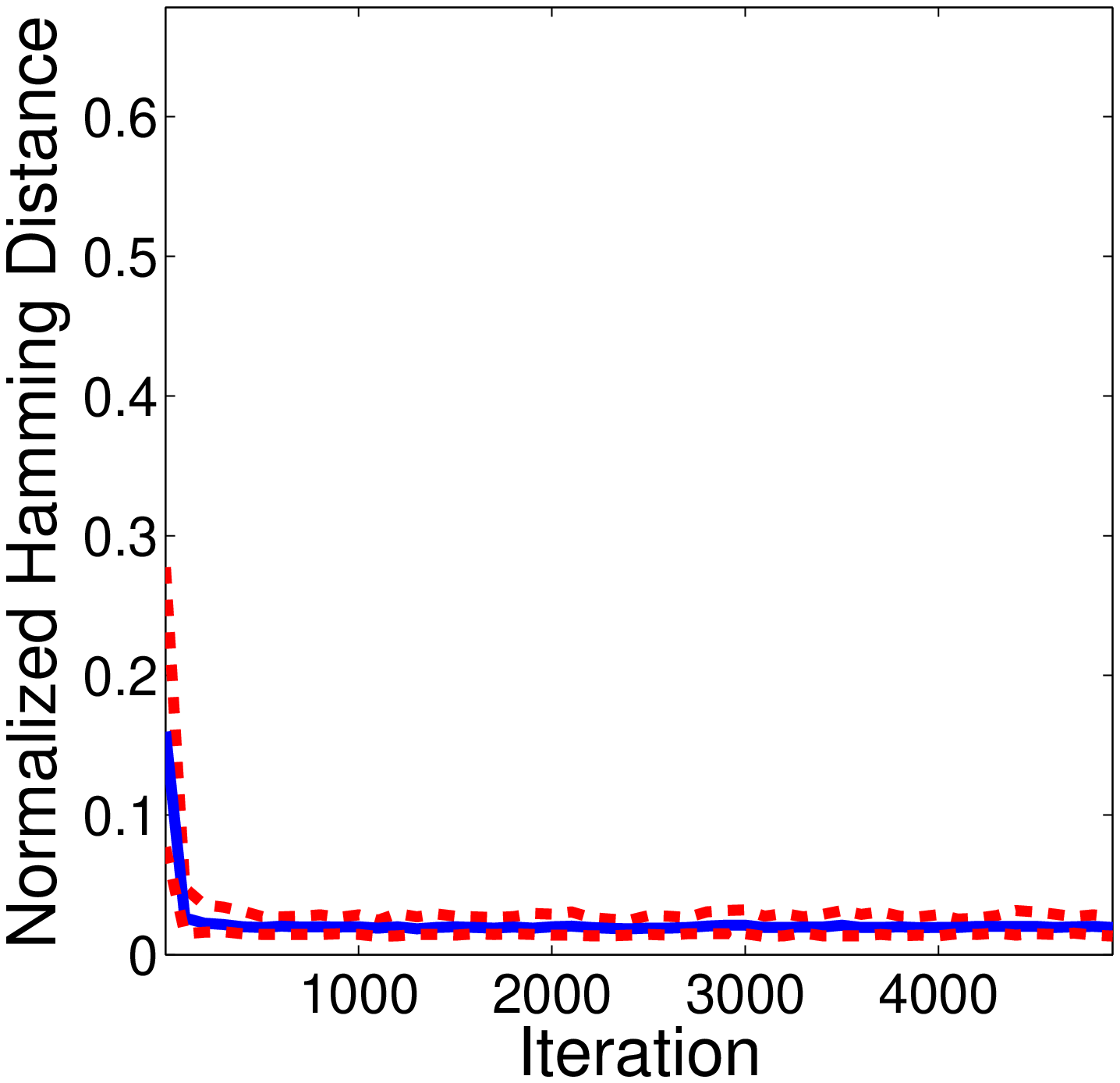} & \hspace{-0.15in}
		\includegraphics[height = 1.1in]{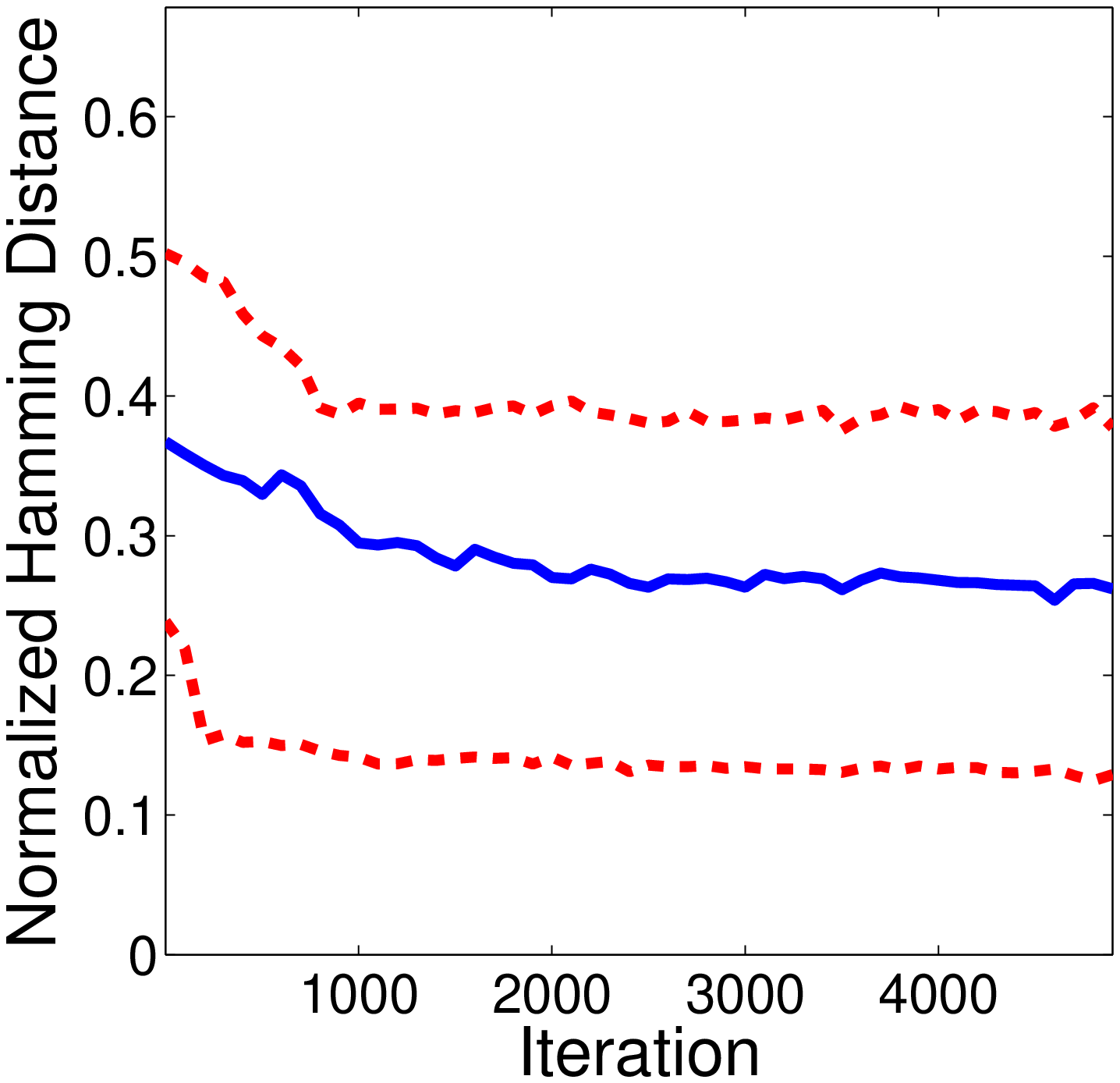}\\
		\hspace{-0.15in}
		\includegraphics[height = 1.1in]{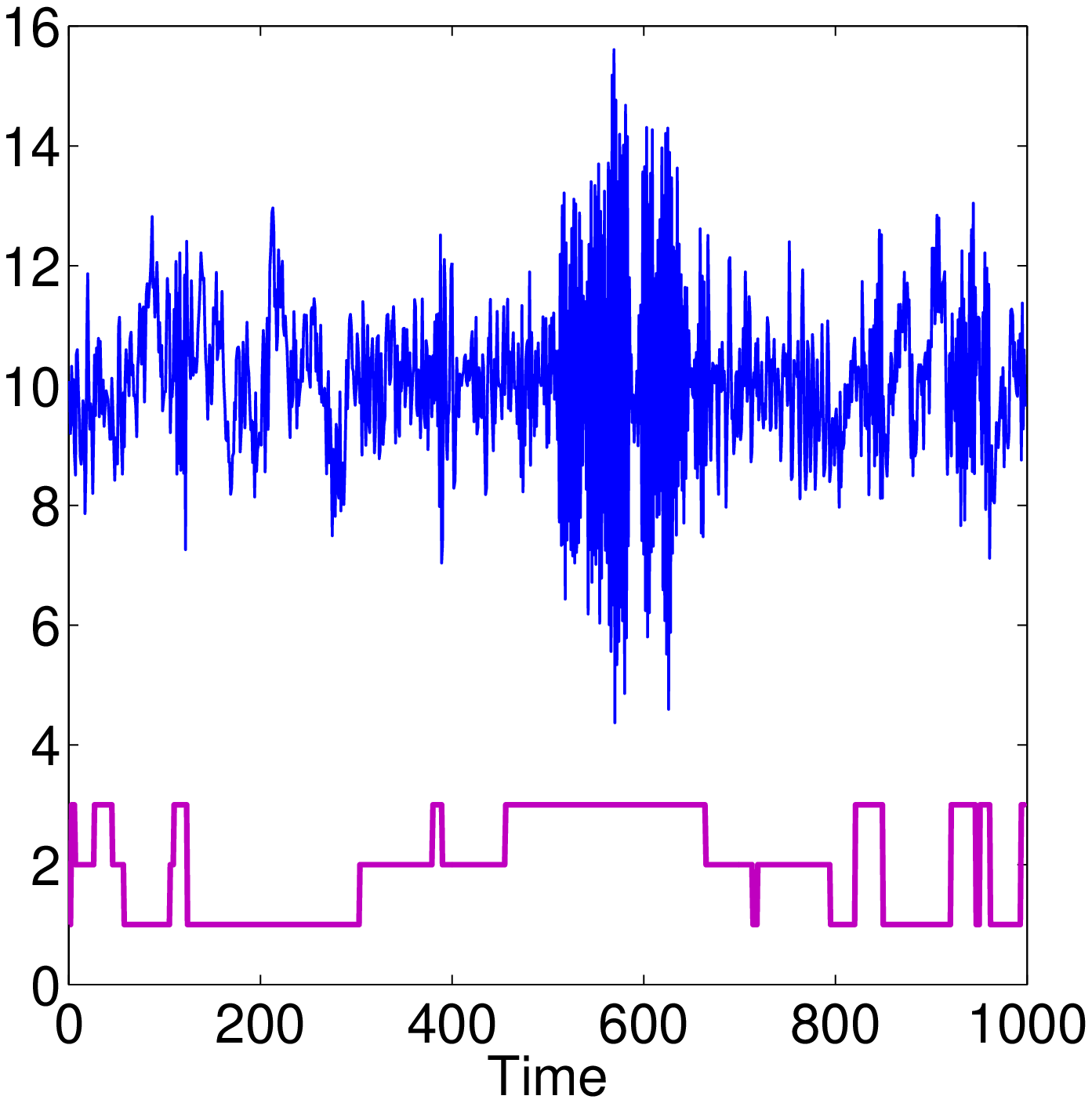} & \hspace{-0.15in}
		\includegraphics[height = 1.1in]{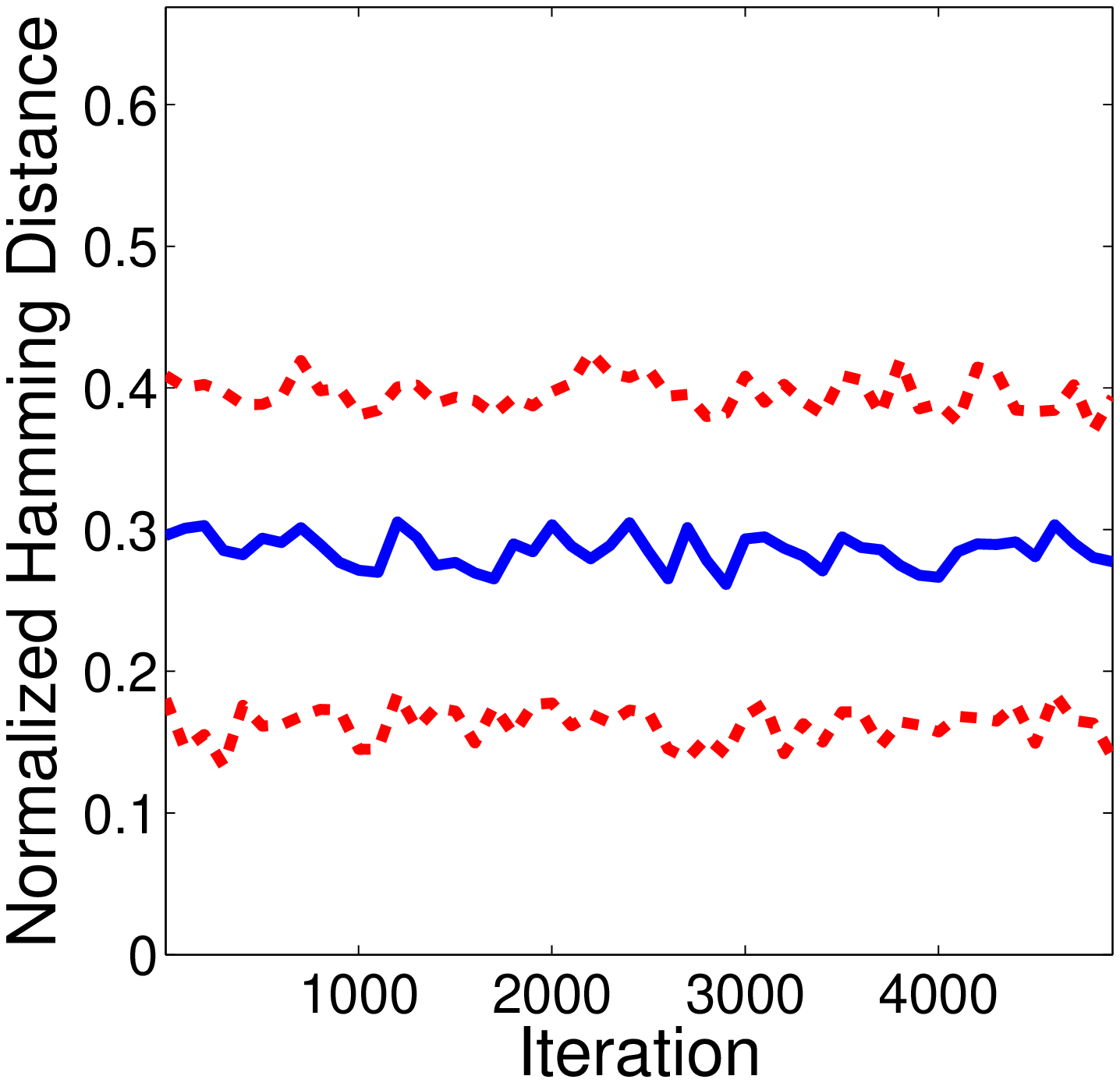} & \hspace{-0.15in}
		\includegraphics[height = 1.1in]{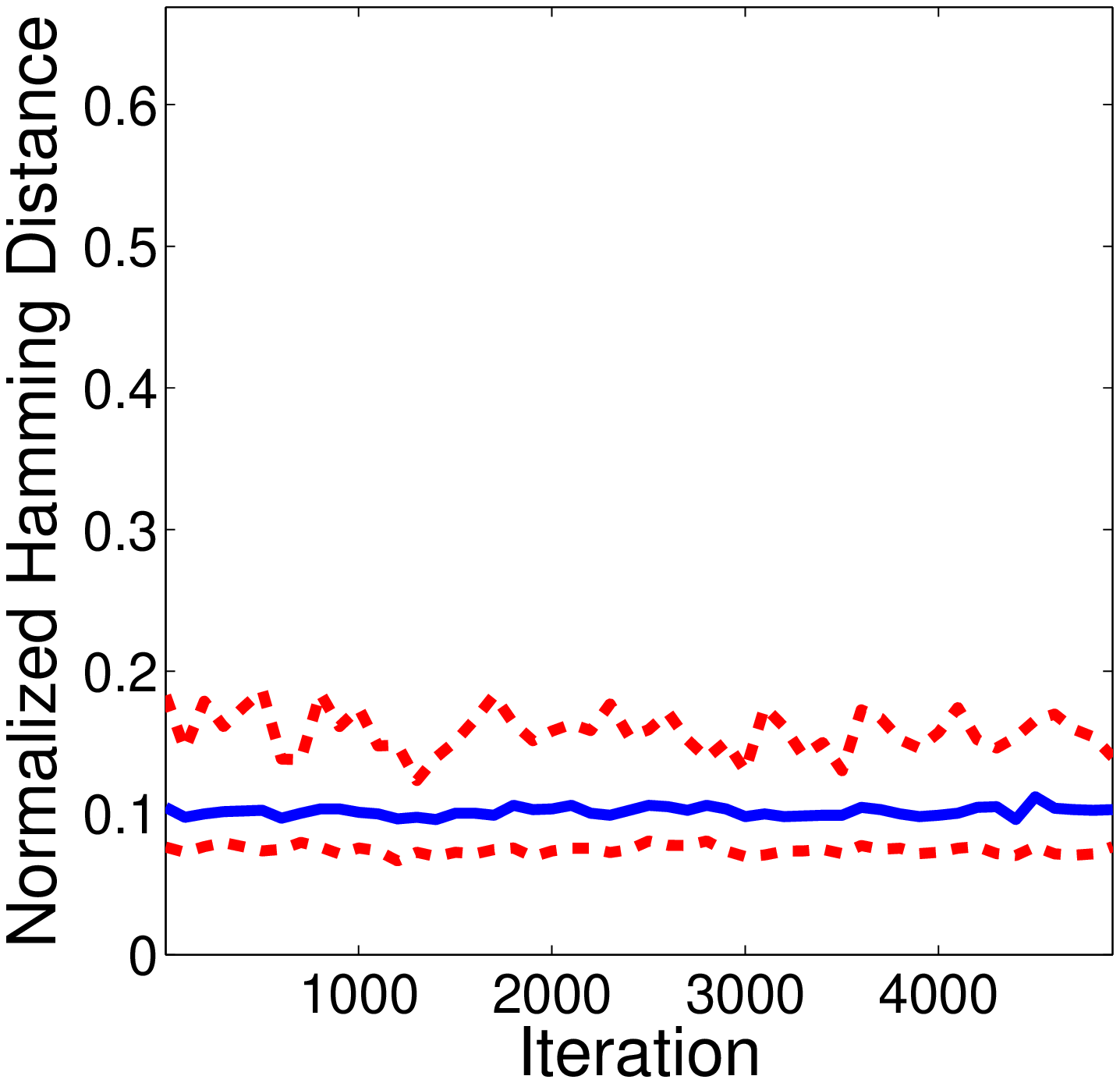} & \hspace{-0.15in} 
		\includegraphics[height = 1.1in]{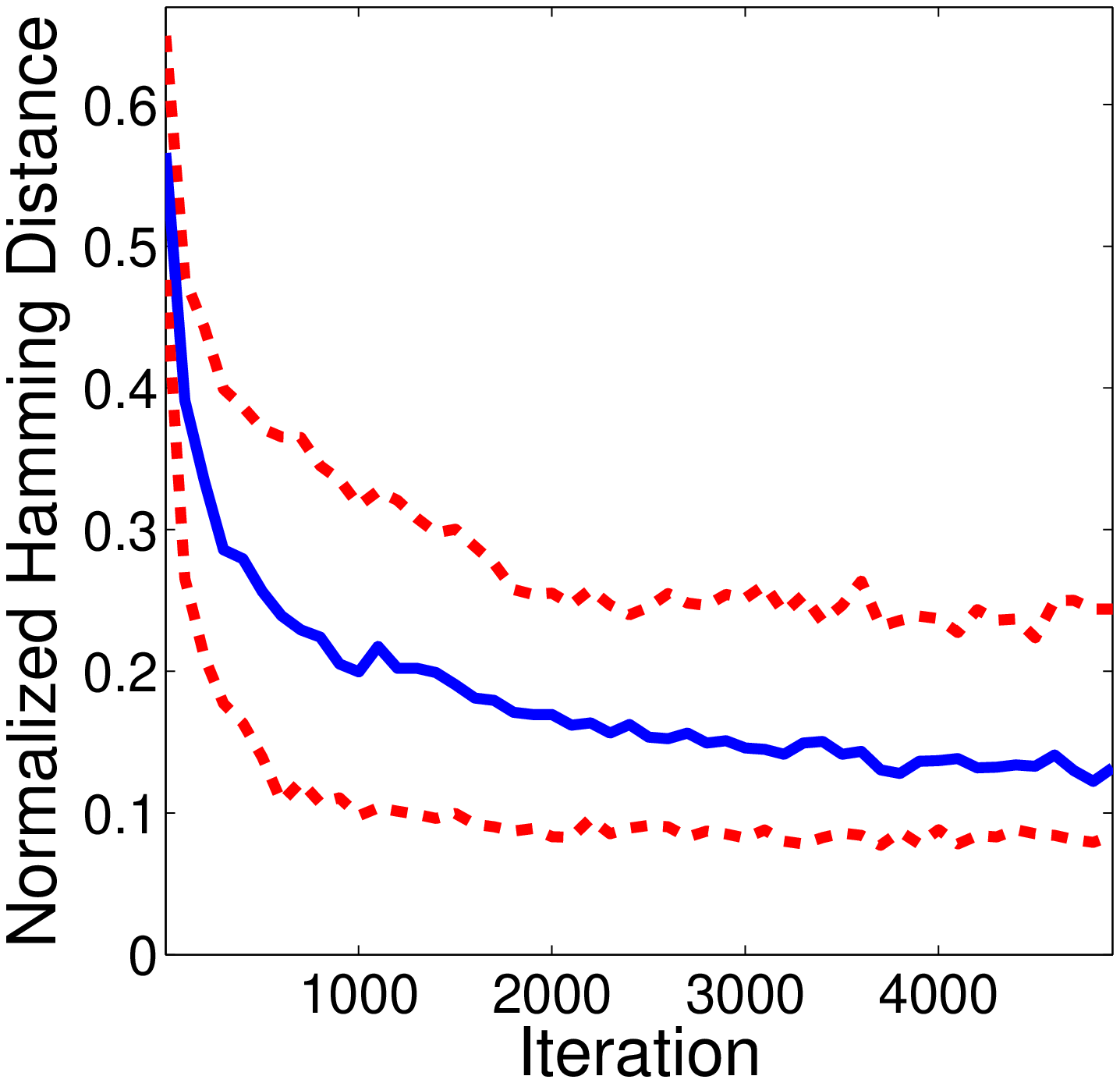} & \hspace{-0.15in}
		\includegraphics[height = 1.1in]{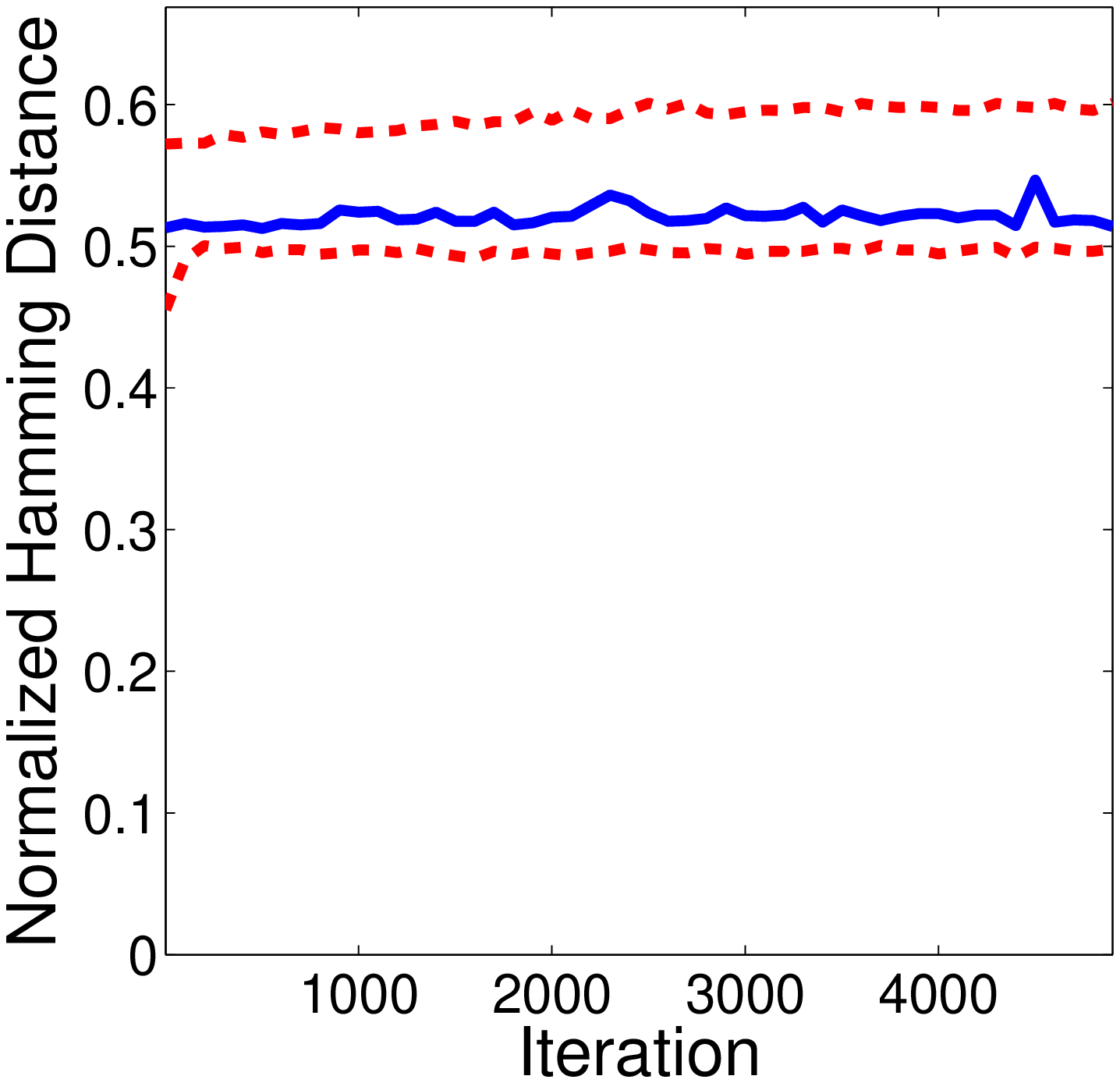}\\		
		%  \hspace{-0.1in}{\small (a)} & \hspace{-0.1in}{\small (b)} & \hspace{-0.1in}{\small
		%  (c)} & \hspace{-0.1in}{\small
		%  (d)}\\
		\hspace{-0.15in}
		\includegraphics[height = 1.1in]{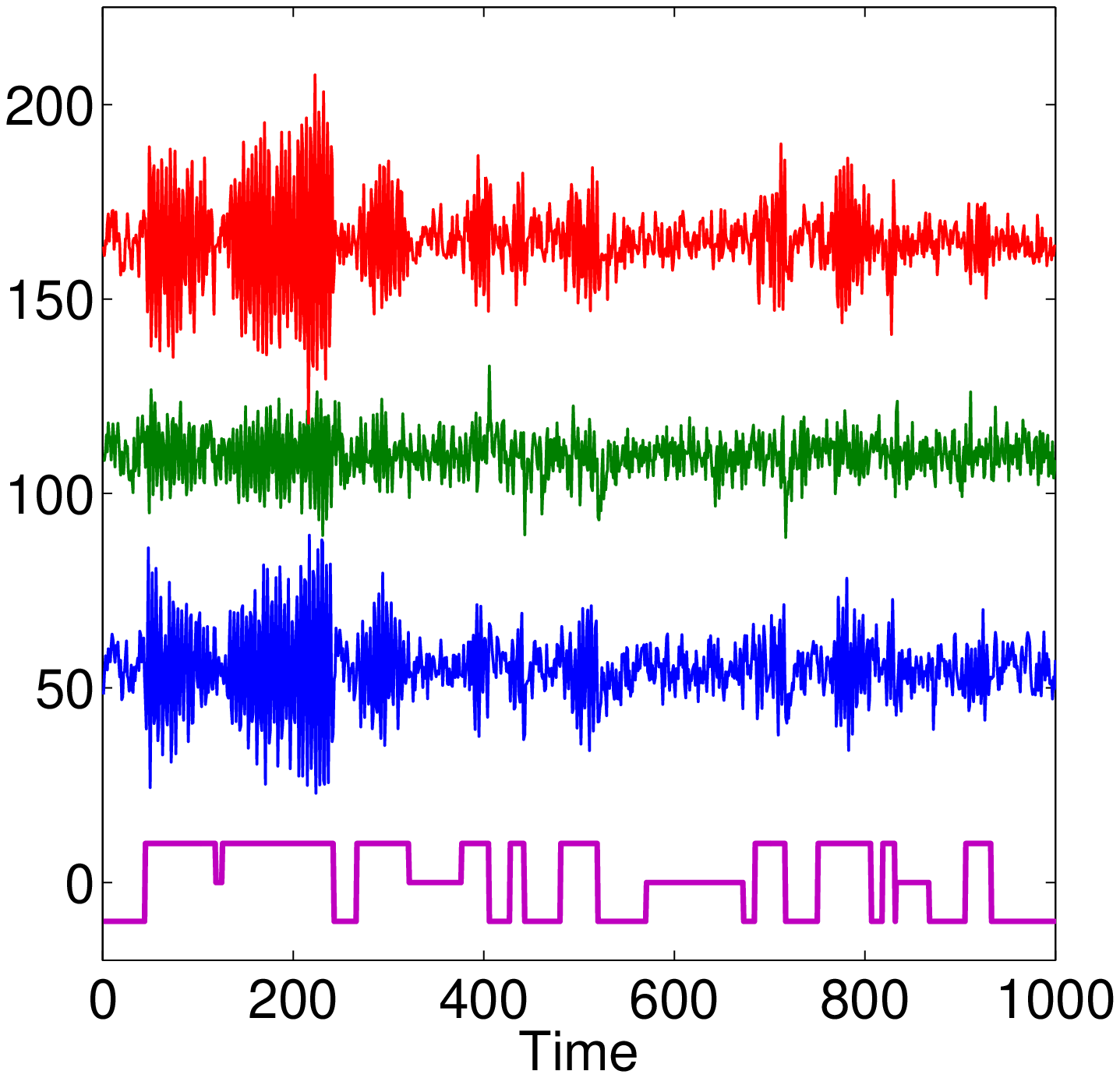} & \hspace{-0.15in}
		\includegraphics[height = 1.1in]{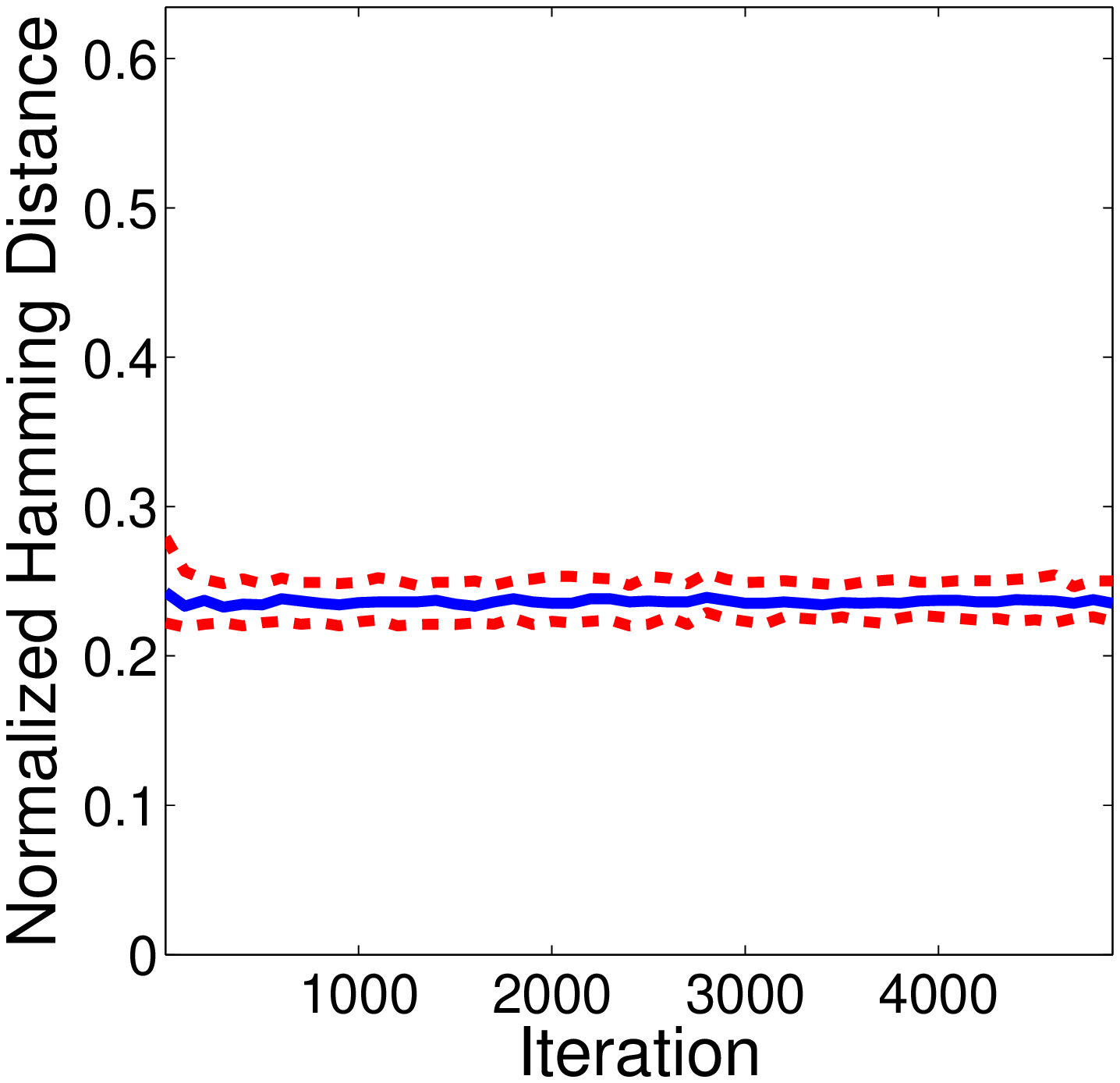} & \hspace{-0.15in}
		\includegraphics[height = 1.1in]{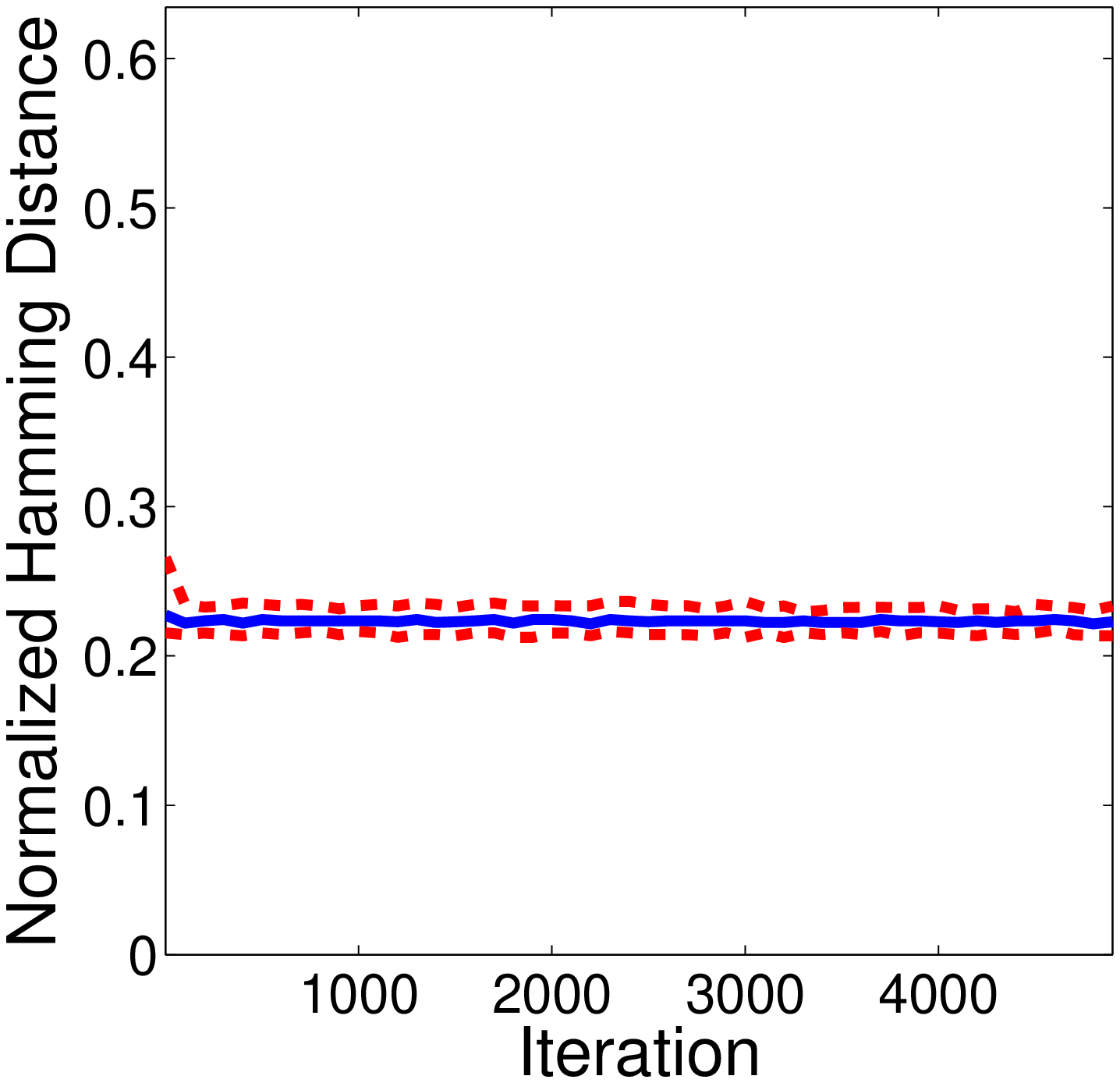} & \hspace{-0.15in}
		\includegraphics[height = 1.1in]{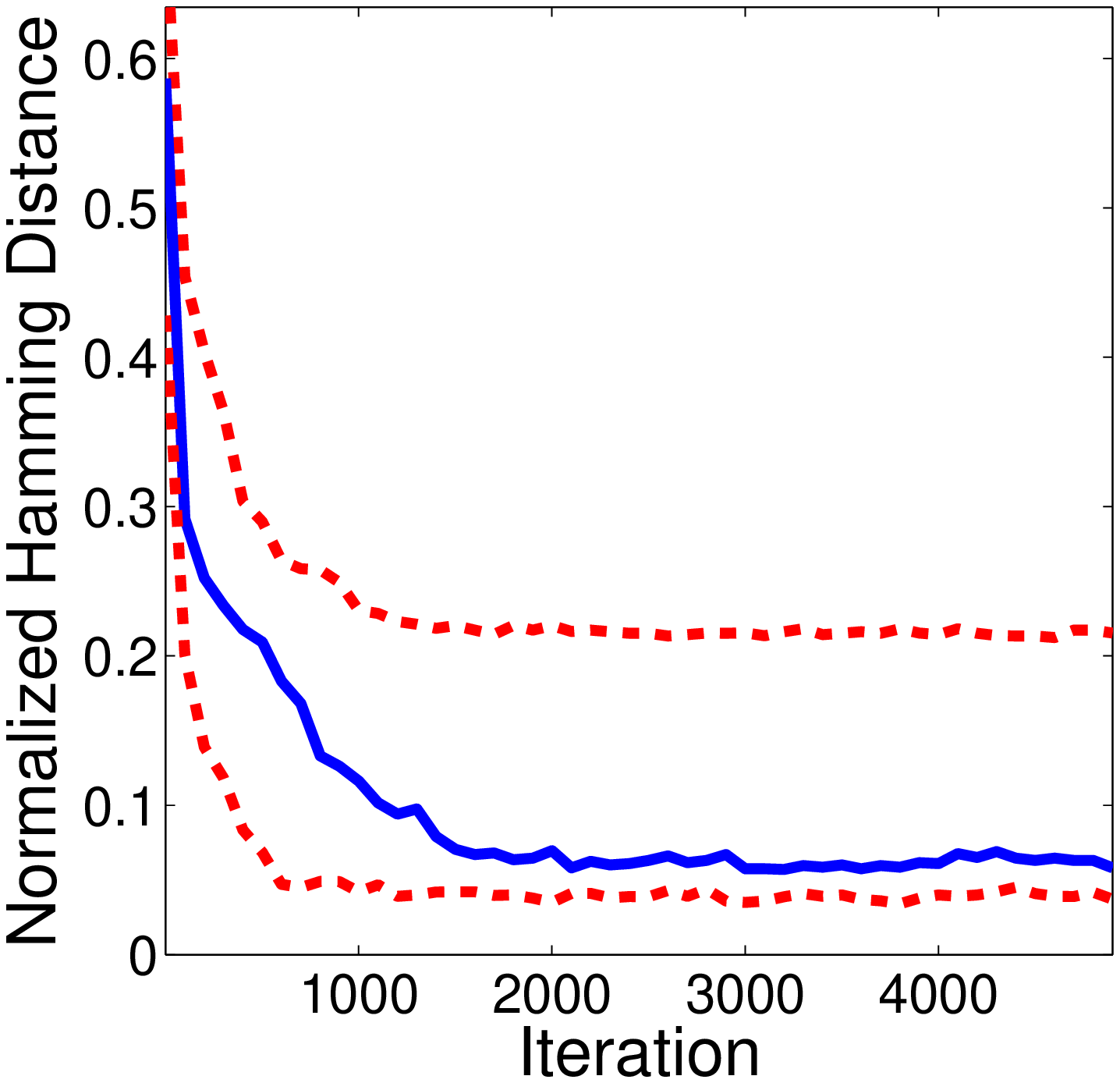} & \hspace{-0.15in}
		\includegraphics[height = 1.1in]{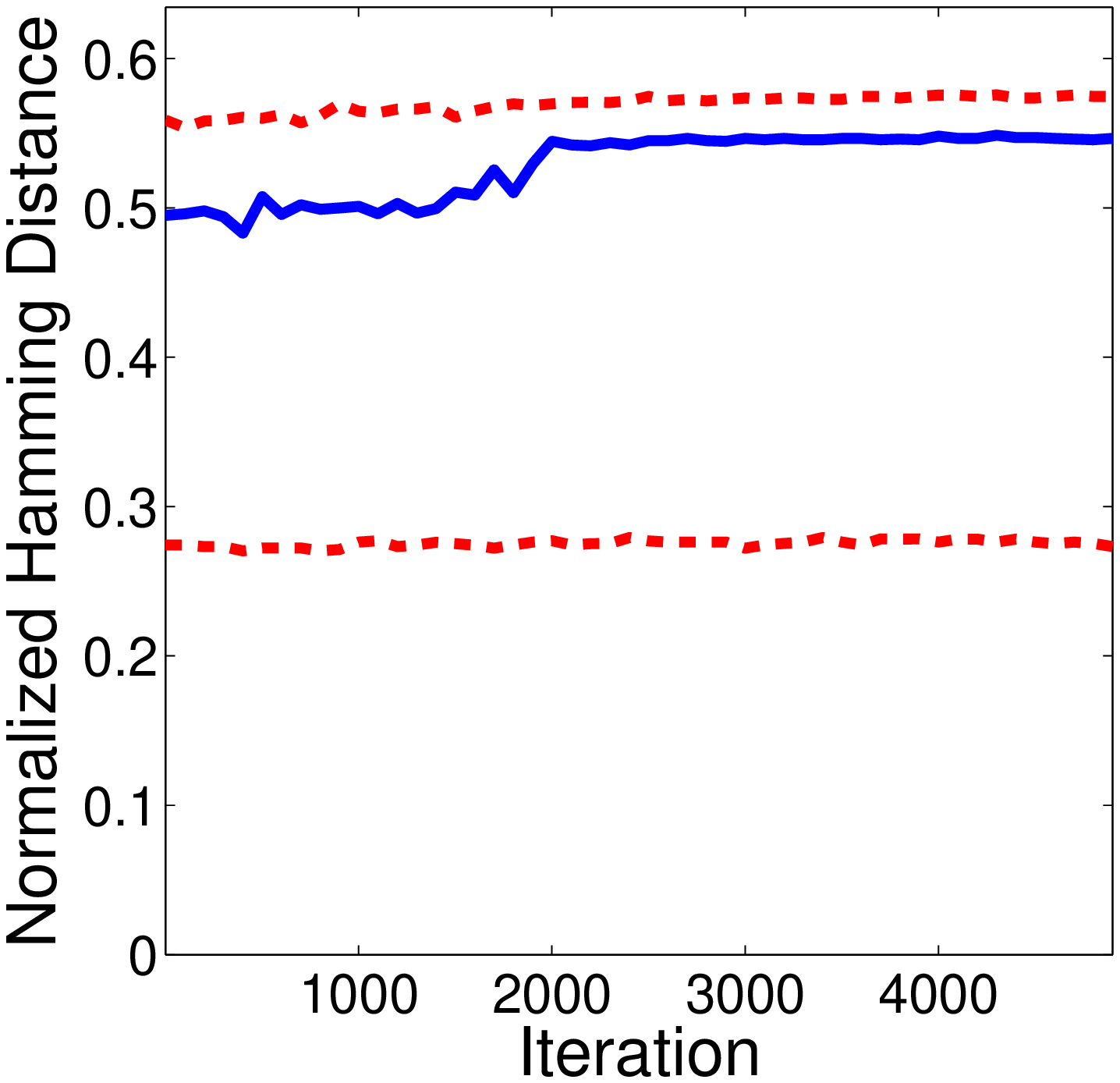}\\ %\vspace{-0.15in}\\
		\hspace{-0.15in} (a) & \hspace{-0.15in} (b) & \hspace{-0.15in} (c) & \hspace{-0.15in} (d) & \hspace{-0.15in} (e) %\vspace{-0.15in}
	\end{tabular}
	\caption[Synthetic data results for the three sequences using the HDP-SLDS, HDP-AR-HMM, and HDP-HMM.] {(a) Observation sequence (blue, green, red) and associated mode sequence
	  (magenta) for a 5-mode switching VAR($1$) process (top), 3-mode switching AR($2$) process (middle), and 3-mode SLDS (bottom). The associated 10th, 50th, and 90th Hamming distance quantiles over 100 trials are shown for the (b) HDP-VAR($1$)-HMM, (c) HDP-VAR($2$)-HMM, (d) HDP-SLDS with $C=I$ (top and bottom) and $C=[1 \ \ 0]$ (middle), and (e) sticky HDP-HMM using first difference observations. }\label{fig:modelmismatch} %\vspace{-0.2in}
\end{figure*}

We begin by examining a set of three synthetic datasets displayed in Fig.~\ref{fig:modelmismatch}(a) in order to analyze the relative modeling power of the HDP-VAR($1$)-HMM\footnote{We use the notation HDP-VAR($r$)-HMM to refer to an order $r$ HDP-AR-HMM with vector observations.}, HDP-VAR($2$)-HMM, and HDP-SLDS using the MNIW prior. We compare to a baseline sticky HDP-HMM using first difference observations, imitating a HDP-VAR($1$)-HMM with $A^{(k)}=I$ for all $k$. In Fig.~\ref{fig:modelmismatch}(b)-(e) we display Hamming distance errors that are calculated by choosing the optimal mapping of indices maximizing overlap between the true and estimated mode sequences.
%
%For all of the scenarios, we set the MNIW hyperparameters from statistics of the data in the following way. We start by assuming the mean matrix $M$ is $\BF{0}$, and setting $K = I_m$. This choice centers the mass of the matrix-normal distribution around stable dynamic matrices while allowing for considerable variability in the matrix values. The inverse-Wishart portion of the prior is given $n_0=m+2$ degrees of freedom. For the HDP-AR-HMM, the scale matrix $S_0$ is set to 0.75 times the empirical covariance of the entire dataset. Setting the prior directly from the data can help move the mass of the distribution to reasonable values of the parameter space. For an HDP-SLDS with $\BF{x}_t \in \RR^n$ and $\BF{y}_t \in \RR^d$, $n>d$ (i.e., larger state dimension than observation dimension,) we need a method for setting the mean of the extra $n-d$ dimensions of the process noise covariance. We use the following heuristic: we set the upper $d\times d$ lefthand quadrant of the scale matrix $S_0$ to be 0.675 times the empirical covariance; the $n-d\times n-d$ lower righthand quadrant is set to be diagonal with determinant equal to that of the upper righthand $d\times d$ block. We then set the inverse-Wishart prior on the measurement noise, $R$, to have $r_0=d+2$ degrees of freedom and a scale matrix equal to 0.075 times the empirical covariance. Although we have chosen this specific heuristic for setting the hyperparameters of the MNIW prior, we have found that the results are fairly robust to various settings.

We place a $\mbox{Gamma}(a,b)$ prior on the sticky HDP-HMM concentration parameters $\alpha+\kappa$ and $\gamma$, and a $\mbox{Beta}(c,d)$ prior on the self-transition proportion parameter $\rho = \kappa/(\alpha+\kappa)$. We choose the weakly informative setting of $a=1$, $b=0.01$, $c=10$, and $d=1$.  The details on setting the MNIW hyperparameters from statistics of the data are discussed in the Appendix.

For the first scenario (Fig.~\ref{fig:modelmismatch}~(top)), the data were generated from a five-mode switching VAR($1$) process with a 0.98 probability of self-transition and equally likely transitions to the other modes. The same mode-transition structure was used in the subsequent two scenarios, as well. The three switching linear dynamical models provide comparable performance since both the HDP-VAR($2$)-HMM and HDP-SLDS with $C=I_3$ contain the class of HDP-VAR($1$)-HMMs. In the second scenario (Fig.~\ref{fig:modelmismatch}~(middle)), the data were generated from a 3-mode switching AR($2$) process. The HDP-AR($2$)-HMM has significantly better performance than the HDP-AR($1$)-HMM while the performance of the HDP-SLDS with $C=[1 \, \, 0]$ performs similarly, but has greater posterior variability because the HDP-AR(2)-HMM model family is smaller. Note that the HDP-SLDS sampler is slower to mix since the hidden, continuous state is also sampled. The data in the third scenario (Fig.~\ref{fig:modelmismatch}~(bottom)) were generated from a three-mode SLDS model with $C=I_3$. Here, we clearly see that neither the HDP-VAR($1$)-HMM nor HDP-VAR($2$)-HMM is equivalent to the HDP-SLDS. Note that all of the switching models yielded significant improvements relative to the baseline sticky HDP-HMM. This input representation is more effective than using raw observations for HDP-HMM learning, but still much less effective than richer models which switch among learned LDS. Together, these results demonstrate both the differences between our models as well as the models' ability to learn switching processes with varying numbers of modes.
% TODO:
%\textcolor{red}{Maybe add results from models with an ARD prior on
%these synthetic datasets?  I have results on the first two
%scenarios, but cannot seem to get the third one to work yet (the
%SLDS).}
%%
%\begin{figure*}[t!] \centering 
%	\begin{tabular}{ccc} 
%		\includegraphics[height = 1.65in]{\figdir/observations_AR1} & 
%		\includegraphics[height = 1.65in]{\figdir/obs_fixed_AR2_2} & 
%		\includegraphics[height = 1.65in]{\figdir/obs_SLDS}\vspace{-0.1in}\\
%		(a) & (b) & (c) \vspace{-0.1in}
%	\end{tabular}
%	\caption[Plots of three synthetic data sequences generated from switching linear dynamical processes.] {(a) Observation sequence (blue, green, red) and associated mode sequence (magenta) for: (a) 5-mode switching VAR($1$) process, (b) 3-mode switching AR($2$) process, and (c) 3-mode SLDS.} \label{fig:modelmismatch_obs} \vspace{-0.2in}
%\end{figure*}
%%
%
\subsection{ARD prior} \label{sec:ARDresults}
We now compare the utility of the ARD prior to the MNIW prior using the HDP-SLDS model when the true underlying dynamical modes have sparse dependencies relative to the assumed model order\footnote{That is, the HDP-SLDS may have dynamical regimes reliant on lower state dimensions, or the HDP-AR-HMM may have modes described by lower order VAR processes.}. We generated data from a two-mode SLDS with 0.98 probability of self-transition and
\begin{align*}
	\bigk{A}{1} = 
	\begin{bmatrix}
		0.8 & -0.2 & 0\\
		-0.2 & 0.8 & 0\\
		0 & 0 & 0
	\end{bmatrix}
	\hspace{0.1in} \bigk{A}{2} = 
	\begin{bmatrix}
		-0.2 & 0 & 0.8\\
		0.8 & 0 & -0.2\\
		0 & 0 & 0
	\end{bmatrix}, 
\end{align*}
with $C=[I_2 \,\, 0]$, $\symk{\Sigma}{1}=\symk{\Sigma}{2}=I_3$, and $R=I_2$. The first dynamical process can be equivalently described by just the first and second state components since the third component is simply white noise that does not contribute to the state dynamics and is not directly (or indirectly) observed. For the second dynamical process, the third state component is once again a white noise process, but \emph{does} contribute to the dynamics of the first and second state components. However, we can equivalently represent the dynamics of this mode as
\begin{equation*}
	\begin{aligned}
		x_{1,t} &= -0.2x_{1,t-1}+\tilde{e}_{1,t}\\
		x_{2,t} &= 0.8x_{1,t-1}+\tilde{e}_{2,t}
	\end{aligned} \hspace{0.25in}
			\bigk{\tilde{A}}{2} = 
			\begin{bmatrix}
				-0.2 & 0 & 0\\
				0.8 & 0 & 0\\
				0 & 0 & 0
			\end{bmatrix},
\end{equation*}
where $\tilde{\BF{e}}_t$ is a white noise term defined by the original process noise combined with $x_{3,t}$, and $\bigk{\tilde{A}}{2}$ is the dynamical matrix associated with this equivalent representation of the second dynamical mode. Notice that this SLDS does not satisfy Criterion~\ref{crit:nested} since the second column of $\bigk{A}{2}$ is zero while the second column of $C$ is not. Nevertheless, because the realization is in our canonical form with $C=[I_2 \,\, 0]$, we still expect to recover the $\symk{\BF{a}_2}{2} = \symk{\BF{a}_3}{2} = 0$ sparsity structure. We set the parameters of the $\mbox{Gamma}(a,b)$ prior on the ARD precisions as $a=|\mathcal{S}_{\ell}|$ and $b=a/1000$, where we recall the definition of $\mathcal{S}_\ell$ from Eq.~\eqref{eqn:alphaPost}. This specification fixes the mean of the prior to 1000 while aiming to provide a prior that is equally informative for various choices of model order (i.e., sizes $|\mathcal{S}_{\ell}|$).

\begin{figure*}[t!] \centering 
	\begin{tabular}{c} 
		\begin{tabular}{ccccc} \hspace{-0.3in}
			\includegraphics[height = 1.15in]{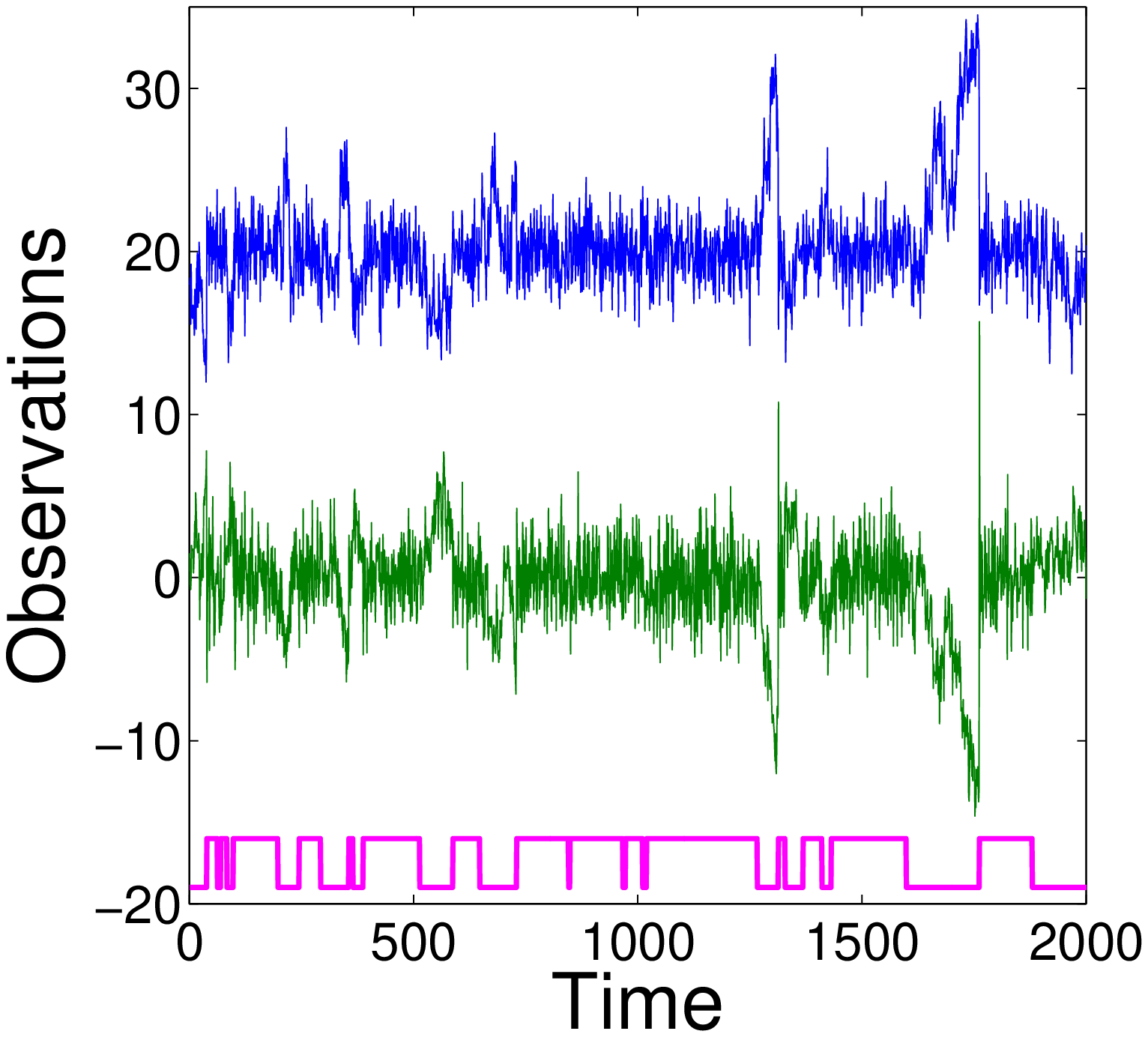} & \hspace{-0.2in}
			\includegraphics[height = 1.15in]{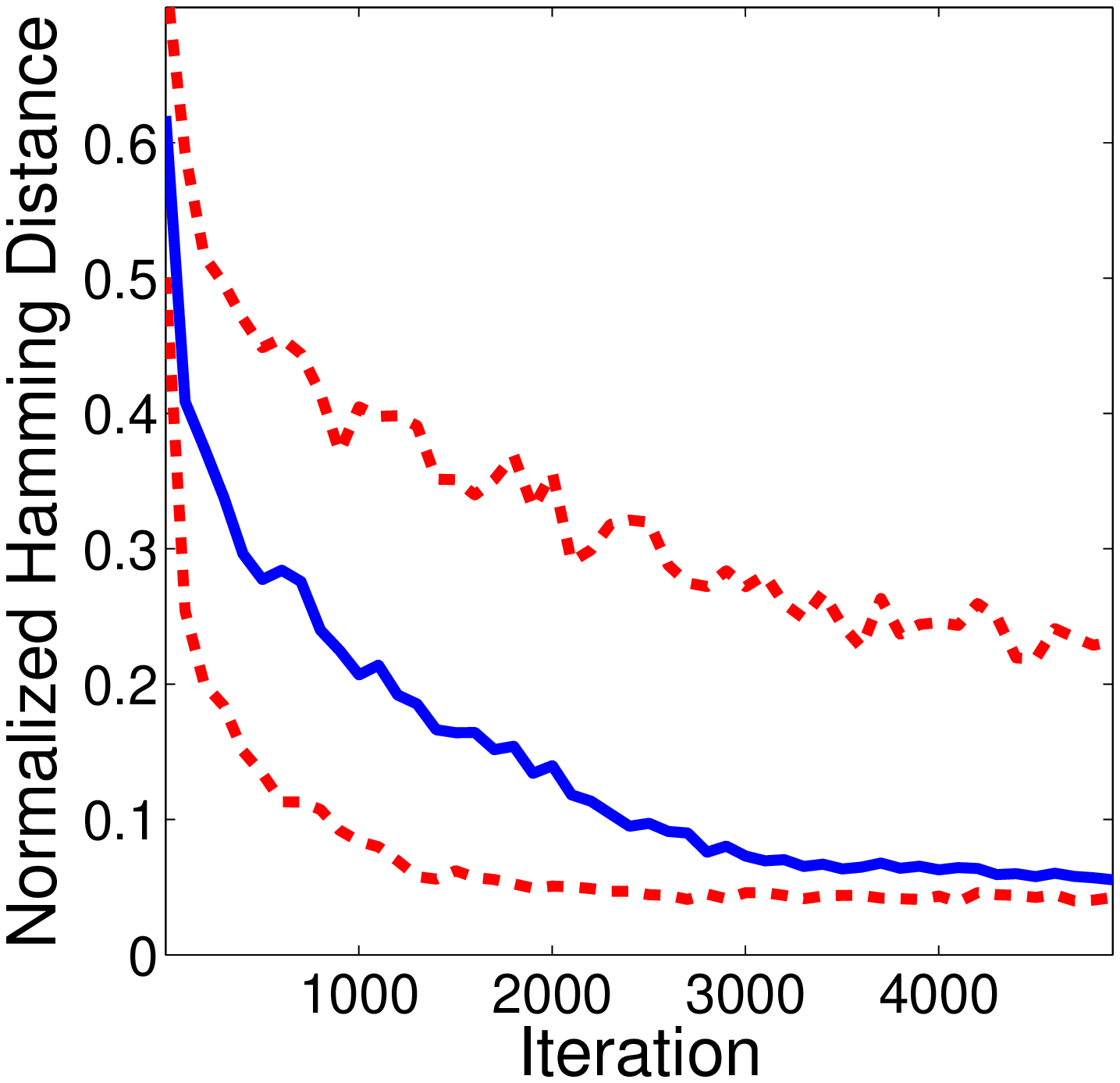} & \hspace{-0.18in}
			\includegraphics[height = 1.15in]{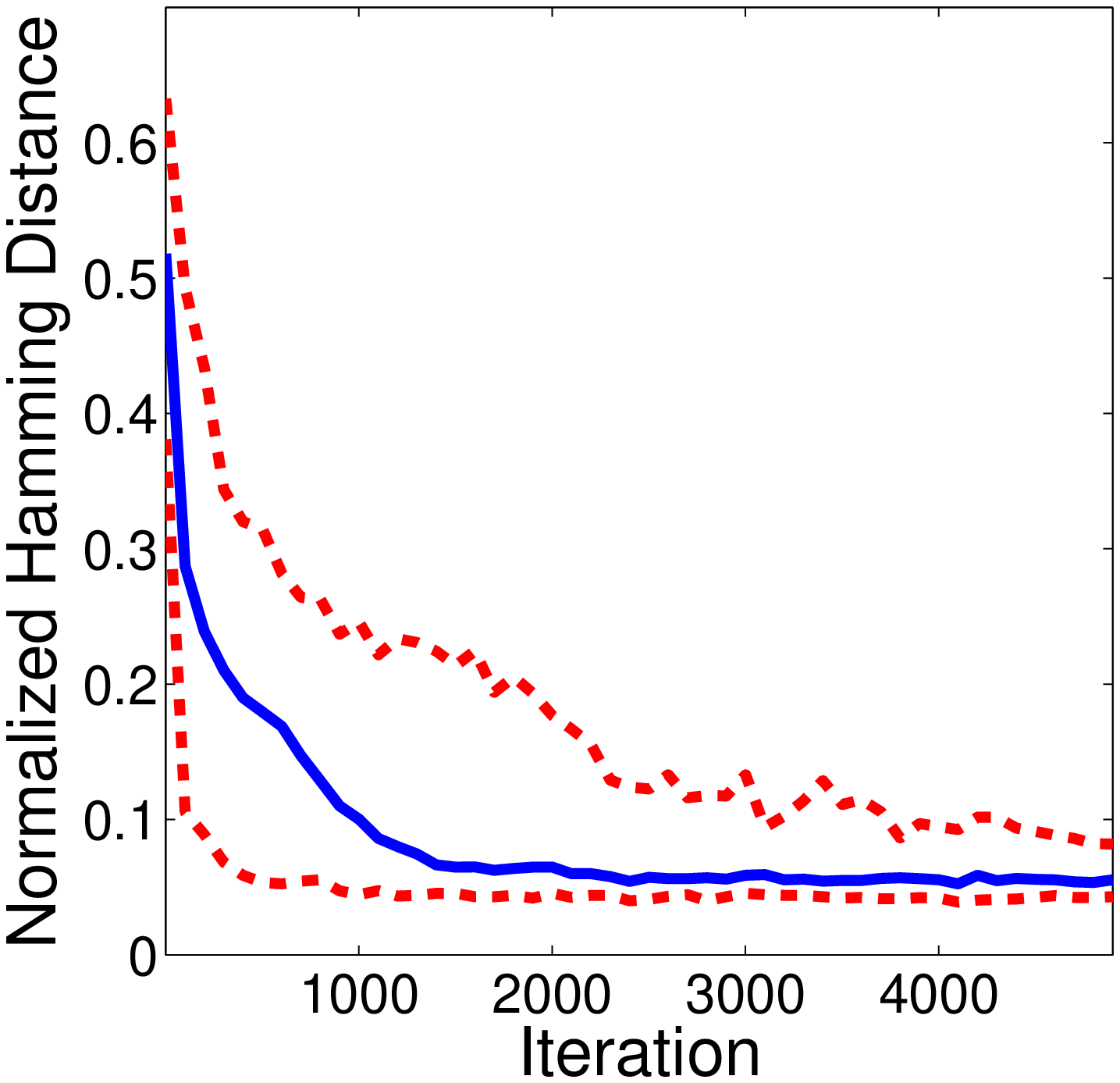} & \hspace{-0.2in}
			\includegraphics[height = 1.15in]{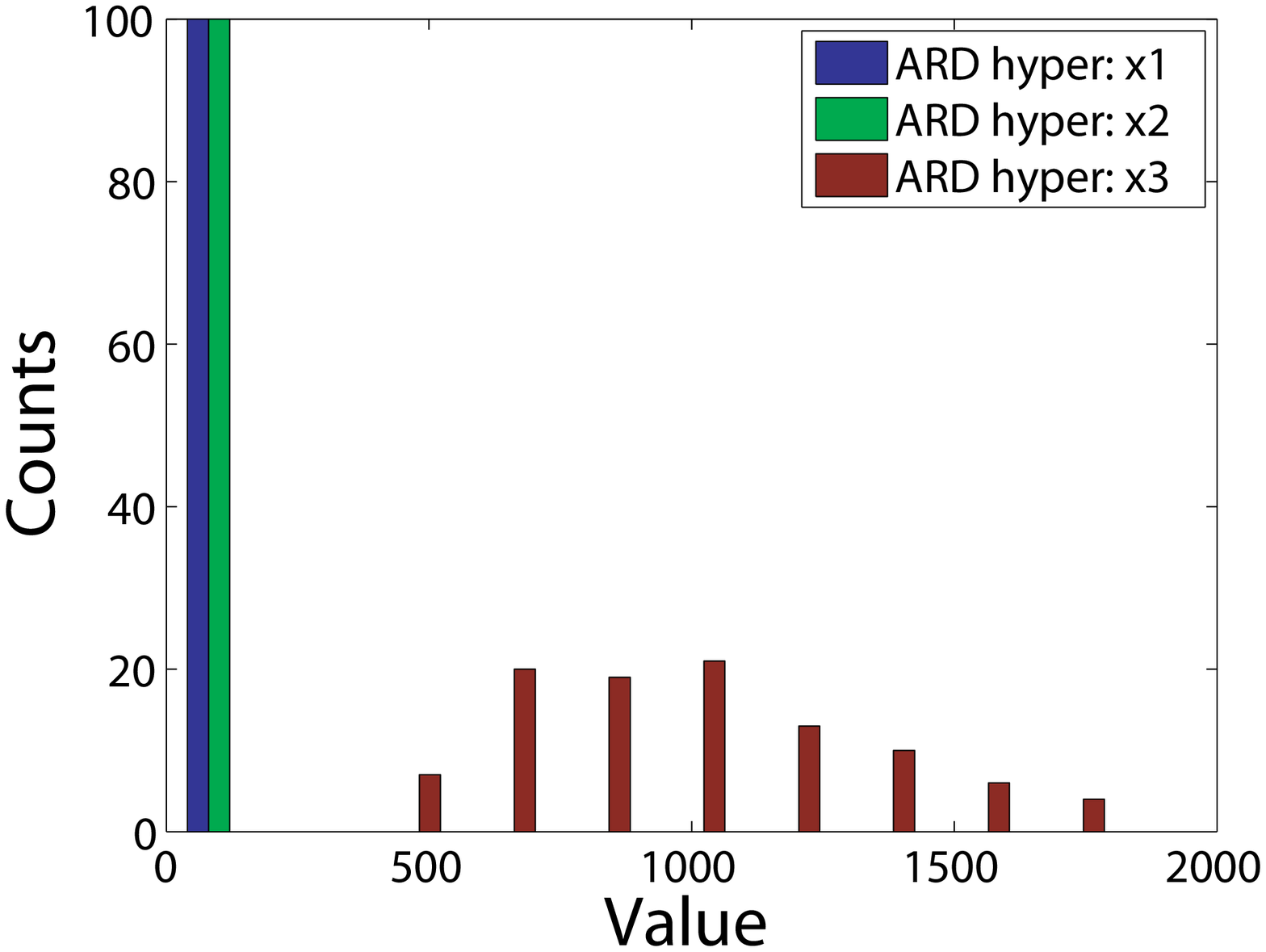} & \hspace{-0.25in}
			\includegraphics[height = 1.15in]{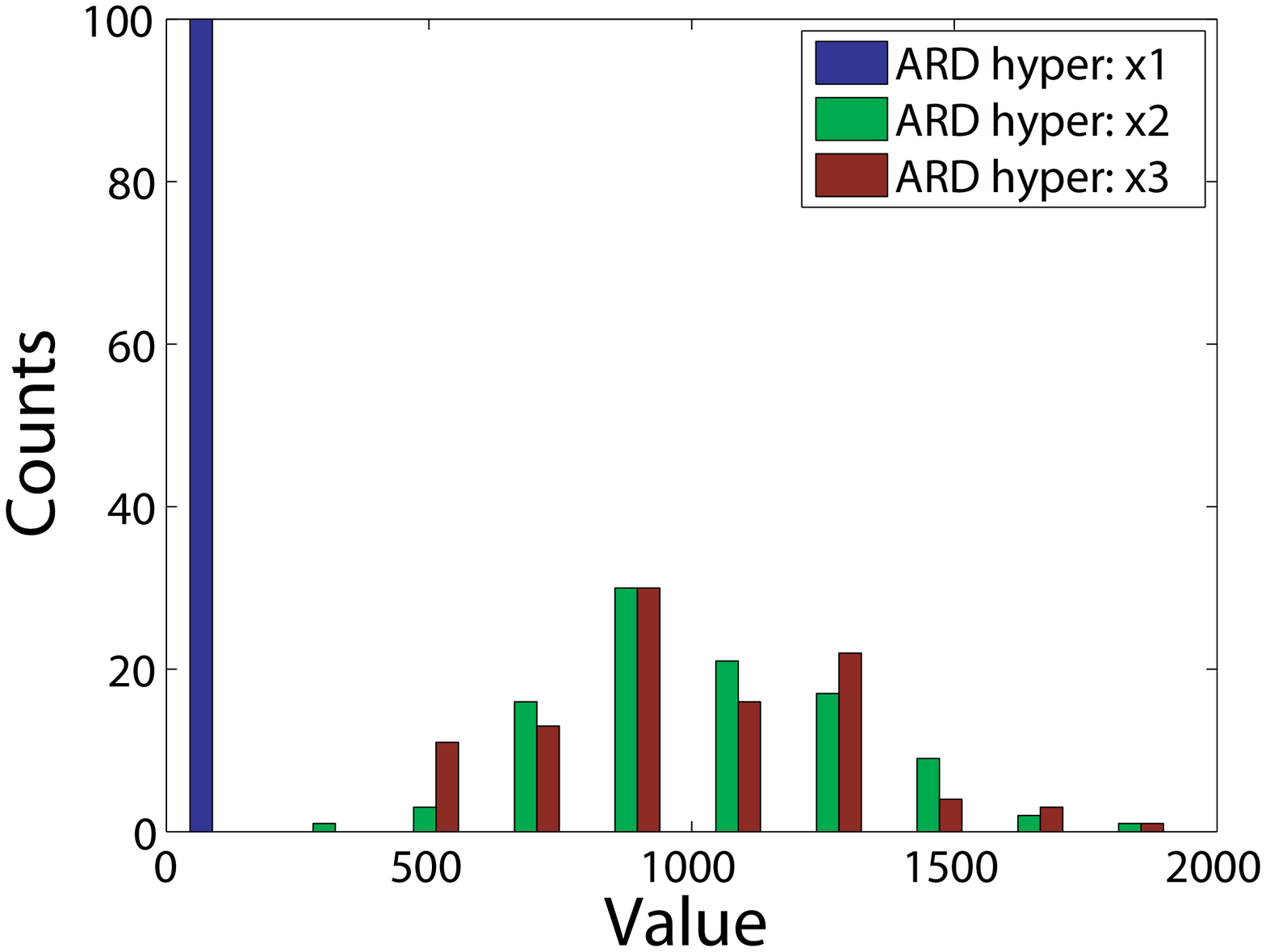}\\%\vspace{-0.1in}\\
			\hspace{-0.3in}(a) & \hspace{-0.2in}(b) & \hspace{-0.18in}(c) & \hspace{-0.2in}(d) & \hspace{-0.25in}(e)%\vspace{-0.1in} 
		\end{tabular}
	\end{tabular}
	\caption[Synthetic data generated from an SLDS with a sparse dynamical matrix, and results comparing the HDP-SLDS with an ARD vs. MNIW prior.] {(a) Observation sequence (green, blue) and mode sequence (magenta) of a 2-mode SLDS, where the first mode can be realized by the first two state components and the second mode solely by the first. The associated 10th, 50th, and 90th Hamming distance quantiles over 100 trials are shown for the (b) MNIW and (c) ARD prior. (d)-(e) Histograms of inferred ARD precisions associated with the first and second dynamical modes, respectively, at the 5000th Gibbs iteration. Larger values correspond to non-dynamical components.} \label{fig:ARD} %\vspace{-0.2in}
\end{figure*}

In Fig.~\ref{fig:ARD}, we see that even in this low-dimensional example, the ARD provides superior mode-sequence estimates, as well as a mechanism for identifying non-dynamical state components. The histograms of the inferred $\BF{\alpha}^{(k)}$ are shown in Fig.~\ref{fig:ARD}(d)-(e). From the clear separation between the sampled dynamic range of $\alpha^{(1)}_3$ and ($\alpha^{(1)}_1, \alpha^{(1)}_2$), and between that of ($\alpha^{(2)}_2, \alpha^{(2)}_3$) and $\alpha^{(2)}_1$, we see that we are able to correctly identify dynamical systems with $\symk{\BF{a}_3}{1} = 0$ and $\symk{\BF{a}_2}{2} = \symk{\BF{a}_3}{2} = 0$.
\subsection{Dancing Honey Bees} \label{sec:BeeResults}
Honey bees perform a set of dances within the beehive in order to communicate the location of food sources. Specifically, they switch between a set of \emph{waggle}, \emph{turn-right}, and \emph{turn-left} dances. During the waggle dance, the bee walks roughly in a straight line while rapidly shaking its body from left to right. The turning dances simply involve the bee turning in a clockwise or counterclockwise direction. We display six such sequences of honey bee dances in Fig.~\ref{fig:dancingbee_trajectories}. The data consist of measurements $\BF{y}_t = [\cos(\theta_t) \ \ \sin(\theta_t) \ \ x_t \ \ y_t]^T$, where $(x_t,y_t)$ denotes the 2D coordinates of the bee's body and $\theta_t$ its head angle\footnote{The data are available at \mbox{http://www.cc.gatech.edu/~borg/ijcv\_psslds/}.}. Both Oh et. al.~\cite{Oh:08} and Xuan and Murphy~\cite{Xuan:07} used switching dynamical models to analyze these honey bee dances. We wish to analyze the performance of our Bayesian nonparametric variants of these models in segmenting the six sequences into the dance labels displayed in Fig.~\ref{fig:dancingbee_trajectories}.
\subsubsection*{MNIW Prior --- Unsupervised} 
\begin{figure*}[t!] \centering 
	\begin{tabular}{cccc} 
		\hspace{-0.15in}\includegraphics[height = 1.5in]{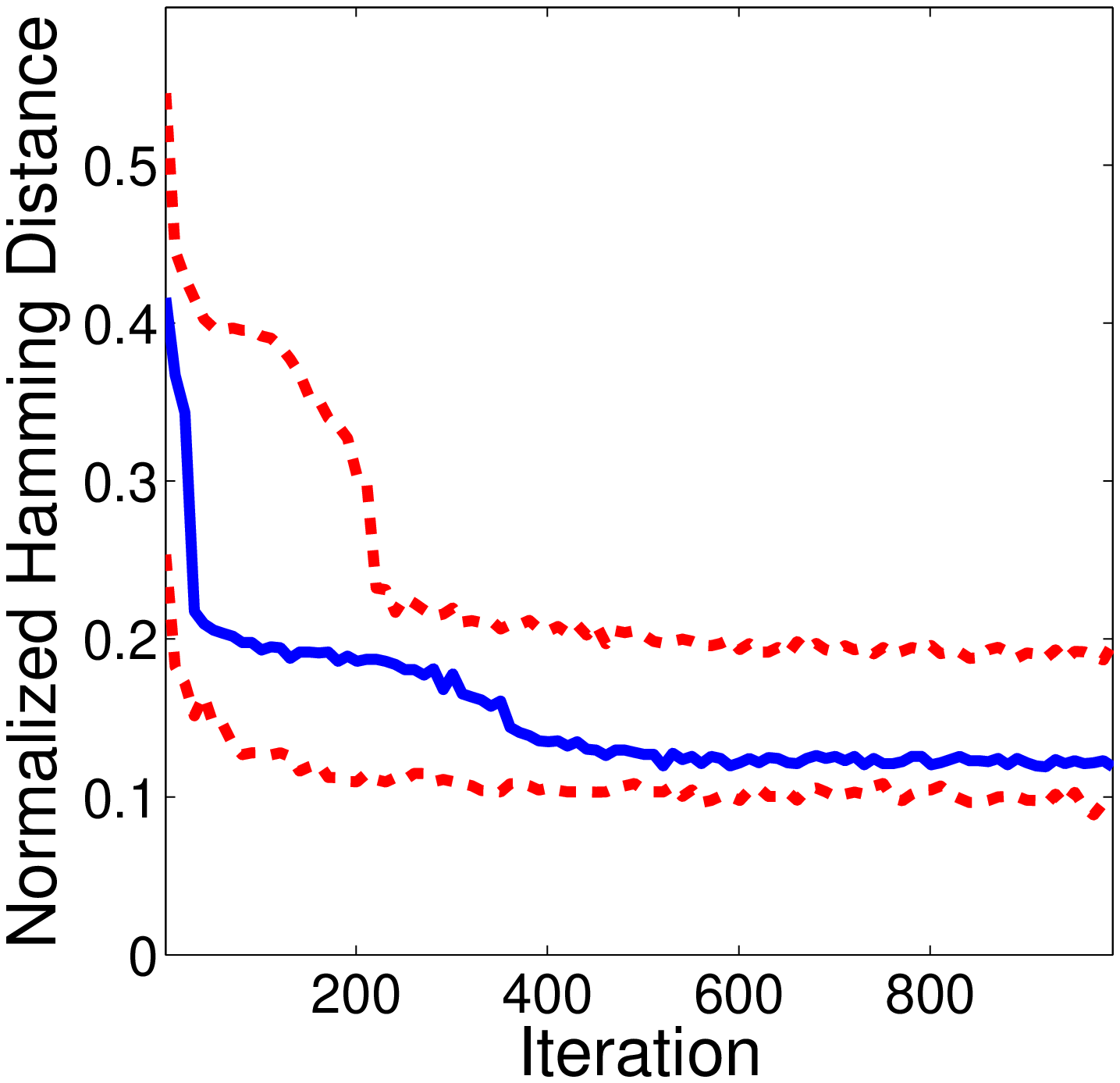}& 
		\includegraphics[height = 1.5in]{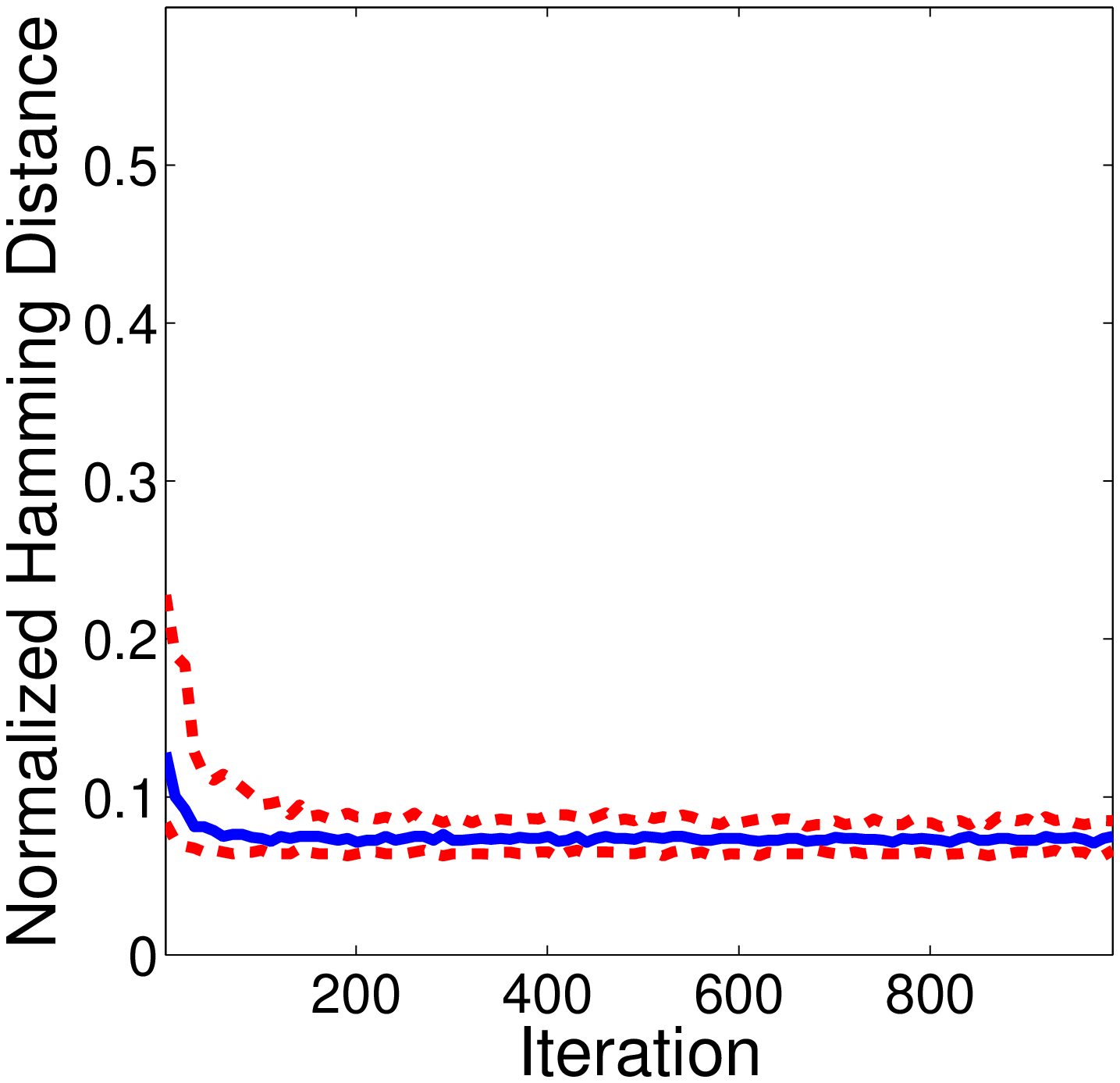}& 
		\includegraphics[height = 1.5in]{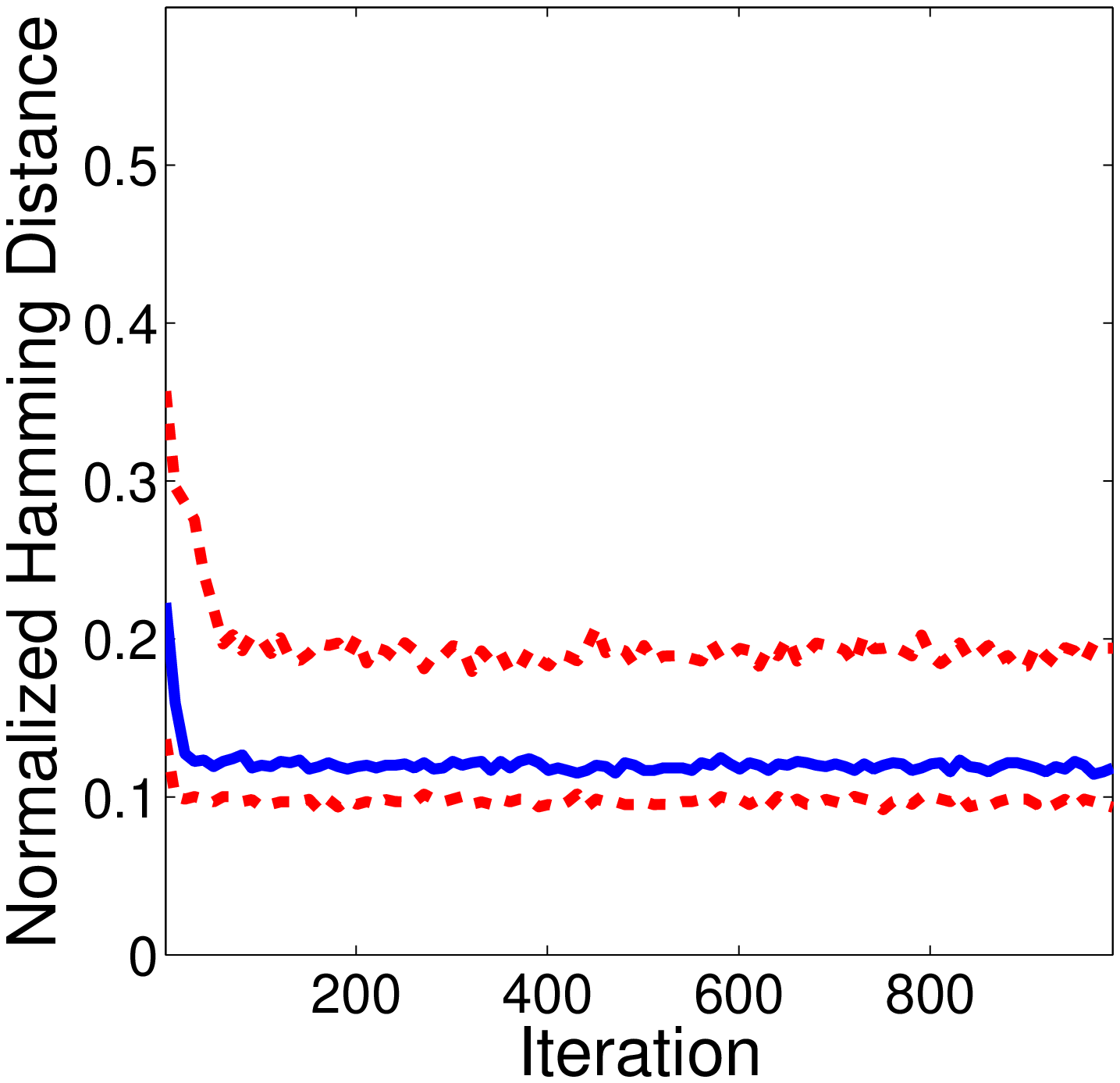}&
		\includegraphics[height = 1.5in]{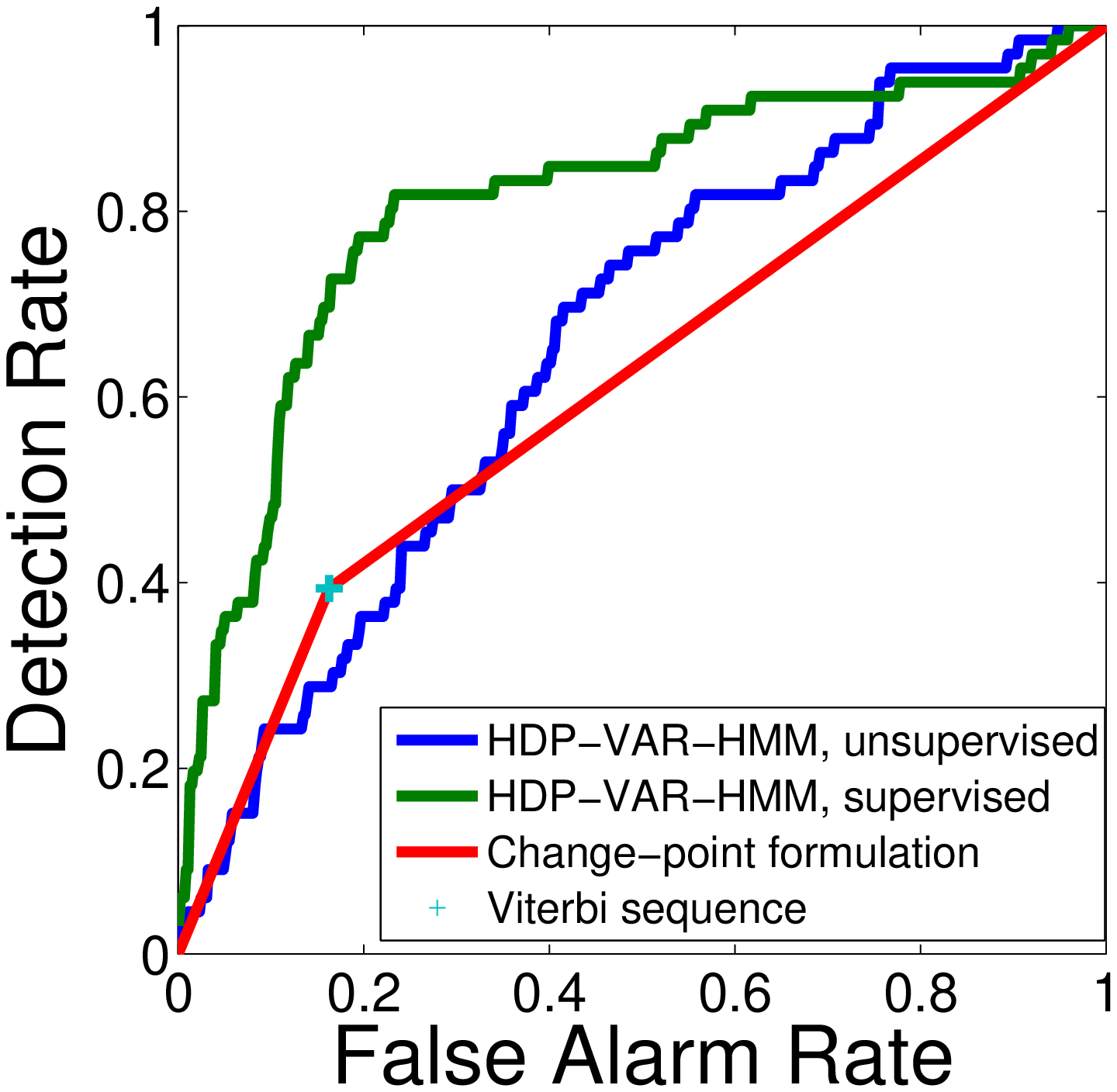}\\%\vspace{-0.1in}\\
		\hspace{-0.15in}(a) & (b) & (c) & (d)\vspace{0.05in}\\
		\hspace{-0.15in}\includegraphics[height = 1.5in]{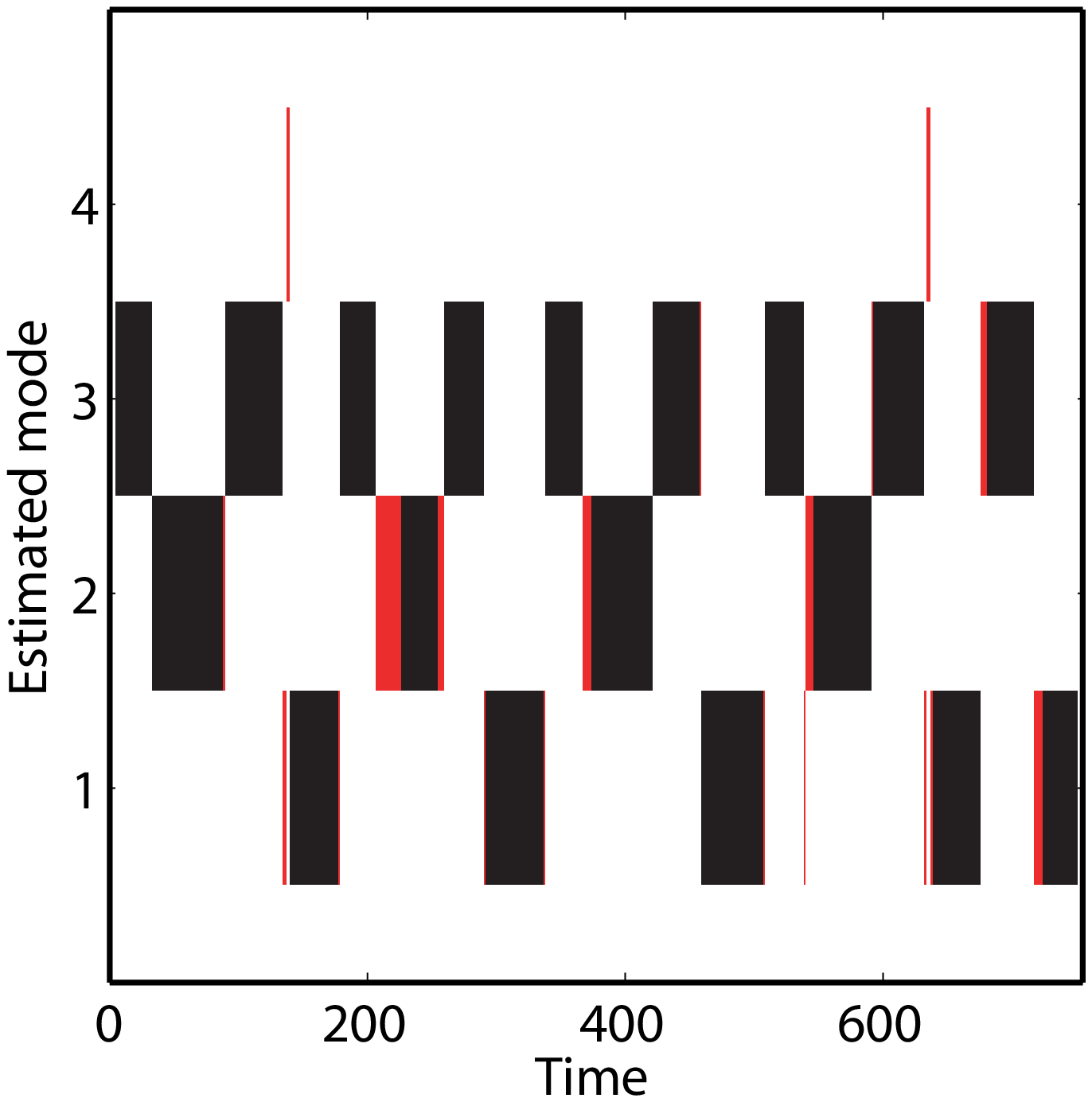}& 
		\includegraphics[height = 1.5in]{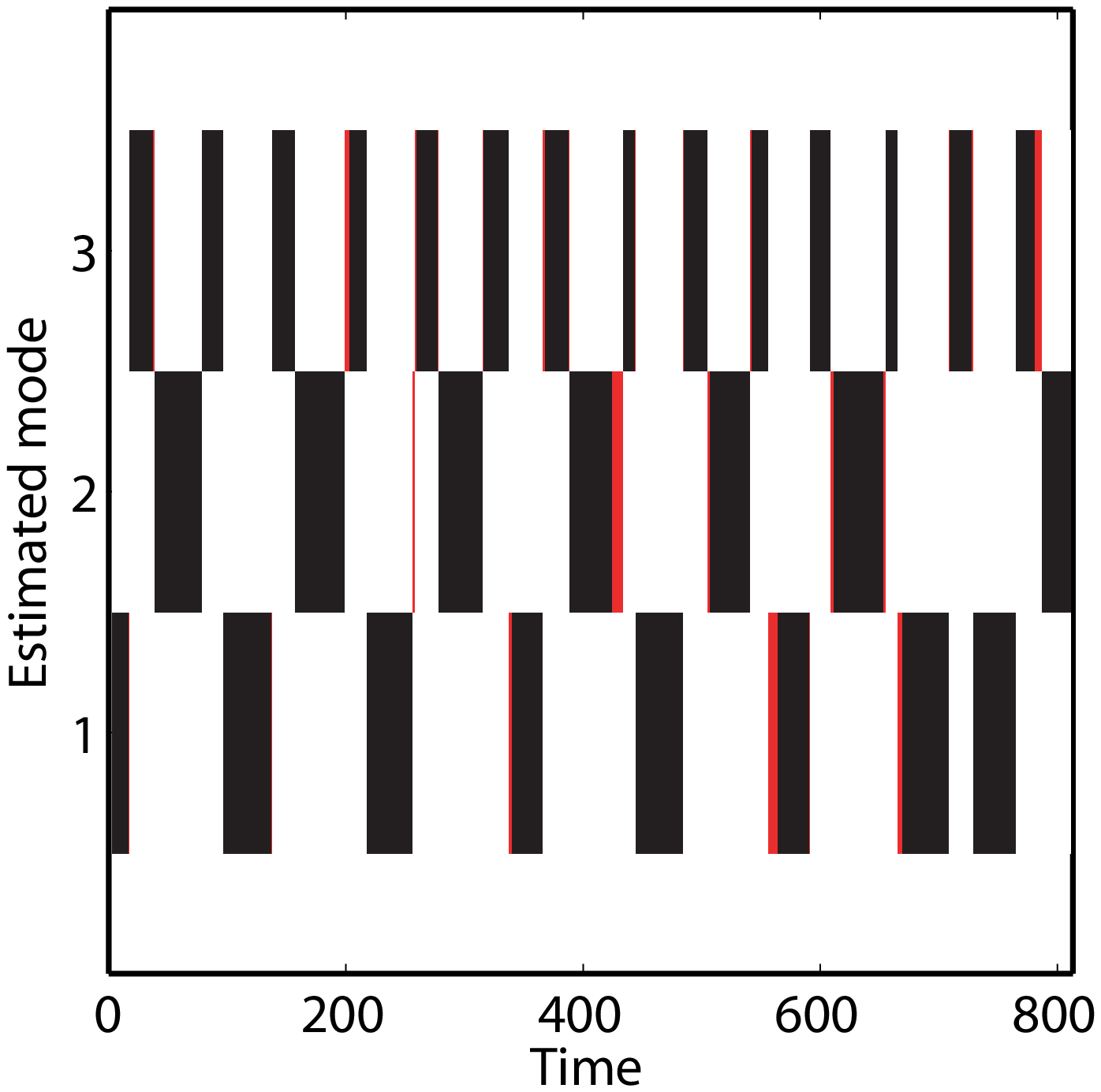}& 
		\includegraphics[height = 1.5in]{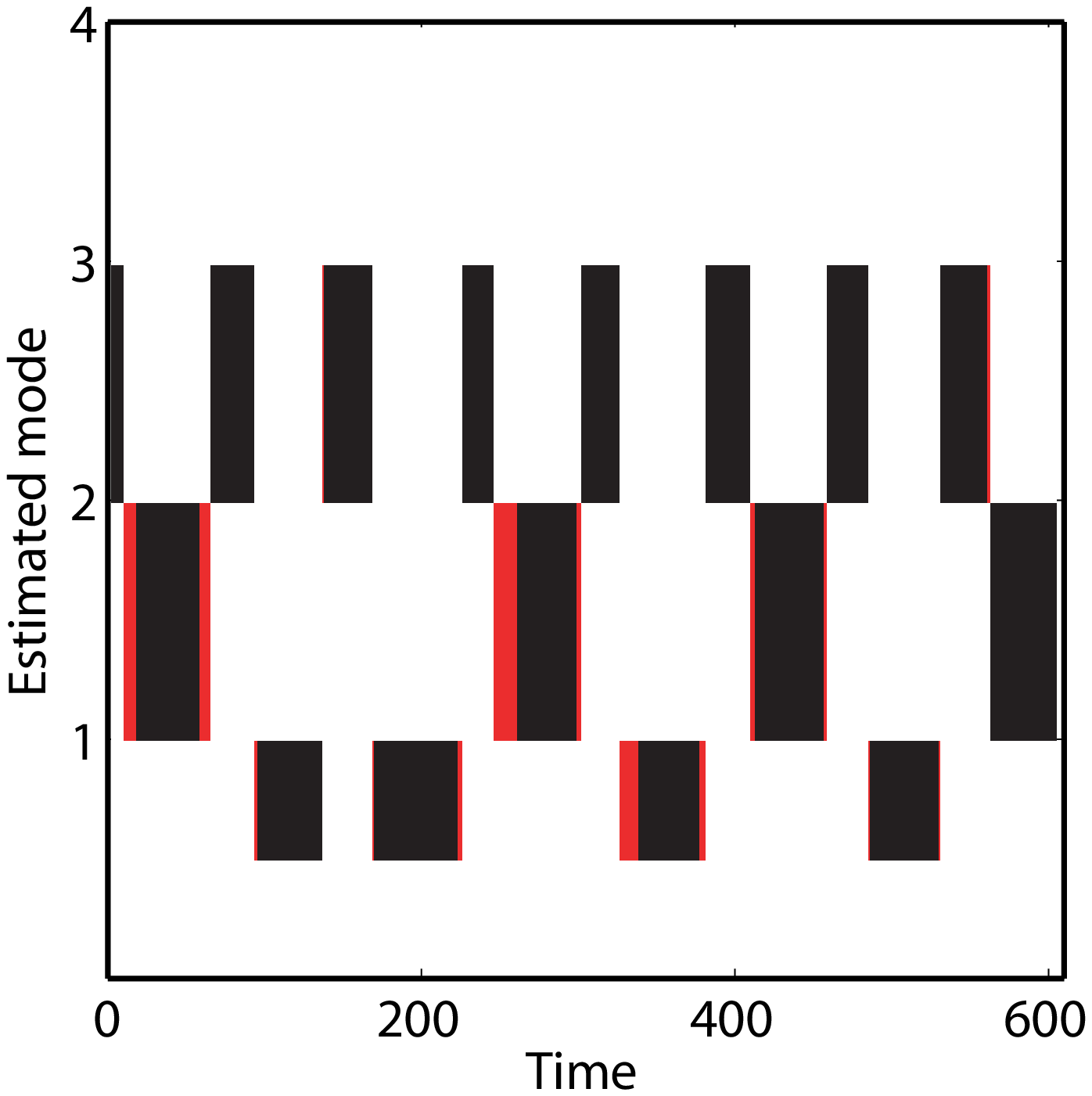}&
		\includegraphics[height = 1.5in]{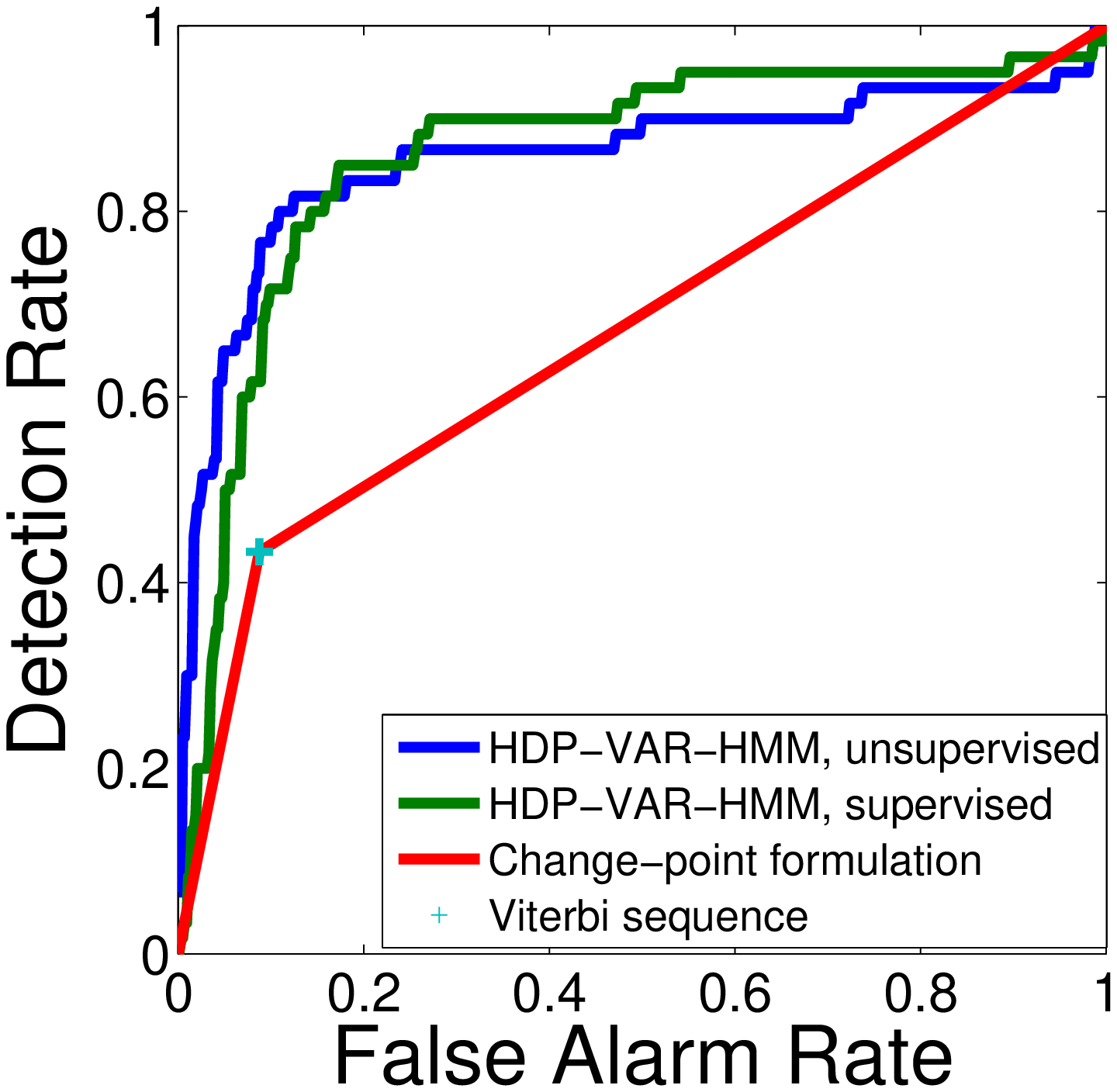}\\%\vspace{-0.1in}\\
		\hspace{-0.15in}(e) & (f) & (g) & (h) %\vspace{-0.1in}
	\end{tabular}
	\caption[Segmentation performance of the HDP-AR-HMM on the six honey bee dance sequences.] {(a)-(c) The 10th, 50th, and 90th Hamming distance quantiles over 100 trials are shown for sequences 4, 5, and 6, respectively. (e)-(g) Estimated mode sequences representing the median error for sequences 4, 5, and 6 at the 200th Gibbs iteration, with errors indicated in red.
(d) and (h) ROC curves for the unsupervised HDP-VAR-HMM, partially supervised HDP-VAR-HMM, and change-point formulation of \cite{Xuan:07} using the Viterbi sequence for segmenting datasets 1-3 and 4-6, respectively.} \label{fig:dancingbee_Hamming} %\vspace{-0.2in}
\end{figure*}

We start by testing the HDP-VAR($1$)-HMM using a MNIW prior. (Note that we did not see performance gains by considering the HDP-SLDS, so we omit showing results for that architecture.) We set the prior distributions on the dynamic parameters and hyperparameters as in Sec.~\ref{sec:MNIWresults} for the synthetic data examples, with the MNIW prior based on a pre-processed observation sequence. %with mean matrix $M=\BF{0}$, $K = 0.1*I_m$, degrees of freedom $n_0=6$, and scale matrix $S_0$ set to 0.75 times the empirical covariance of a pre-processed observation sequence. 
The pre-processing involves centering the position observations around $0$ and scaling each component of $\BF{y}_t$ to be within the same dynamic range. %As in the synthetic data examples, we set the $\mbox{Gamma}(a,b)$ priors on the concentration parameters $\alpha+\kappa$ and $\gamma$ to have $a=1$ and $b=0.01$, and the $\mbox{Beta}(c,d)$ prior on $\rho$ to have $c=10$ and $d=1$.
\begin{figure*}[t!] \centering 
	\begin{tabular}{cccccc} 
		\hspace{-0.2in} \includegraphics[width = 0.15\columnwidth]{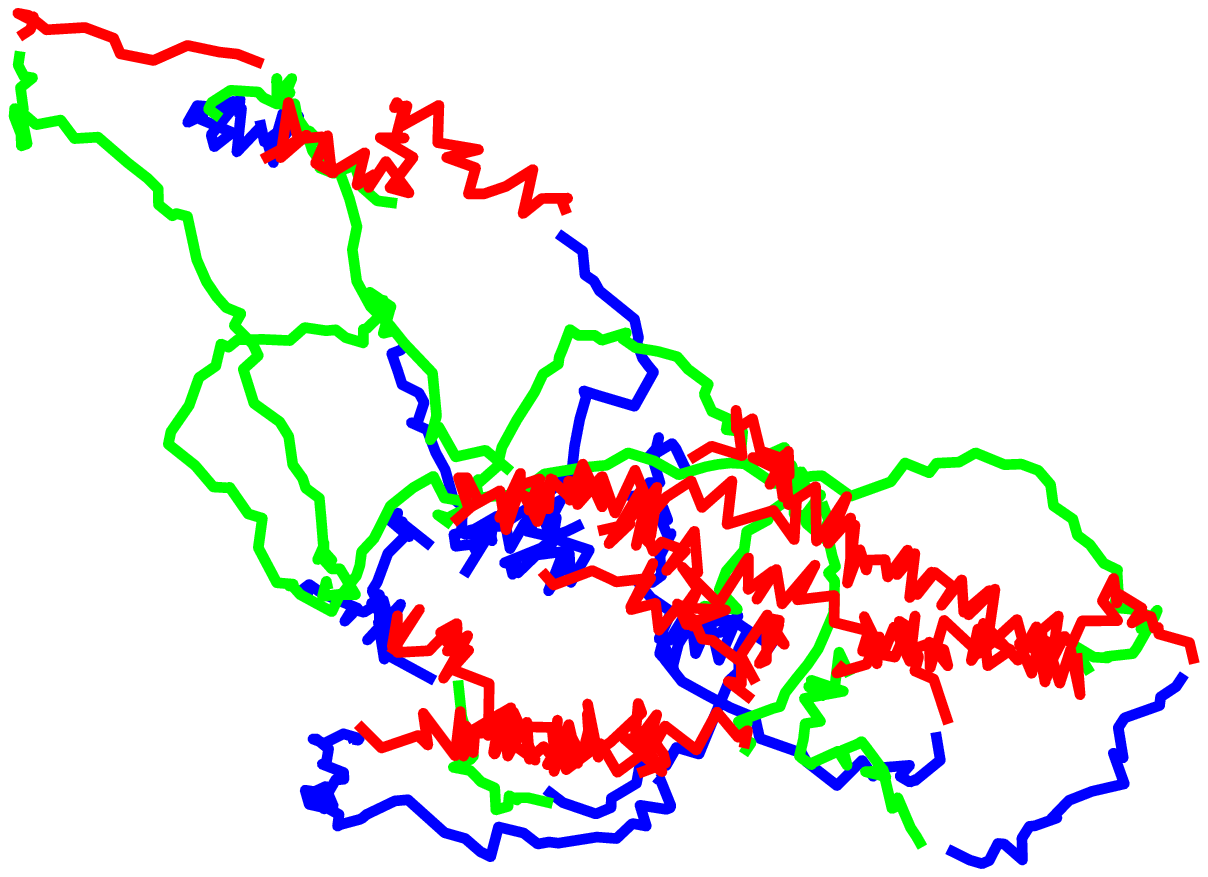} & 
		\hspace{-0.05in} \includegraphics[width = 0.15\columnwidth]{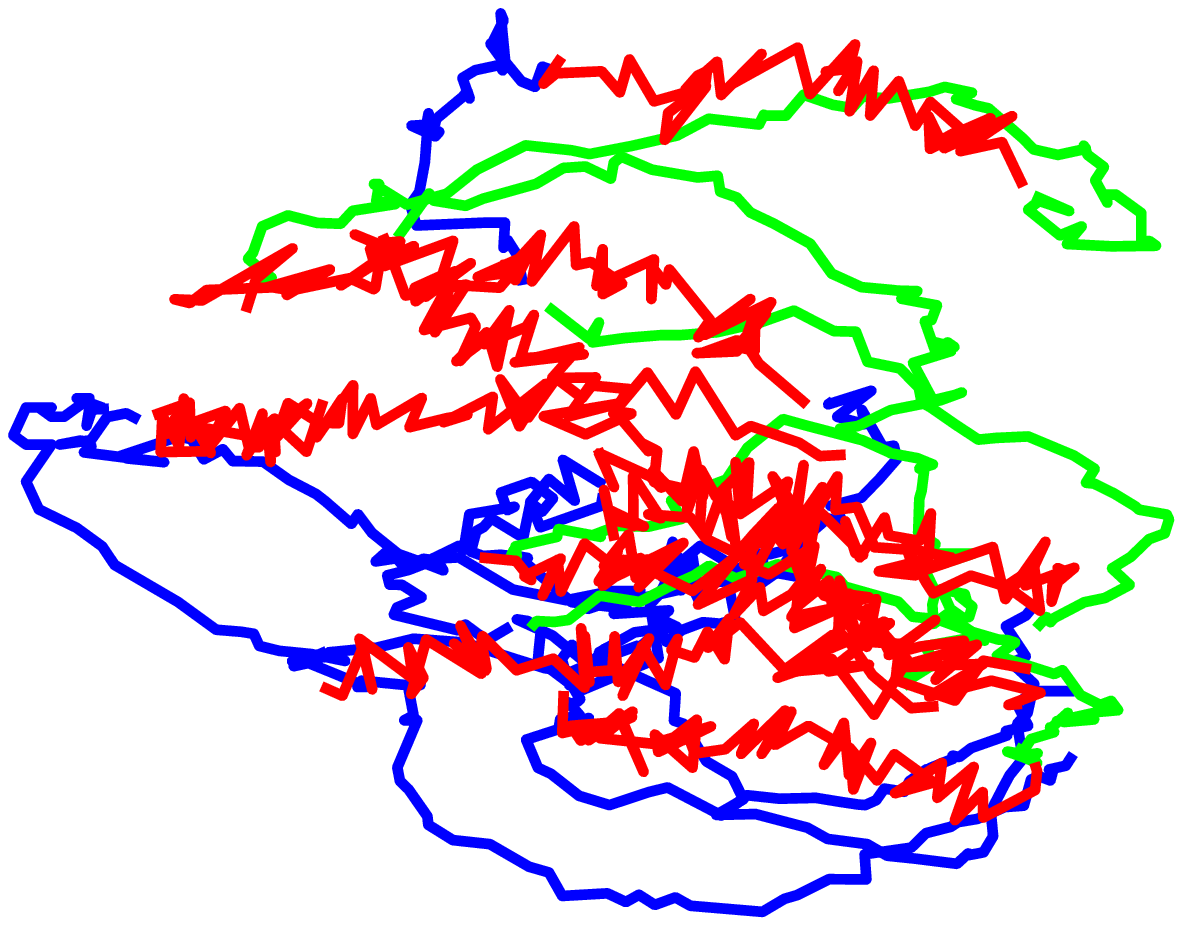} & 
		\hspace{-0.05in} \includegraphics[width = 0.15\columnwidth]{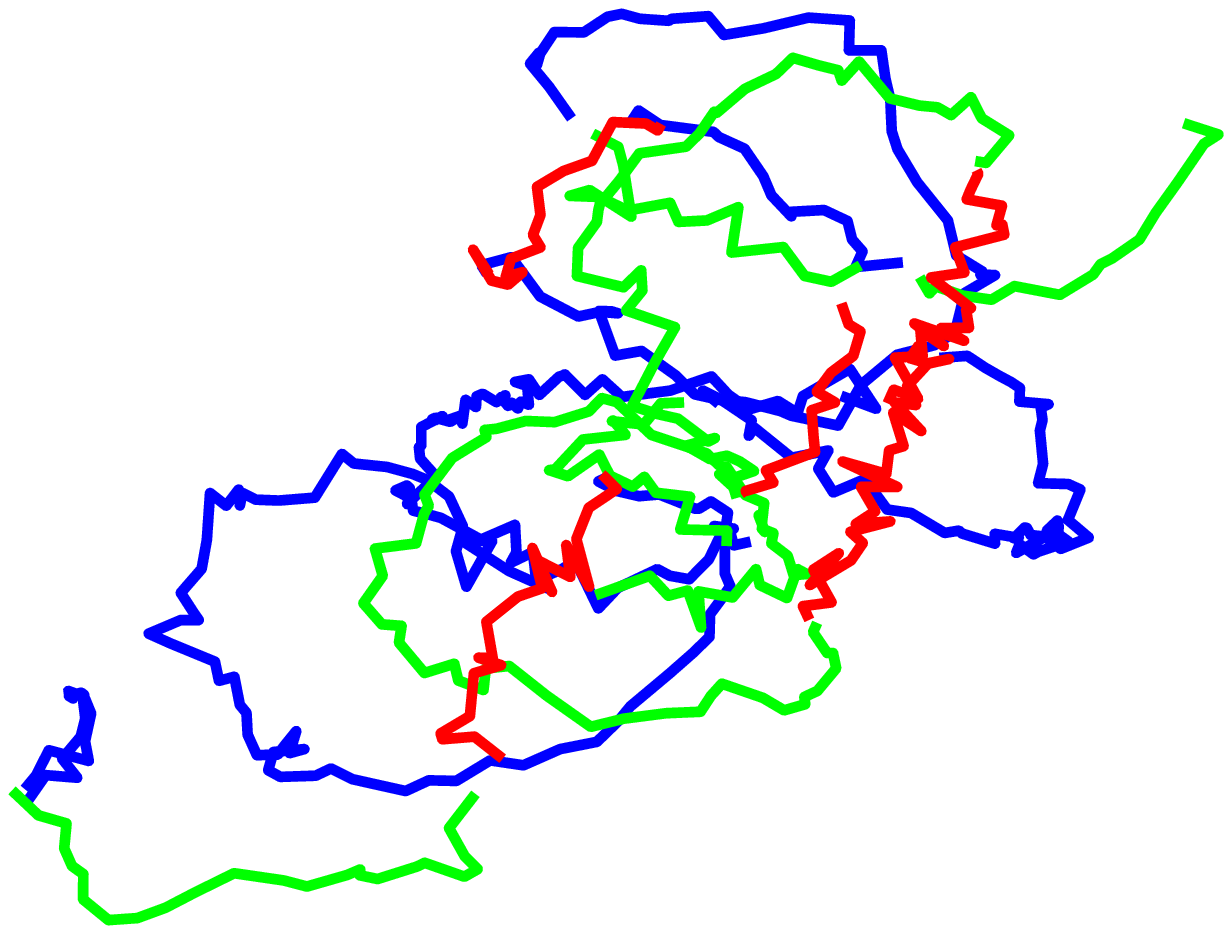} &
		\hspace{-0.05in} \includegraphics[width = 0.15\columnwidth]{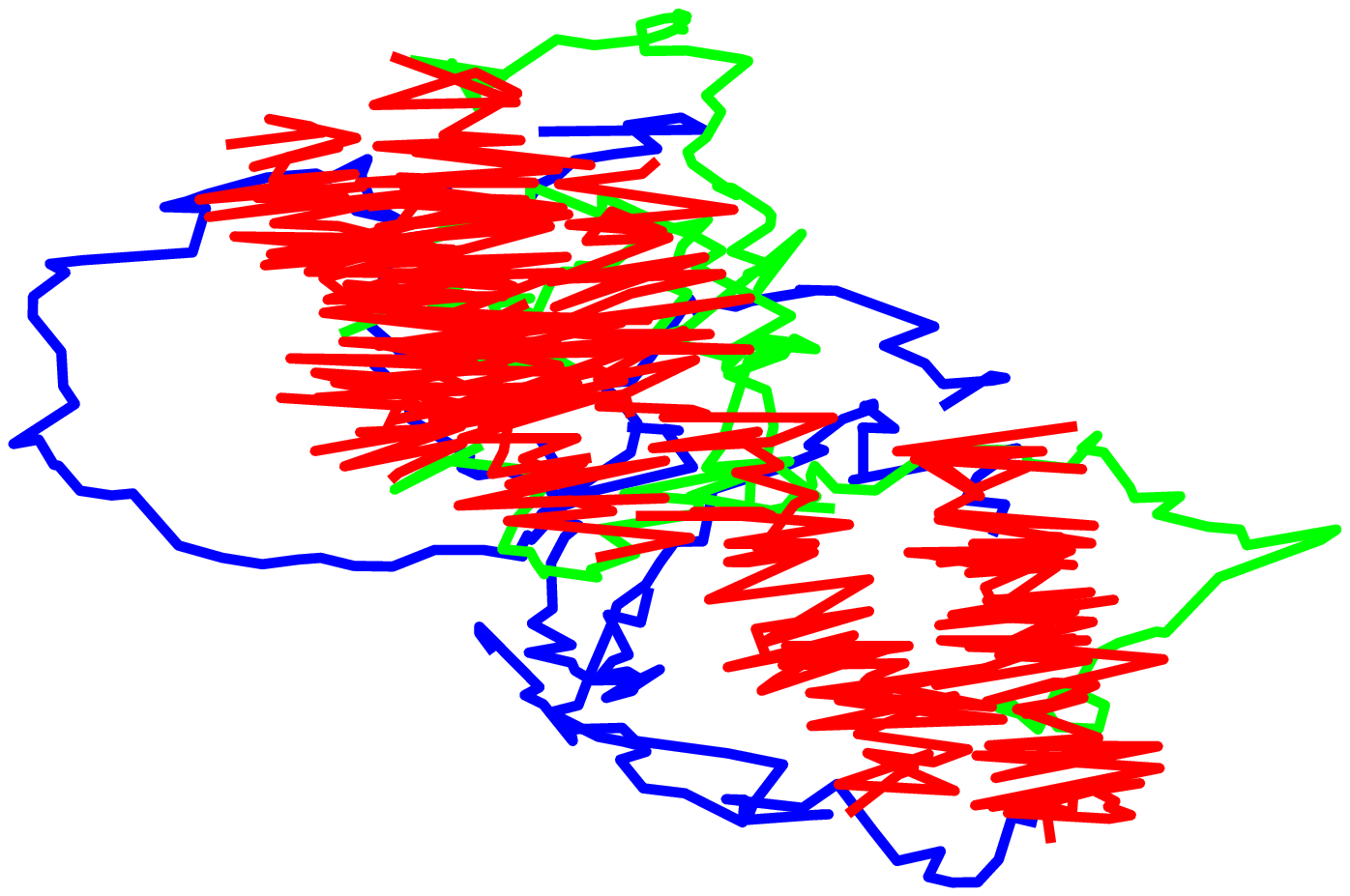} & 
		\hspace{-0.05in} \includegraphics[width = 0.15\columnwidth]{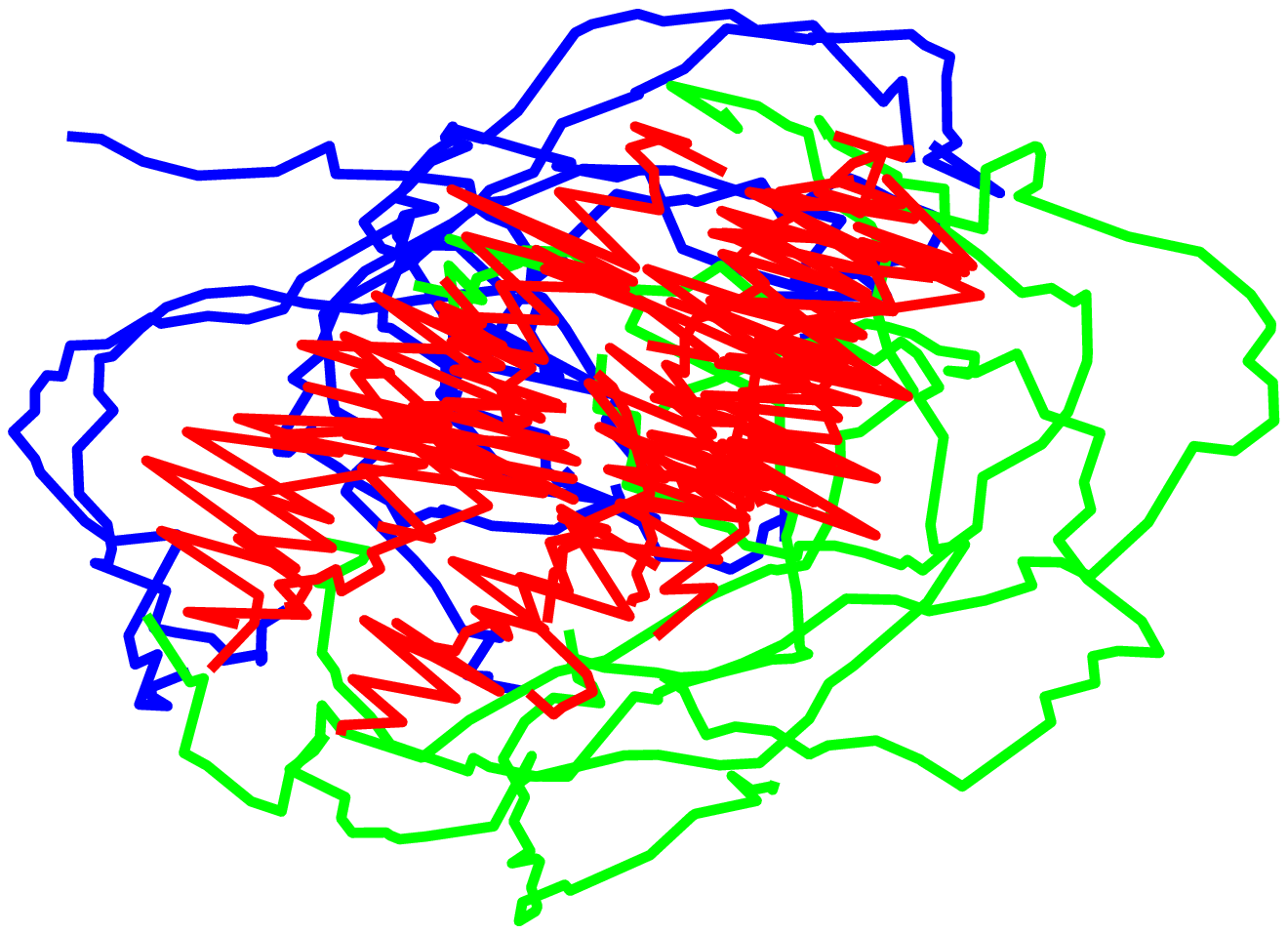} & 
		\hspace{-0.05in} \includegraphics[width = 0.15\columnwidth]{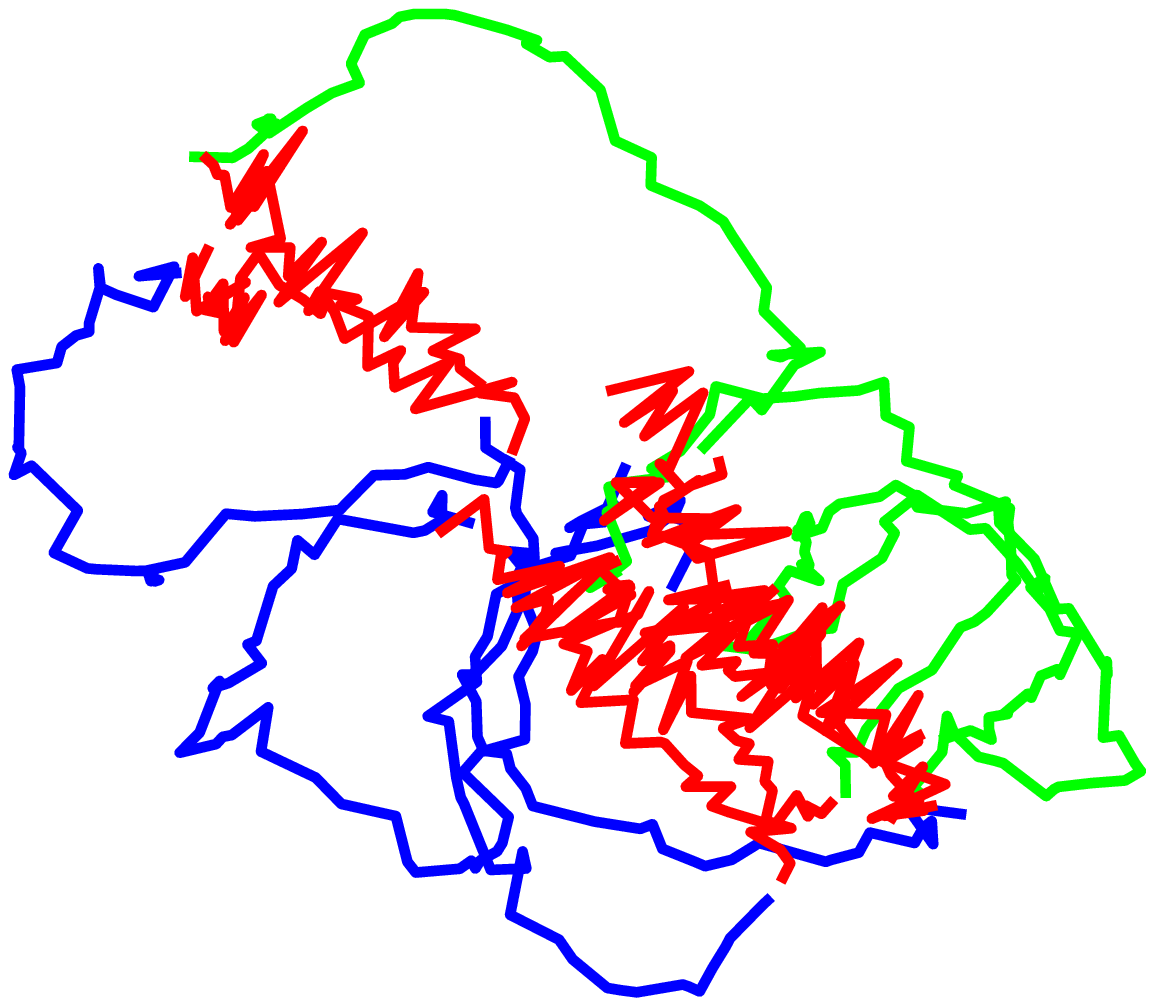}\\
		\hspace{-0.2in} \includegraphics[width = 0.15\columnwidth]{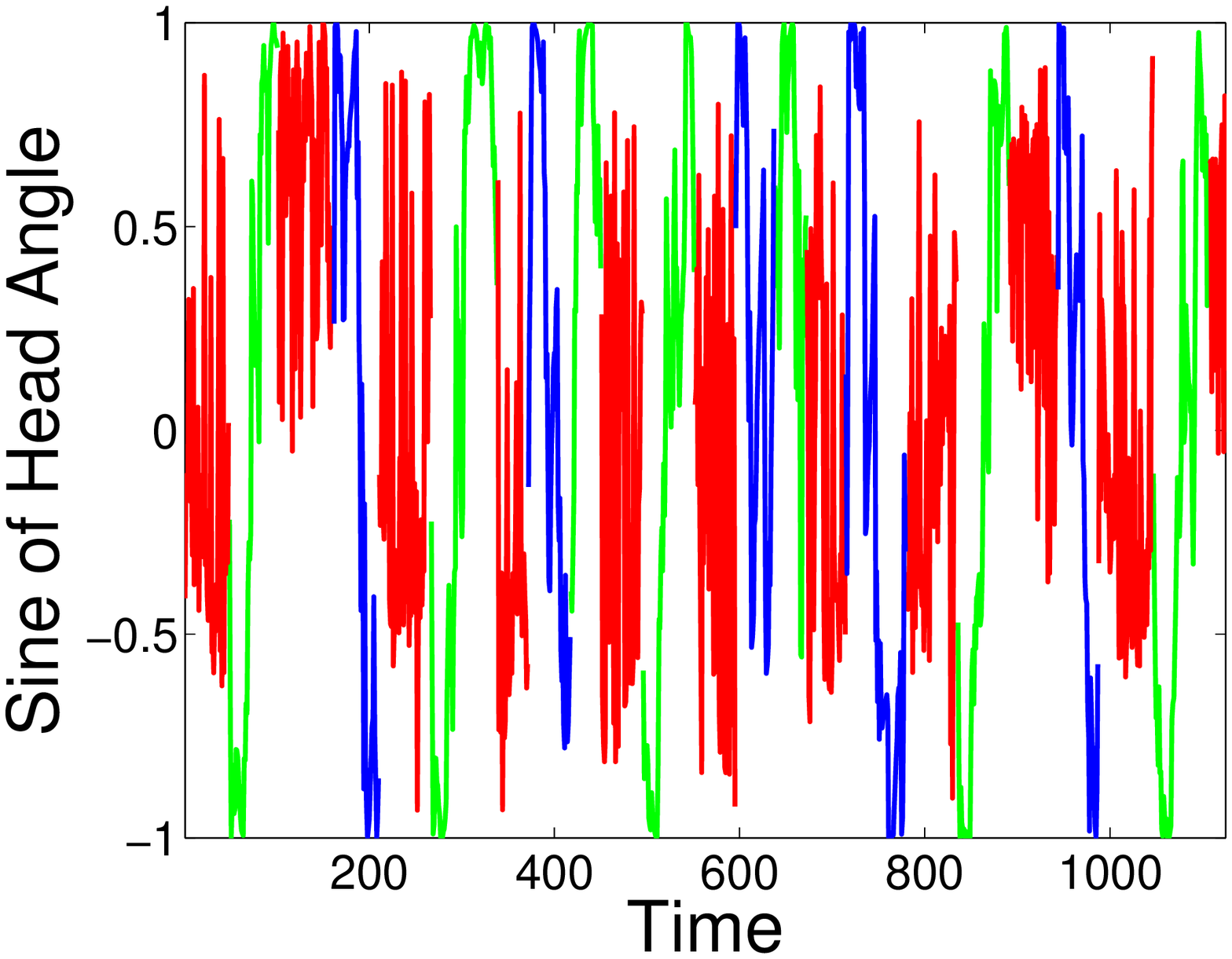} & 
		\hspace{-0.05in} \includegraphics[width = 0.15\columnwidth]{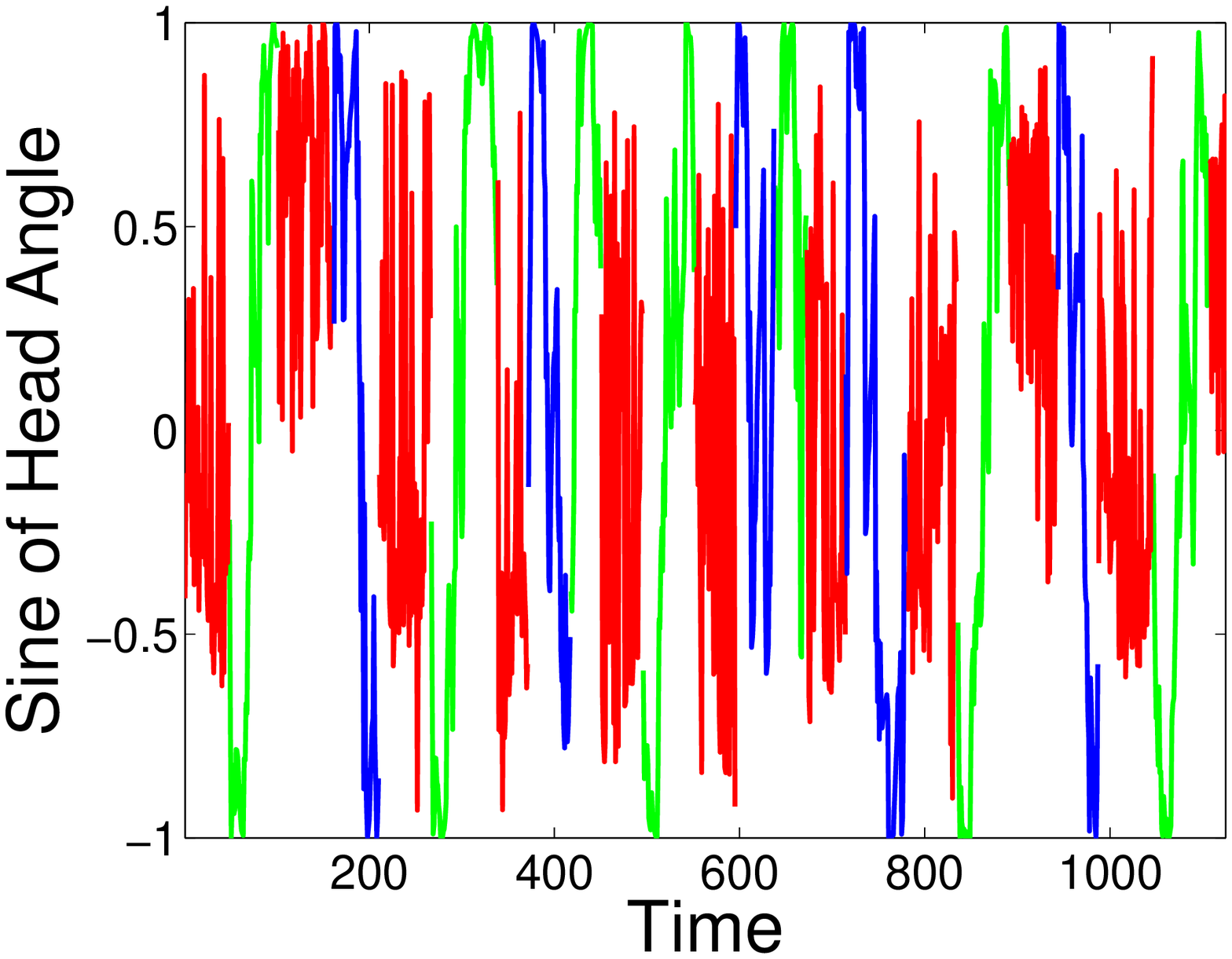} & 
		\hspace{-0.05in} \includegraphics[width = 0.15\columnwidth]{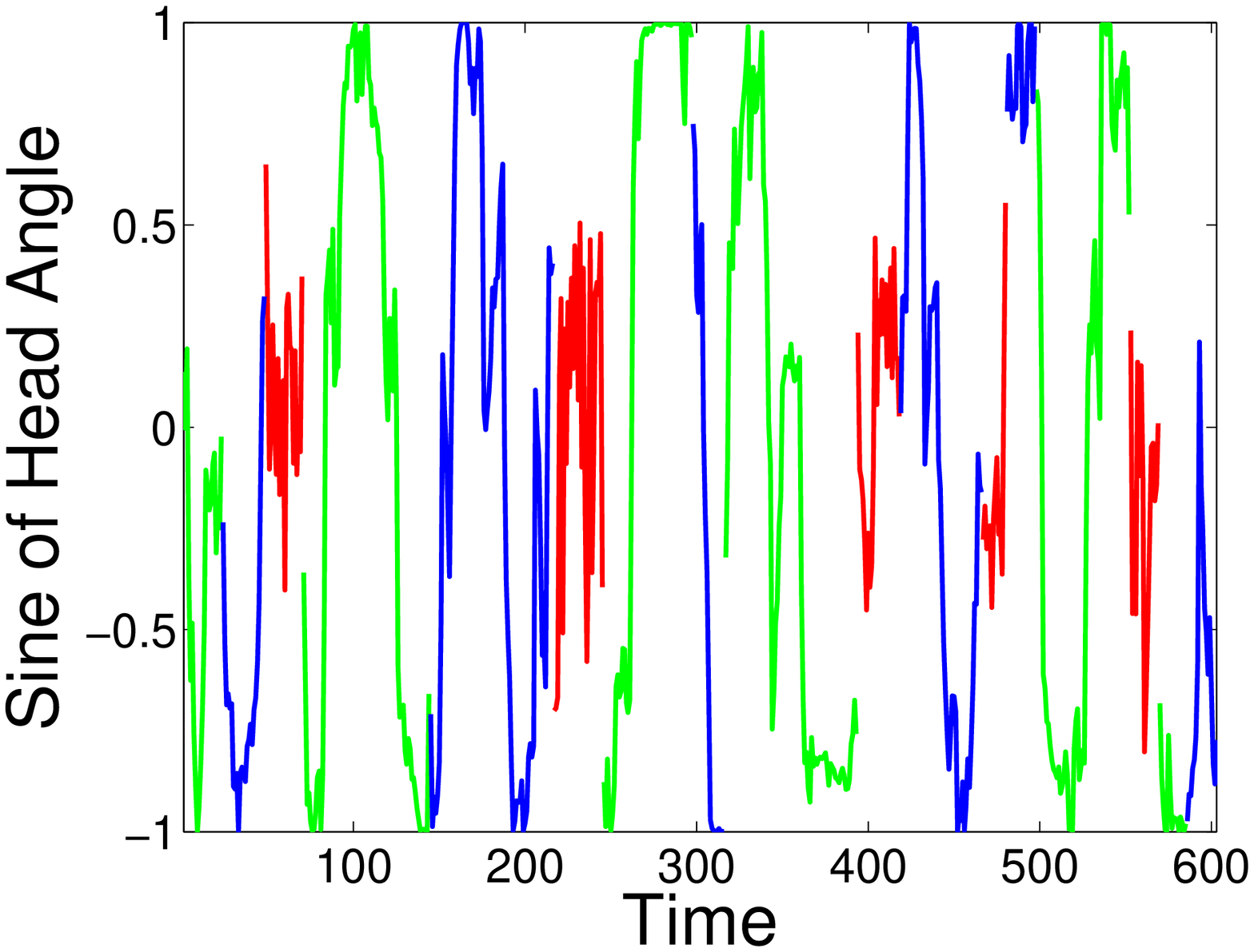} &
		\hspace{-0.05in} \includegraphics[width = 0.15\columnwidth]{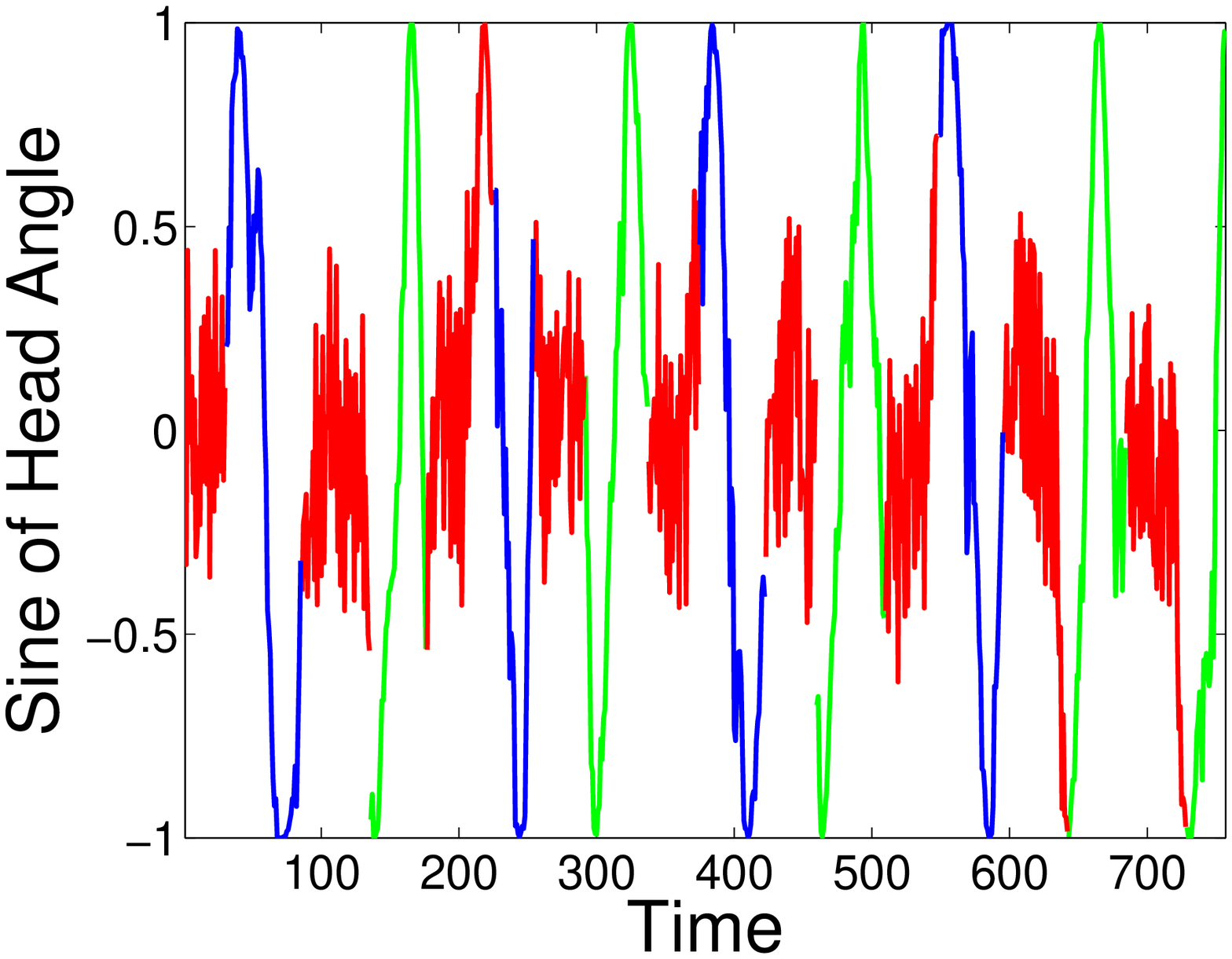} & 
		\hspace{-0.05in} \includegraphics[width = 0.15\columnwidth]{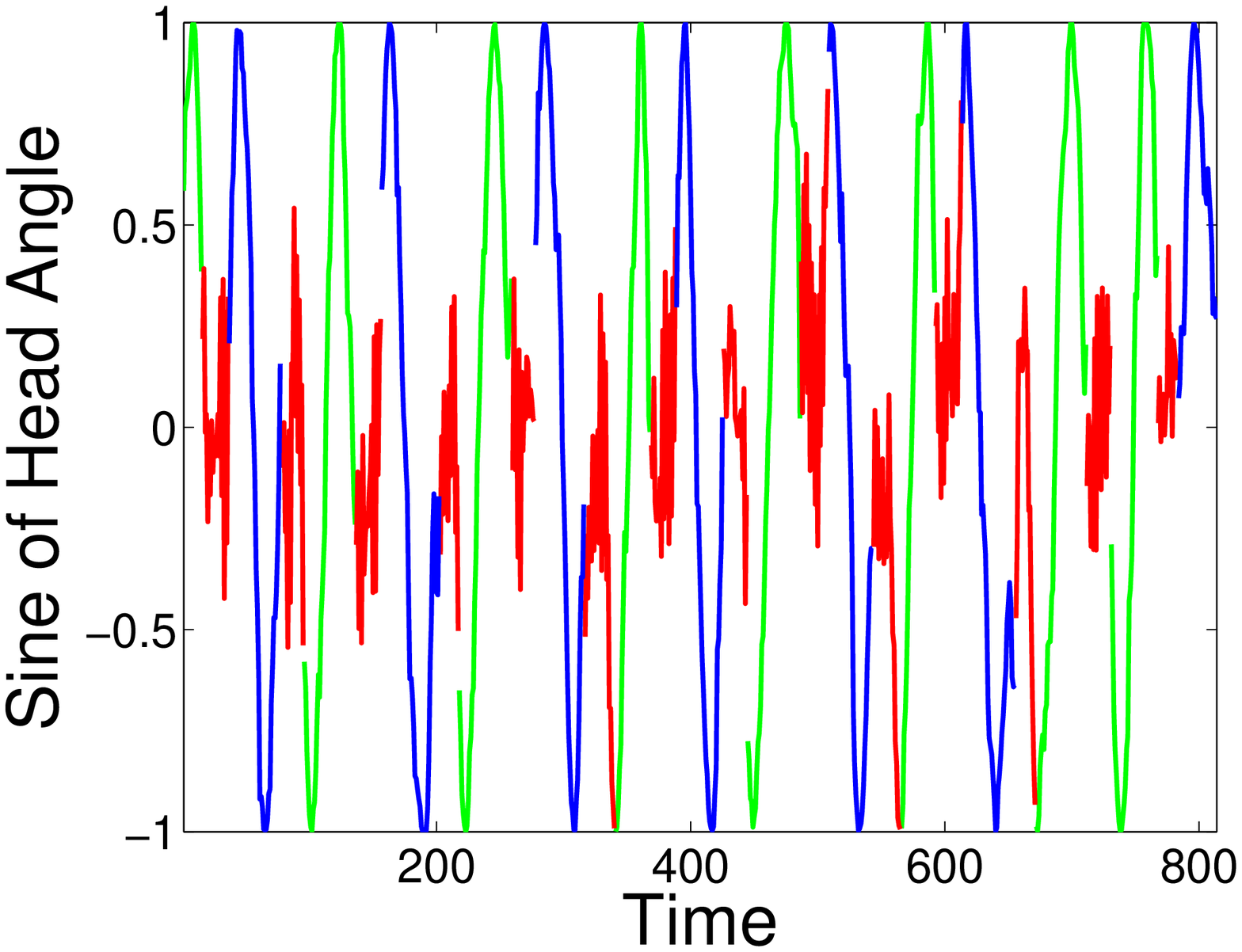} & 
		\hspace{-0.05in} \includegraphics[width = 0.15\columnwidth]{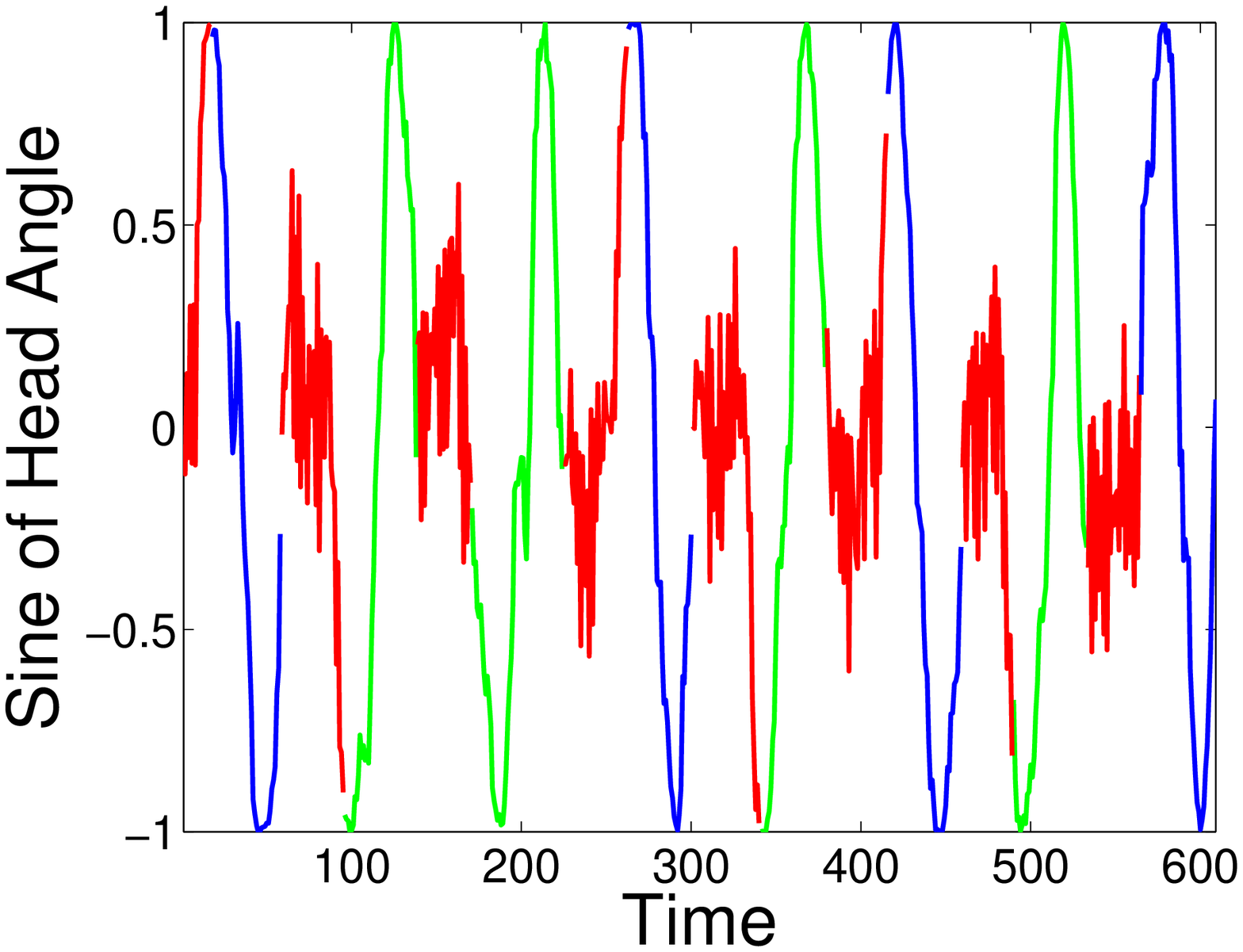}\\ %\vspace{-0.1in}\\
		\hspace{-0.2in} (1) &\hspace{-0.05in} (2) &\hspace{-0.05in} (3) &\hspace{-0.05in} (4) &\hspace{-0.05in} (5) &\hspace{-0.05in} (6) %\vspace{-0.1in}
	\end{tabular}
	\caption{\textit{Top:} Trajectories of the dancing honey bees for sequences 1 to 6, colored by \emph{waggle} (red), \emph{turn right} (blue), and \emph{turn left} (green) dances. \textit{Bottom:} Sine of the bee's head angle measurements colored by ground truth labels.} \label{fig:dancingbee_trajectories} %\vspace{-0.2in}
\end{figure*}
We compare our results to those of Xuan and Murphy~\cite{Xuan:07}, who used a change-point detection technique for inference on this dataset. As shown in Fig.~\ref{fig:dancingbee_Hamming}(d) and (h), our model achieves a superior segmentation compared to the change-point formulation in almost all cases, while also identifying modes which reoccur over time.
%%
%\begin{figure*}[t!] \centering 
%	\begin{tabular}{cc} 
%		\includegraphics[height = 1.75in]{\figdir/ROC_1to3} & 
%		\includegraphics[height = 1.75in]{\figdir/ROC_4to6} \vspace{-0.1in}\\
%		(a) & (b) \vspace{-0.1in}
%	\end{tabular}
%	\caption[Change-point detection performance of the HDP-AR-HMM on the six honey bee dance sequences as compared to the method of Xuan and Murphy~\cite{Xuan:07}.]{ROC curves for the %unsupervised HDP-VAR-HMM, partially supervised HDP-VAR-HMM, and change-point formulation of Xuan and Murphy~\cite{Xuan:07} using the Viterbi sequence for segmenting datasets (a) 1-3 and (b) %4-6.} \label{fig:dancingbee_changepoint} 
%\end{figure*}
%%
Oh et. al.\cite{Oh:08} also presented an analysis of the honey bee data, using an SLDS with a fixed number of modes. Unfortunately, that analysis is not directly comparable to ours, because Oh et. al.~\cite{Oh:08} used their SLDS in a supervised formulation in which the ground truth labels for all but one of the sequences are employed in the inference of the labels for the remaining held-out sequence, and in which the kernels used in the MCMC procedure depend on the ground truth labels. (The authors also considered a ``parameterized segmental SLDS (PS-SLDS),'' which makes use of domain knowledge specific to honey bee dancing and requires additional supervision during the learning process.) Nonetheless, in Table~\ref{table:bee} we report the performance of these methods as well as the median performance (over 100 trials) of the unsupervised HDP-VAR($1$)-HMM in order to provide a sense of the level of performance achievable without detailed, manual supervision. As seen in Table~\ref{table:bee}, the HDP-VAR($1$)-HMM yields very good performance on sequences 4 to 6 in terms of the learned segmentation and number of modes (see Fig.~\ref{fig:dancingbee_Hamming}); the performance approaches that of the supervised method.
For sequences 1 to 3---which are much less regular than sequences 4 to 6---the performance of the unsupervised procedure is substantially worse. In Fig.~\ref{fig:dancingbee_trajectories}, we see the extreme variation in head angle during the waggle dances of sequences 1 to 3.\footnote{From Fig.~\ref{fig:dancingbee_trajectories}, we also see that even in sequences 4 to 6, the ground truth labeling appear to be inaccurate at times. Specifically, certain time steps are labeled as waggle dances (red) that look more typical of a turning dance (green, blue).} As noted by Oh, the tracking results based on the vision-based tracker are noisier for these sequences and the patterns of switching between dance modes is more irregular. This dramatically affects our performance since we do not use domain-specific information. Indeed, our learned segmentations consistently identify turn-right and turn-left modes, but often create a new, sequence-specific waggle dance mode. Many of our errors can be attributed to creating multiple waggle dance modes within a sequence. Overall, however, we are able to achieve reasonably good segmentations without having to manually input domain-specific knowledge.
\subsubsection*{MNIW Prior --- Partially Supervised} The discrepancy in performance between our results and the supervised approach of Oh et. al.~\cite{Oh:08} motivated us to also consider a partially supervised variant of the HDP-VAR($1$)-HMM in which we fix the ground truth mode sequences for five out of six of the sequences, and jointly infer both a combined set of dynamic parameters and the left-out mode sequence. This is equivalent to informing the prior distributions with the data from the five fixed sequences, and using these updated posterior distributions as the prior distributions for the held-out sequence. As we see in Table~\ref{table:bee}, this partially supervised approach considerably improves performance for these three sequences, especially sequences 2 and 3. Here, we hand-aligned sequences so that the waggle dances tended to have head angle measurements centered about $\pi/2$ radians. Aligning the waggle dances is possible by looking at the high frequency portions of the head angle measurements. Additionally, the pre-processing of the unsupervised approach is not appropriate here as the scalings and shiftings are dance-specific, and such transformations modify the associated switching VAR(1) model. Instead, to account for the varying frames of reference (i.e., point of origin for each bee body) we allowed for a mean $\symkB{\mu}{k}$ on the process noise, and placed an independent $\mathcal{N}(0,\Sigma_0)$ prior on this parameter. See the Appendix for details on how the hyperparameters of these prior distributions are set.  %We set $\Sigma_0$ to 0.75 times the scale matrix $S_0$ of the inverse-Wishart prior on $\symk{\Sigma}{k}$. Since we are not shifting and scaling the observations, we set the scale matrix $S_0$ to 0.75 times the empirical covariance of the $\emph{first difference}$ observations. We also use $n_0 = 10$ degrees of freedom, making the distribution around the expected covariance tighter than in the unsupervised case. Examining first differences is appropriate since the bee's dynamics are better approximated as a random walk than as i.i.d. observations. Using raw observations in the unsupervised approach creates a larger expected covariance matrix making the prior on the dynamic matrix less informative, which is useful in the absence of other labeled data.
\begin{table}
	\centering 
	\begin{tabular}
		{|c|c|c|c|c|c|c|} \hline		
		Sequence & 1 & 2 & 3 & 4 & 5 & 6 \\
		\hline \hline HDP-VAR($1$)-HMM unsupervised & 45.0 & 42.7 & 47.3 & 88.1 & 92.5 & 88.2 \\
		\hline HDP-VAR($1$)-HMM partially supervised & 55.0 & 86.3 & 81.7 & 89.0 & 92.4 & 89.6 \\
		\hline SLDS DD-MCMC & 74.0 & 86.1 & 81.3 & 93.4 & 90.2 & 90.4 \\
		\hline PS-SLDS DD-MCMC & 75.9 & 92.4 & 83.1 & 93.4 & 90.4 & 91.0 \\
		\hline 
	\end{tabular}
	\caption[Median label accuracy of the HDP-AR-HMM compared to accuracy of the approach of Oh. et. al.~\cite{Oh:08}.]{Median label accuracy of the HDP-VAR($1$)-HMM using unsupervised and partially supervised Gibbs sampling, compared to accuracy of the supervised PS-SLDS and SLDS procedures, where the latter algorithms were based on a supervised MCMC procedure (DD-MCMC)~\cite{Oh:08}.}\label{table:bee} \vspace{-0.3in}
\end{table}
\subsubsection*{ARD Prior}
%%
%\begin{figure*}[t!] \centering 
%	\begin{tabular}{ccc} 
%		\includegraphics[height = 1.5in]{\figdir/ARDhypers_beeSeq4_turn1_square_new} & 
%		\includegraphics[height = 1.5in]{\figdir/ARDhypers_beeSeq4_turn2_square_new} & 
%		\includegraphics[height = 1.5in]{\figdir/ARDhypers_beeSeq4_waggle_square_new}\vspace{-0.06in}\\
%		\includegraphics[height = 1.5in]{\figdir/ARDhypers_beeSeq5_turn1_square_new} & 
%		\includegraphics[height = 1.5in]{\figdir/ARDhypers_beeSeq5_turn2_square_new} & 
%		\includegraphics[height = 1.5in]{\figdir/ARDhypers_beeSeq5_waggle_square_new}\vspace{-0.06in}\\
%		\includegraphics[height = 1.5in]{\figdir/ARDhypers_beeSeq6_turn1_square_new} & 
%		\includegraphics[height = 1.5in]{\figdir/ARDhypers_beeSeq6_turn2_square_new} & 
%		\includegraphics[height = 1.5in]{\figdir/ARDhypers_beeSeq6_waggle_square_new}\vspace{-0.1in}\\
%		\hspace{-0.1in}(a) & (b) & (c)\vspace{-0.1in} 
%	\end{tabular}
%	\caption[Inferred ARD hyperparameters for the learned honey bee dance modes.]{(a)-(c) Histograms of the inferred ARD hyperparameters for the learned \emph{turn right}, \emph{turn left}, and \emph{waggle} dance modes, respectively, at the 400th Gibbs iteration for the trials with Hamming distance below the median. Larger values correspond to unnecessary lag components. Note the horizontal axis scale in column (c).} \label{fig:dancingbee_ARD} \vspace{-0.2in}
%\end{figure*}
%%
%
\begin{figure*}[t!] \centering 
	\begin{tabular}{ccc} 
		\includegraphics[height = 1.5in]{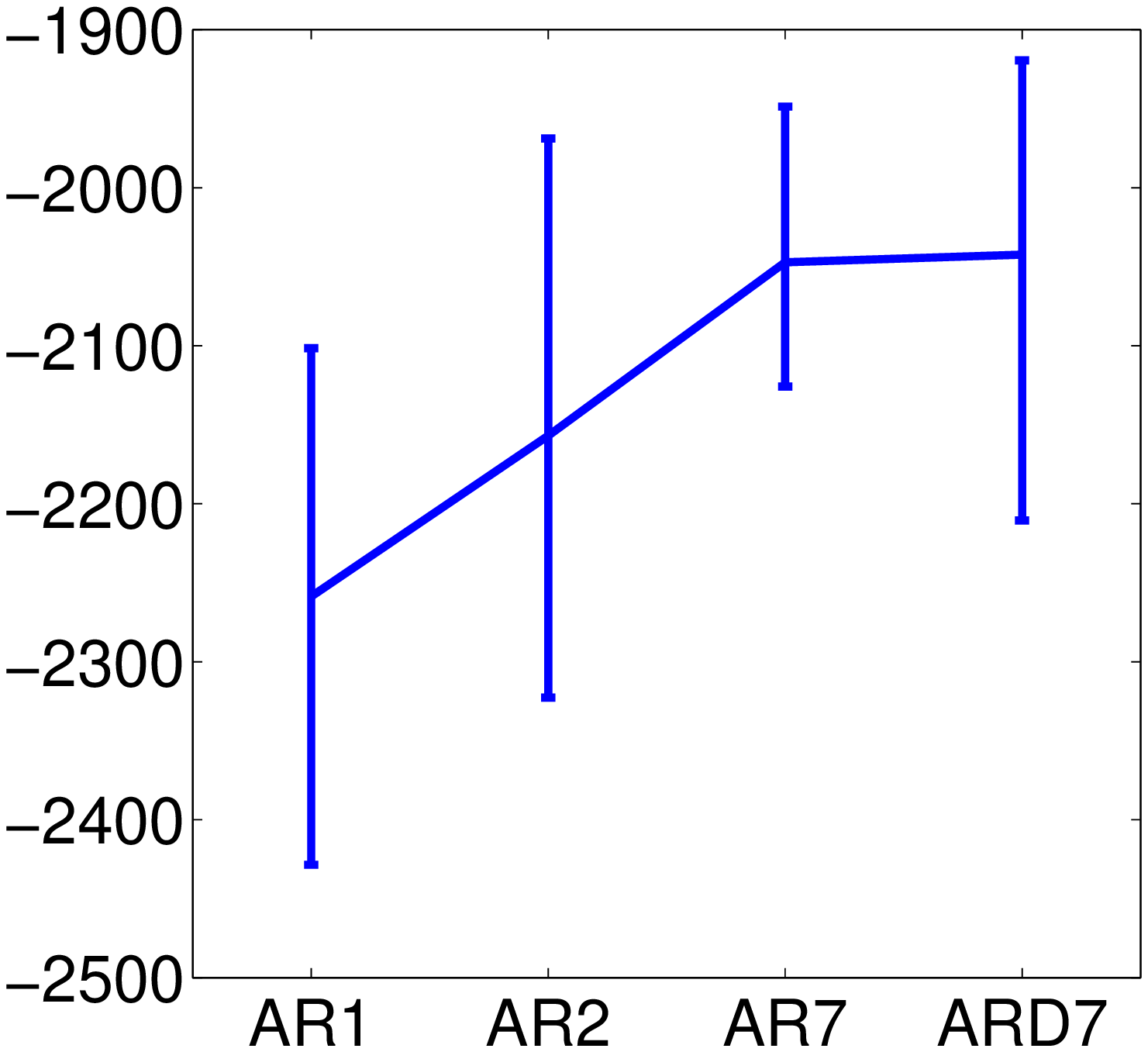} & 
		\includegraphics[height = 1.5in]{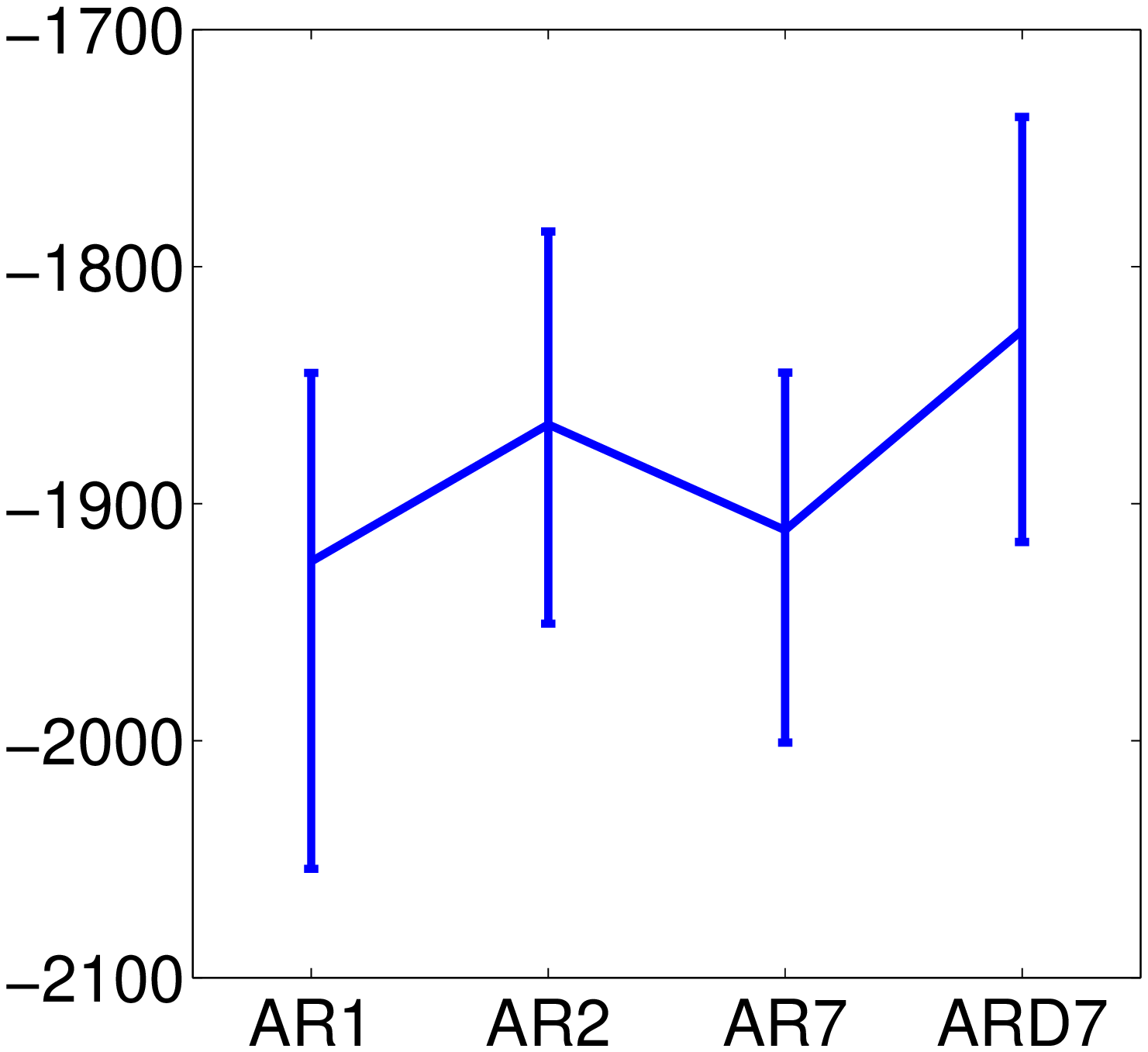} & 
		\includegraphics[height = 1.5in]{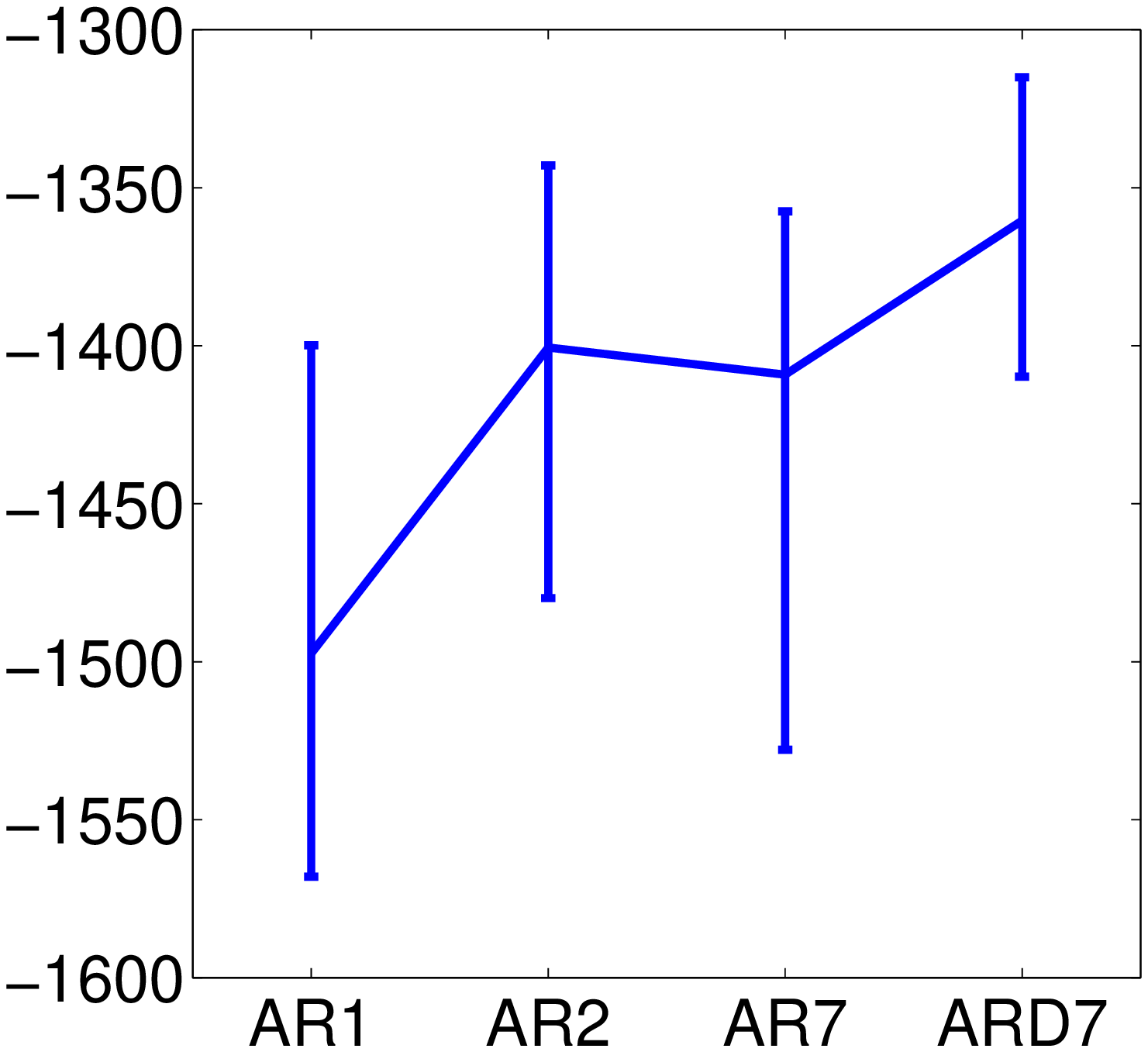}\\%\vspace{-0.1in}\\
		\hspace{-0.1in}(a) & (b) & (c)%\vspace{-0.1in} 
	\end{tabular}
	\caption{For an order 1, 2, and 7 HDP-AR-HMM with a MNIW prior and an order 7 HDP-AR-HMM with an ARD prior, we plot the shortest intervals containing 95\% of the held-out log-likelihoods calculated based on a set of Gibbs samples taken at iteration 1000 from 100 chains. (a) Log-likelihood of the second half of honey bee dance sequence 4 based on model parameters inferred from the first half of the sequence. (b)-(c) Similarly for sequences 5 and 6, respectively.} \label{fig:dancingbee_ARD} %\vspace{-0.2in}
\end{figure*}
Using the cleaner sequences 4 to 6, we investigate the affects of the sparsity-inducing ARD prior by assuming a higher order switching VAR model and computing the likelihood of the second half of each dance sequence based on parameters inferred from Gibbs sampling using the data from the first half of each sequence.  In Fig.~\ref{fig:dancingbee_ARD}, we specifically compare the performance of an HDP-VAR($r$)-HMM with a conjugate MNIW prior for $r=1,2,7$ to that of an HDP-VAR(7)-HMM with an ARD prior.  We use the same approach to setting the hyperparameters as in Sec.~\ref{sec:ARDresults}.  We see that assuming a higher order model improves the predictive likelihood performance, but only when combined with a regularizing prior (e.g., the ARD) that avoids over-fitting in the presence of limited data.  Although not depicted here (see instead~\cite{Fox:PhD}), the ARD prior also informs us of the variable-order nature of this switching dynamical process. When considering an HDP-VAR(2)-HMM with an ARD prior, the posterior distribution of the ARD hyperparameters for the first and second order lag components associated with each of the three dominant inferred dances clearly indicates that two of the turning dances simply rely on the first lag component while the other dance relies on both lag components. To verify these results, we provided the data and ground truth labels to MATLAB's {\tt lpc} %\footnote{{\tt lpc} computes AR coefficients for scalar data, so we analyzed each component of the observation vector independently. The order was consistent across these components.} 
implementation of Levinson's algorithm, which indicated that the turning dances are well approximated by an order 1 process, while the waggle dance relies on an order 2 model. Thus, our learned orders for the three dances match what is indicated by Levinson's algorithm on ground-truth segmented data.
%
%Using the cleaner sequences 4 to 6, we investigate the honey bee dance's variable order structure by assuming a higher order switching VAR model and employing the ARD prior. Namely, we choose an HDP-VAR($2$)-HMM and use the same approach to setting the hyperparameters as in Sec.~\ref{sec:ARDresults}. Although not depicted here, the Hamming distance plots for the HDP-VAR($2$)-HMM with the ARD prior are indistinguishable from those of Fig.~\ref{fig:dancingbee_Hamming}(a)-(c) using the HDP-VAR($1$)-HMM with the MNIW prior. Thus, the information in the first lag component is sufficient for the segmentation problem. However, the ARD prior informs us of the variable-order nature of this switching dynamical process. From Fig.~\ref{fig:dancingbee_ARD}(a)-(c), we see that the turning dances simply rely on the first lag component while the waggle dance relies on both lag components. To verify these results, we provided the data and ground truth labels to MATLAB's {\tt lpc}\footnote{{\tt lpc} computes AR coefficients for scalar data, so we analyzed each component of the observation vector independently. The order was consistent across these components.} implementation of Levinson's algorithm, which indicated that the turning dances are well approximated by an order 1 process, while the waggle dance relies on an order 2 model. Thus, our learned orders for the three dances match what is indicated by Levinson's algorithm on ground-truth segmented data.
%
\section{Model Variants} \label{sec:ModelVariants}
There are many variants of the general SLDS and switching VAR models that are pervasive in the literature. One important example is when the dynamic matrix is shared between modes; here, the dynamics are instead distinguished based on a switching mean, such as the Markov switching stochastic volatility (MSSV) model. In the maneuvering target tracking community, it is often further assumed that the dynamic matrix is shared and \emph{known} (due to the understood physics of the target). We explore both of these variants in the following sections.
\subsection{Shared Dynamic Matrix, Switching Driving Noise} \label{sec:shared}
In many applications, the dynamics of the switching process can be described by a shared linear dynamical system matrix $A$; the dynamics within a given mode are then determined by some external force acting upon this LDS, and it is how this force is exerted that is mode-specific. The general form for such an SLDS is given by
\begin{equation}
	\begin{aligned}
		\begin{array}{c}
		z_t\mid z_{t-1} \sim \pi_{z_{t-1}}\\
		\BF{x}_t = A\BF{x}_{t-1}+\BF{e}_t(z_t) \hspace{0.2in} \BF{y}_t = C\BF{x}_t + \BF{w}_t,
		\end{array} 
	\end{aligned}
	\label{eqn:SLDS_sharedA} 
\end{equation}
with process and measurement noise $\BF{e}_t(k) \sim \mathcal{N}(\symkB{\mu}{k},\symk{\Sigma}{k})$ and $\BF{w}_t \sim \mathcal{N}(0,R)$, respectively. 
In this scenario, the data are generated from one dynamic matrix, $A$, and multiple process noise covariance matrices, $\symk{\Sigma}{k}$. Thus, one cannot place a MNIW prior jointly on these parameters (conditioned on $\symkB{\mu}{k}$) due to the coupling of the parameters in this prior. We instead consider independent priors on $A$, $\symk{\Sigma}{k}$, and $\symkB{\mu}{k}$. We will refer to the choice of a normal prior on $A$, inverse-Wishart prior on $\symk{\Sigma}{k}$, and normal prior on $\symkB{\mu}{k}$ as the \emph{N-IW-N} prior. See~\cite{Fox:PhD} for details on deriving the resulting posterior distributions given these independent priors.
\subsubsection*{Stochastic Volatility}
An example of an SLDS in a similar form to that of Eq.~\eqref{eqn:SLDS_sharedA} is the Markov switching stochastic volatility (MSSV) model~\cite{Hamilton:89,Kim:94,So:98}. The MSSV assumes that the log-volatilities follow an AR($1$) process with a Markov switching mean. This underlying process is observed via conditionally independent and normally distributed daily returns. Specifically, let $y_t$ represent, for example, the daily returns of a stock index. The state $x_t$ is then given the interpretation of log-volatilities and the resulting state space is given by~\cite{Carvalho:06}
\begin{equation}
	\begin{aligned}
		\begin{array}{c}
		z_t\mid z_{t-1} \sim \pi_{z_{t-1}}\\
		x_t = ax_{t-1} + e_t(z_t) \hspace{0.2in} y_t = u_t(x_t),
		\end{array} 
	\end{aligned}
	\label{eqn:MSSV} 
\end{equation}
with $e_t(k) \sim \mathcal{N}(\symk{\mu}{k},\sigma^2)$ and $u_t(x_t) \sim \mathcal{N}(0,\exp(x_t))$. Here, only the mean of the process noise is mode-specific. Note, however, that the measurement equation is non-linear in the state $x_t$. Carvalho and Lopes~\cite{Carvalho:06} employ a particle filtering approach to cope with these non-linearities. In~\cite{So:98}, the MSSV is instead modeled in the log-squared-daily-returns domain such that
\begin{align}
	\log(y_t^2) &= x_t + w_t, 
	\label{eqn:logdaily}
\end{align}
where $w_t$ is additive, non-Gaussian noise. This noise is sometimes approximated by a moment-matched Gaussian~\cite{Harvey:94}, while So et. al.~\cite{So:98} use a mixture of Gaussians approximation. The MSSV is then typically bestowed a fixed set of two or three regimes of volatility.
%%
%\begin{figure*}[t!] \centering 
%	\includegraphics[height = 2in]{\figdir/IBOVESPA_dailyreturns}\vspace{-0.2in} \caption{IBOVESPA stock index daily returns from 01/03/1997 to 01/16/2001.} \label{fig:IBOVESPA_obs} %\vspace{-0.2in} 
%\end{figure*}
%%

\begin{figure*}[t!] \centering 
	\begin{tabular}{ccc}
		\hspace{-0.05in}
		\includegraphics[height = 1.47in]{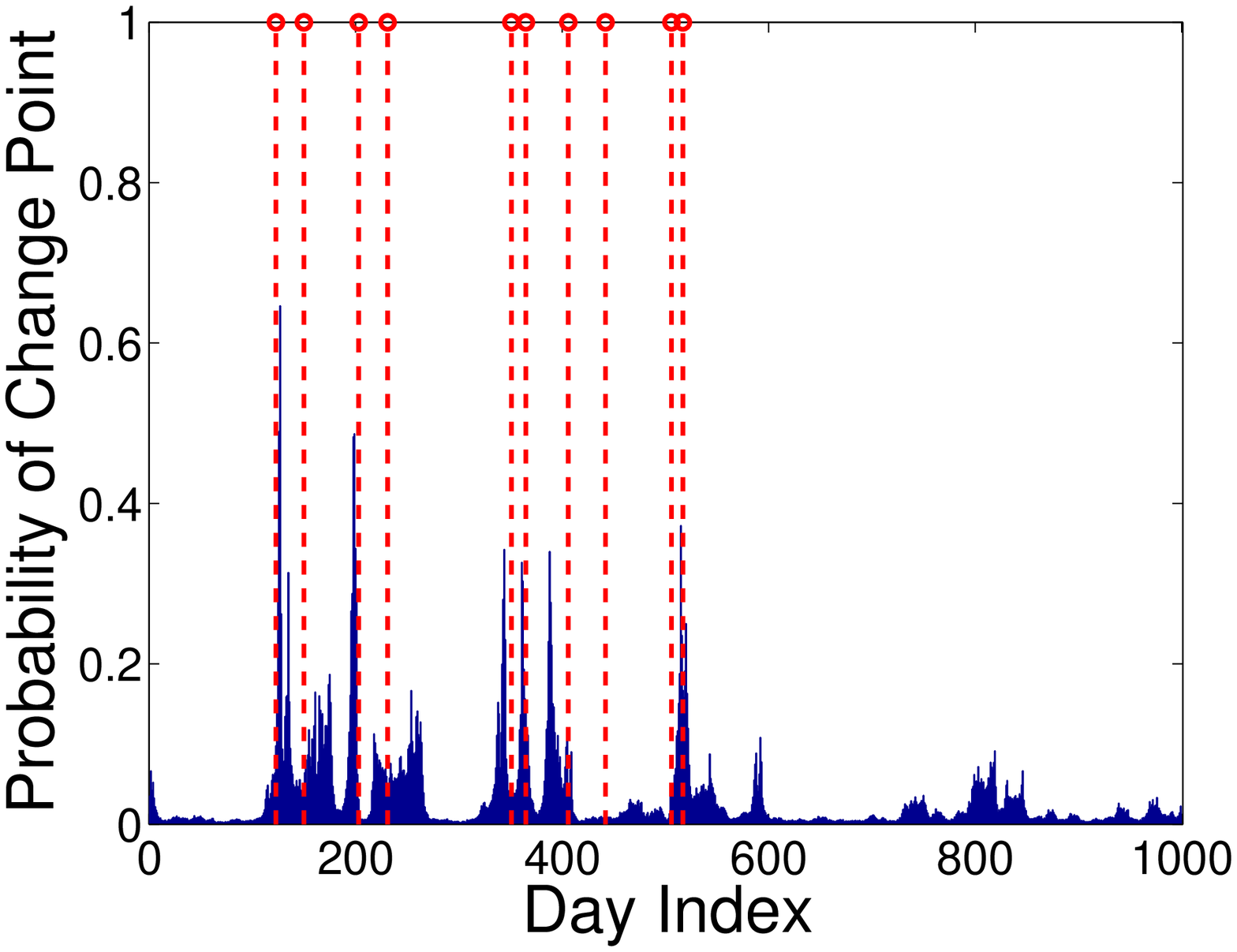} & \hspace{-0.2in}
		\includegraphics[height = 1.47in]{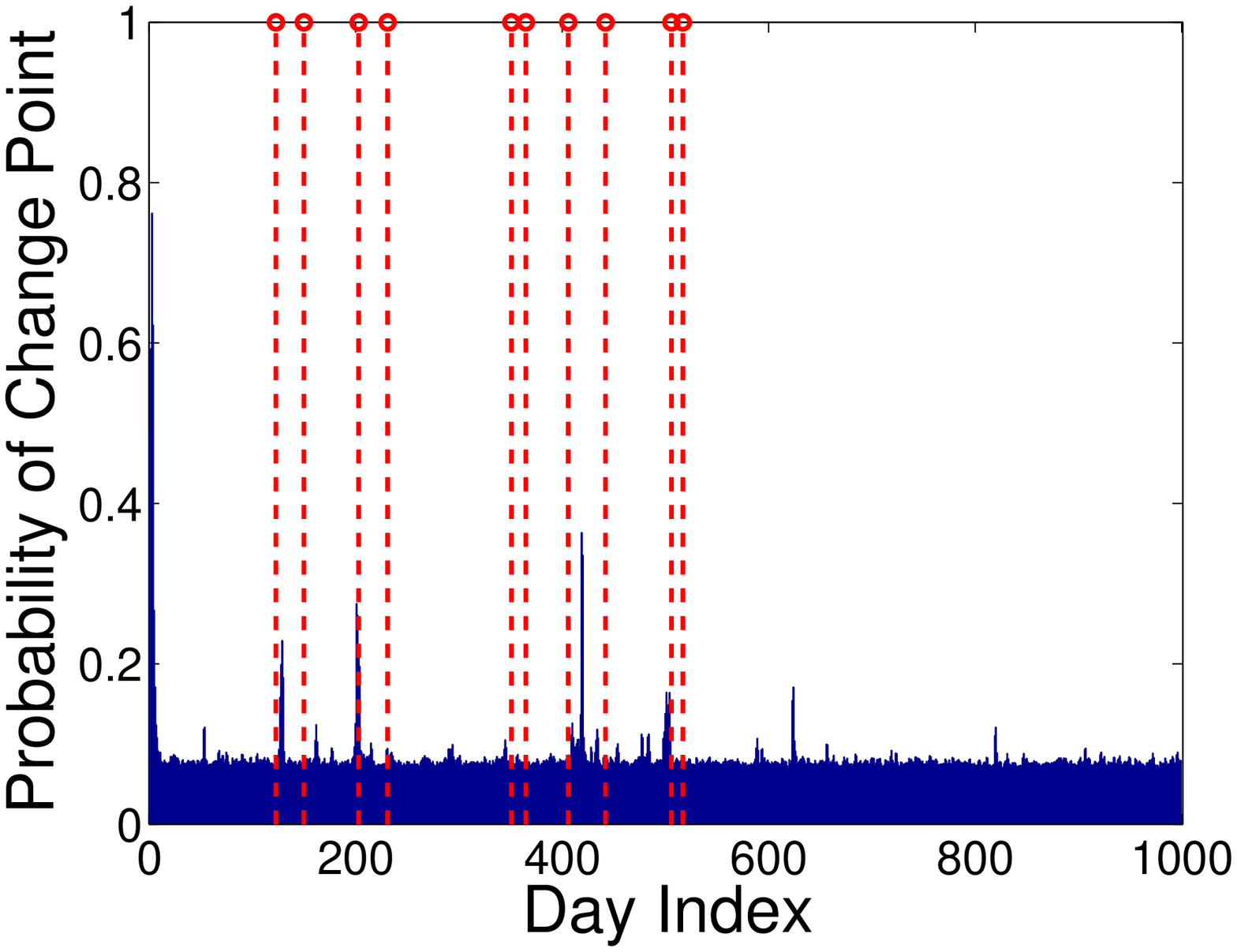} & \hspace{-0.2in}
		\includegraphics[height = 1.47in]{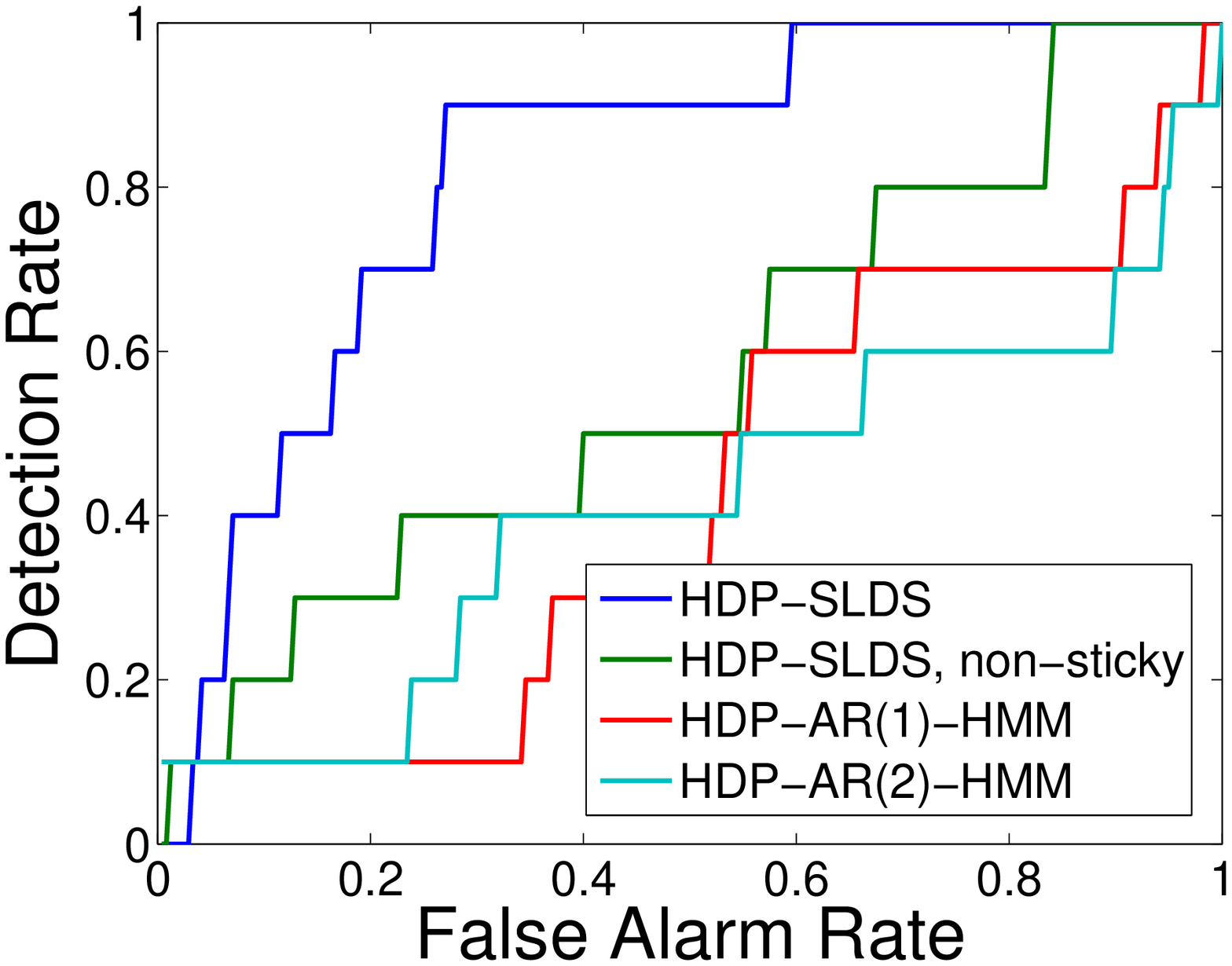}\\ %\vspace{-0.1in}\\
		\hspace{-0.05in}(a) & \hspace{-0.2in}(b) & \hspace{-0.2in}(c)\\ %\vspace{0.05in}\\
		\hspace{-0.05in}
		\includegraphics[height = 1.47in]{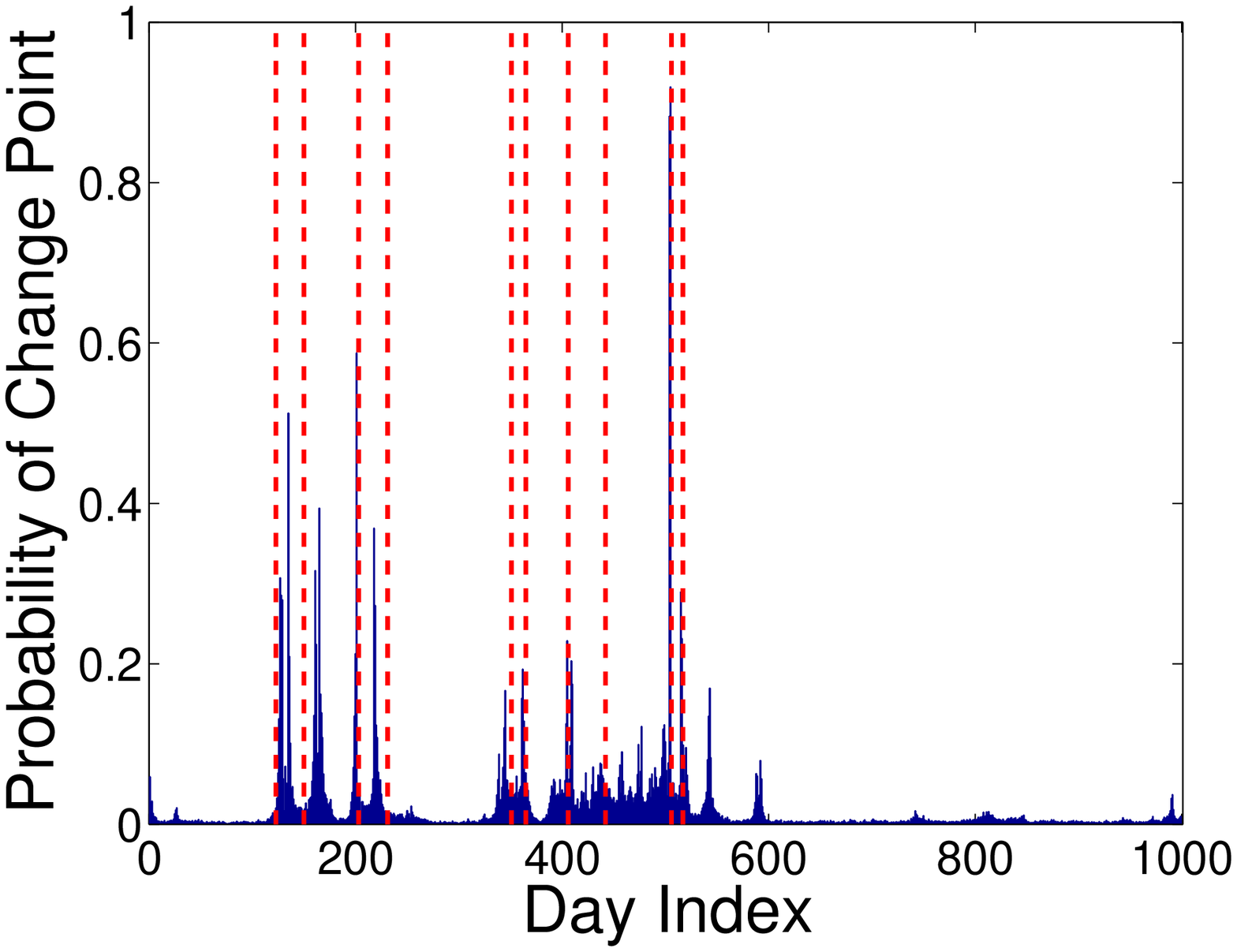} & \hspace{-0.2in}
		\includegraphics[height = 1.47in]{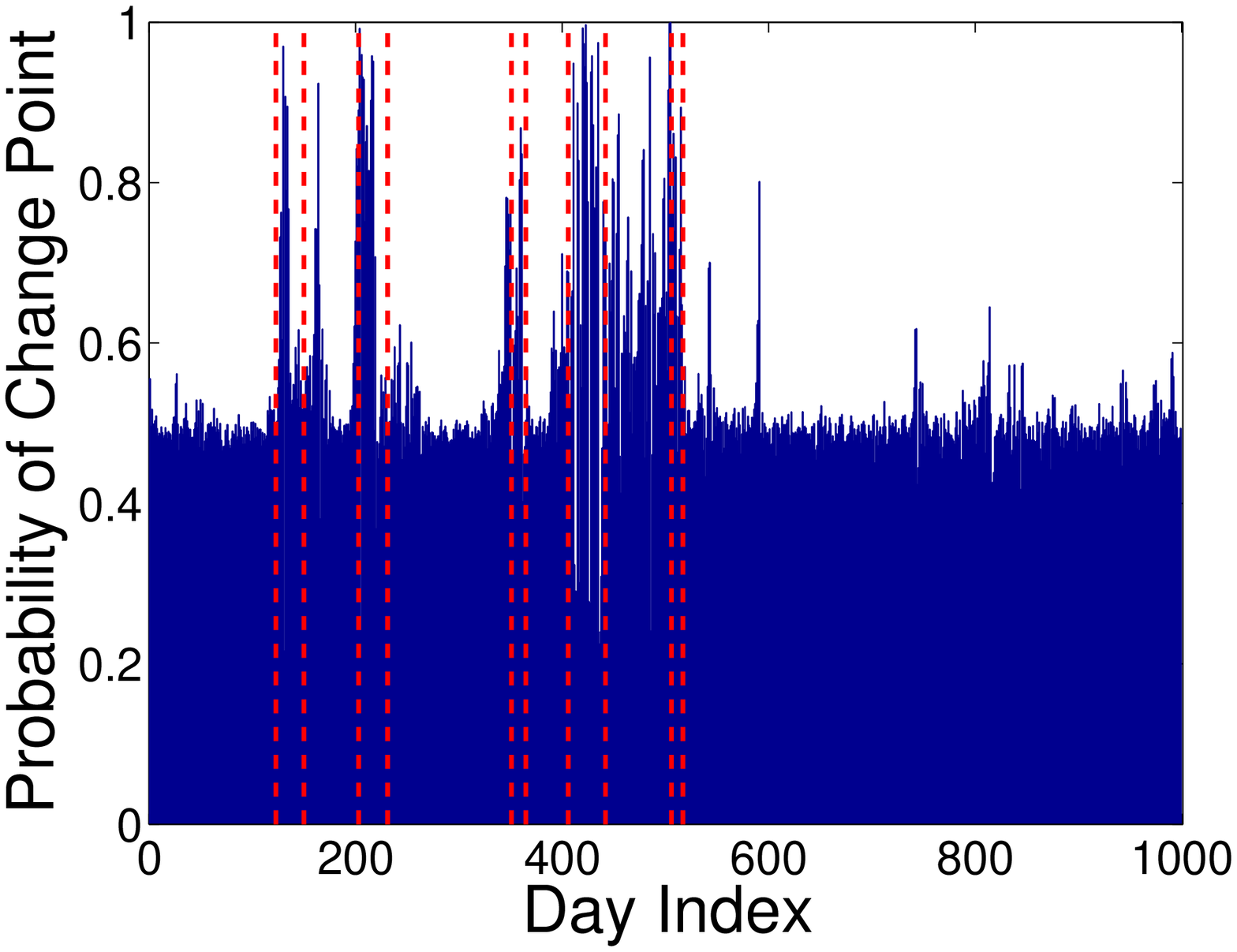} & \hspace{-0.2in}
		\includegraphics[height = 1.47in]{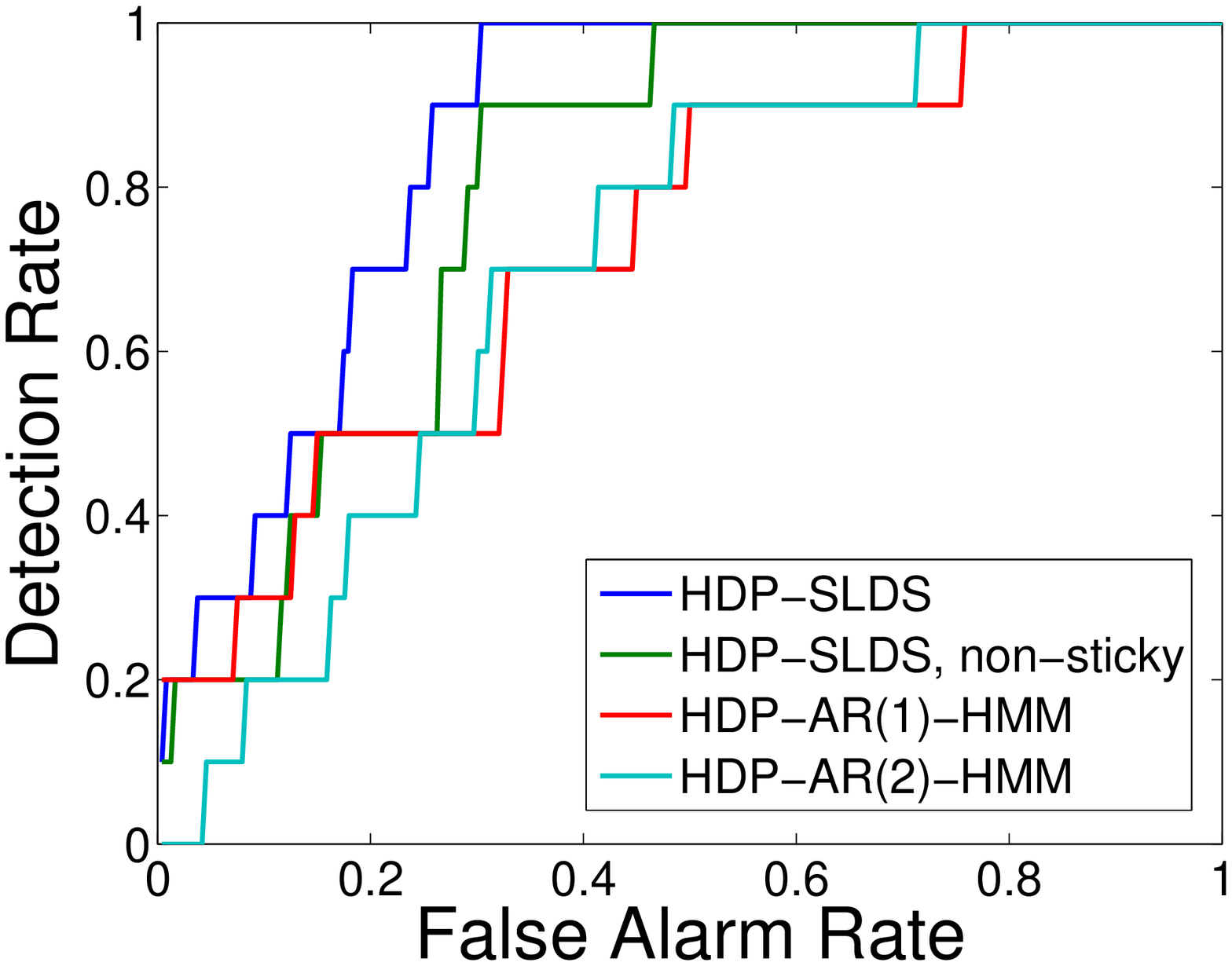}\\ %\vspace{-0.1in}\\
		\hspace{-0.05in}(d) & \hspace{-0.2in}(e) & \hspace{-0.2in}(f) %\vspace{-0.1in}
	\end{tabular}
	\caption[IBOVESPA stock index change-point detection performance for two variants of the HDP-SLDS.]{(a) Plot of the estimated probability of a change point on each day using 3,000 Gibbs samples for a MSSV variant of the HDP-SLDS using a shared dynamic matrix and allowing a mean on the mode-specific process noise and a mixture of Gaussian measurement noise model. The observations are log-squared dialy return measurements, and the 10 key events are indicated with red lines. (b) Similar plot for the \emph{non-sticky} HDP-SLDS with no bias towards self-transitions. (c) ROC curves for the HDP-SLDS, non-sticky HDP-SLDS, HDP-AR($1$)-HMM, and HDP-AR($2$)-HMM. (d)-(f) Analogous plots for the HDP-SLDS of Table~\ref{table:models} using raw daily return measurements.} \label{fig:IBOVESPA} %\vspace{-0.2in}
\end{figure*}

We examine the IBOVESPA stock index (Sao Paulo Stock Exchange) over the period of 01/03/1997 to 01/16/2001, during which ten key world events are cited in \cite{Carvalho:06} as affecting the emerging Brazilian market during this time period. %The daily returns are displayed in Fig.~\ref{fig:IBOVESPA_obs} and the key world event are summarized in Table~\ref{table:IBOVESPA} and shown in the plots of Fig.~\ref{fig:IBOVESPA}. 
The key world events are summarized in Table~\ref{table:IBOVESPA} and shown in the plots of Fig.~\ref{fig:IBOVESPA}. Use of this dataset was motivated by the work of Carvalho and Lopes~\cite{Carvalho:06}, in which a two-mode MSSV model is assumed. We consider a variant of the HDP-SLDS to match the MSSV model of Eq.~\eqref{eqn:MSSV}. Specifically we examine log-squared daily returns, as in Eq.~\eqref{eqn:logdaily}, and use a DP mixture of Gaussians to model the measurement noise:
\begin{equation}
	\begin{aligned}
		\begin{array}{c}
	e_t(k) \sim \mathcal{N}(\symk{\mu}{k},\symk{\Sigma}{k})\\
	w_t \sim \sum_{\ell=1}^\infty \omega_\ell\mathcal{N}(0,R_\ell) \quad 
	\BF{\omega} \sim \mbox{GEM}(\sigma_r), \quad R_\ell \sim \mbox{IW}(n_r,S_r).
		\end{array}
	\end{aligned}
	\label{eqn:MSSVnoise}
\end{equation}
We truncate the measurement noise DP mixture to 10 components. For the HDP concentration hyperparameters, $\alpha$, $\gamma$, and $\kappa$, we use the same prior distributions as in Sec.~\ref{sec:MNIWresults}-\ref{sec:BeeResults}. For the dynamic parameters, we rely on the N-IW-N prior described in Sec.~\ref{sec:shared} and once again set the hyperparameters of this prior from statistics of the data as described in the Appendix. Since we allow for a mean on the process noise and examine log-squared daily returns, we do not preprocess the data. 
\begin{table}
	\centering 
	\begin{tabular}
		{|l|l|} \hline Date & Event\\
		\hline \hline {\small 07/02/1997} & {\small Thailand devalues the Baht by as much as 20\%}\\
		\hline {\small 08/11/1997} & {\small IMF and Thailand set a rescue agreement}\\
		\hline {\small 10/23/1997} & {\small Hong Kongís stock index falls 10.4\%. South Korea won starts to weaken}\\
		\hline {\small 12/02/1997} & {\small IMF and South Korea set a bailout agreement}\\
		\hline {\small 06/01/1998} & {\small Russiaís stock market crashes}\\
		\hline {\small 06/20/1998} & {\small IMF gives final approval to a loan package to Russia}\\
		\hline {\small 08/19/1998} & {\small Russia officially falls into default}\\
		\hline {\small 10/09/1998} & {\small IMF and World Bank joint meeting to discuss global economic crisis. The Fed cuts interest rates}\\
		\hline {\small 01/15/1999} & {\small The Brazilian government allows its currency, the Real, to float freely by lifting exchange controls}\\
		\hline {\small 02/02/1999} & {\small Arminio Fraga is named President of Brazilís Central Bank}\\
		\hline 
	\end{tabular}
	\caption[Table of 10 key world events affecting the IBOVESPA stock index from 01/03/1997 to 01/16/2001.]{Table of 10 key world events affecting the IBOVESPA stock index (Sao Paulo Stock Exchange) over the period of 01/03/1997 to 01/16/2001, as cited by Carvalho and Lopes~\cite{Carvalho:06}.}\label{table:IBOVESPA} \vspace{-0.3in}
\end{table}
%
%We rely on the N-IW-N prior described in Sec.~\ref{sec:shared}. Since we are allowing for a mean on the process noise and dealing with log-squared daily returns, we do not perform any pre-processing of the data. For the normal prior on the dynamic parameter $a$, we set the mean to $0$ and the covariance to 0.75 times the empirical covariance of the observations. This matches with the moments on $a$ defined by our MNIW prior when fixing the process noise covariance to its expected value. Our normal prior on $\symk{\mu}{k}$ is also given a $0$ mean with covariance equal to 0.75 times the empirical covariance. The IW prior on the process noise $\symk{\Sigma}{k}$ was given 3 degrees of freedom and an expected value of 0.75 times the empirical covariance. Finally, each component of the mixture-of-Gaussian measurement noise was given an IW prior with 3 degrees of freedom and an expected value of $5*\pi^2$, which matches with the moment-matching technique of Harvey et. al.~\cite{Harvey:94}. For the HDP-AR(r)-HMM's to which we compare in Fig.~\ref{fig:IBOVESPA}, we place a zero-mean normal prior on the dynamic parameter $a$ with covariance set to the expected noise covariance, which in this case is equal to 0.75 times the empirical covariance plus $5*\pi^2$ (to emulate the cumulative noise in the HDP-SLDS.) This IW prior on the noise parameter is given 3 degrees of freedom. The mean parameter $\symk{\mu}{k}$ is given a normal, zero-mean prior with covariance equal to 0.75 times the empirical covariance.

The posterior probability of an HDP-SLDS inferred change point is shown in Fig.~\ref{fig:IBOVESPA}(a), and in Fig.~\ref{fig:IBOVESPA}(b) we display the corresponding plot for a non-sticky variant (i.e., with $\kappa=0$ so that there is no bias towards mode self-transitions.) The HDP-SLDS is able to infer very similar change points to those presented in \cite{Carvalho:06}. Without the sticky extension, the non-sticky model variant over-segments the data and rapidly switches between redundant states leading to many inferred change points that do not align with any world event. 
In Fig.~\ref{fig:IBOVESPA}(c), the overall change-point detection performance of the HDP-SLDS is compared to that of the HDP-AR($1$)-HMM, HDP-AR($2$)-HMM, and non-sticky HDP-SLDS. The ROC curves shown are calculated by windowing the time axis and taking the maximum probability of a change point in each window. These probabilities are then used as the confidence of a change point in that window. From this plot, we clearly see the advantage of using an SLDS model combined with the sticky HDP-HMM prior on the mode sequence.

We also analyzed the performance of an HDP-SLDS as defined in Table~\ref{table:models}. We used raw daily-return observations, and first pre-processed the data in the same manner as the honey bee data by centering the observations around $0$ and scaling the data to be roughly within a $[-10,10]$ dynamic range.  We then took a MNIW prior on the dynamic parameters, as outlined in the Appendix. % with $M=0$, $K=1$, and $n_0=3$ degrees of freedom. The expected process noise covariance was set to 0.75 times the empirical covariance of the data. The IW prior on the measurement noise covariance, $R$, was given $r_0 = 100$ degrees of freedom and an expected covariance of $25$. Our sampler initializes parameters from the prior, and we found it useful to set the prior around large values of $R$ in order to avoid initial samples chattering between dynamical regimes caused by the state sequence having to account for the noise in the observations. After accounting for the residuals of the data in the posterior distribution, we typically learned $R\approx 10$. 
Overall, although the state of this HDP-SLDS does not have the interpretation of log-volatilities, we see are still able to capture regime-changes in the dynamics of this stock index and find changepoints that align better with the true world events than in the MSSV HDP-SLDS model.
\subsection{Fixed Dynamic Matrix, Switching Driving Noise} \label{sec:MTT}
There are some cases in which the dynamical model is well-defined through knowledge of the physics of the system being observed, such as simple kinematic motion. More complicated motions can typically be modeled using the same fixed dynamical model, but using a more complex description of the driving force. A generic LDS driven by an unknown control input $\BF{u}_t$ can be represented as
\begin{equation}
	\begin{aligned}
		\BF{x}_{t} = A\BF{x}_{t-1} + B\BF{u}_t + \BF{v}_t \hspace{0.2in} \BF{y}_t = C\BF{x}_t + D\BF{u}_t + \BF{w}_t, 
	\end{aligned}
\end{equation}
where $\BF{v}_t \sim \mathcal{N}(0,Q)$ and $\BF{w}_t \sim \mathcal{N}(0,R)$. It is often appropriate to assume $D=0$, as we do herein.
\subsubsection*{Maneuvering Target Tracking}
%
%The methods for modeling a maneuvering target can be primarily classified into three categories: (1) methods which approximate the non-random but unobserved control input sequence $\BF{u}_{1:T}$ as a deterministic unknown, (2) methods which model $\BF{u}_{1:T}$ as a random process, and (3) methods which use a set of dynamic systems to model typical target trajectories. For a thorough survey of maneuvering target tracking, see~\cite{RongLi:Dynamic,RongLi:MM}.
%
%Inference on systems modeled with deterministic unknown inputs is computationally complex, and thus modeling the control input as a random process is a common simplifying assumption. The most basic of these models is to take the control to be white noise, such as the constant velocity (CV) and constant acceleration (CA) models with random walks on velocity and acceleration, respectively. 
Target tracking provides an application domain in which one often assumes that the dynamical model is known.  One method of describing a maneuvering target is to consider the control input as a random process~\cite{RongLi:Dynamic}. For example, a \emph{jump-mean} Markov process~\cite{Moose:79} yields dynamics described as
\begin{equation}
	\begin{aligned}
		\begin{array}{c}
			z_t\mid z_{t-1} \sim \pi_{z_{t-1}}\\
			\BF{x}_t = A\BF{x}_{t-1} + B\BF{u}_t(z_t) + \BF{v}_t \hspace{0.2in} \BF{y}_t = C\BF{x}_t + \BF{w}_t\\
			\BF{u}_t(k) \sim \mathcal{N}(\symkB{\mu}{k},\symk{\Sigma}{k}) \hspace{0.2in} \BF{v}_t \sim \mathcal{N}(0,Q) \hspace{0.2in} \BF{w}_t \sim \mathcal{N}(0,R).
		\end{array}
	\end{aligned}
	\label{eqn:SLDS_fixedA} 
\end{equation}
Classical approaches rely on defining a fixed set of dynamical modes and associated transition distributions. The state dynamics of Eq.~\eqref{eqn:SLDS_fixedA} can be equivalently described as
\begin{align}
	\BF{x}_t &= A\BF{x}_{t-1} + \BF{e}_t(z_t)\\
	\BF{e}_t(k) &\sim \mathcal{N}(B\symkB{\mu}{k},B\symk{\Sigma}{k}B^T+Q). 
\end{align}
This model can be captured by our HDP-SLDS formulation of Eq.~\eqref{eqn:SLDS_sharedA} with a fixed dynamic matrix (e.g., constant velocity or constant acceleration models~\cite{RongLi:Dynamic}) and mode-specific, non-zero mean process noise. Such a formulation was explored in~\cite{Fox:Fusion07} along with experiments that compare the performance to that of standard multiple model techniques, demonstrating the flexibility of the Bayesian nonparametric approach.  Fox et. al.~\cite{Fox:Fusion07} also present an alternative sampling scheme that harnesses the fact that the control input may be much lower-dimensional than the state and sequentially block-samples $(z_t,\BF{u}_t)$ analytically marginalizing over the state sequence $\BF{x}_{1:T}$.  Note that this variant of the HDP-SLDS can be viewed as an extension of the work by Caron et. al.~\cite{Caron:06} in which the exogenous input is modeled as an independent noise process (i.e., no Markov structure on $z_t$) generated from a DP mixture model.
\section{Conclusion}
\label{sec:chap4discussion}	
In this paper, we have addressed the problem of learning switching linear dynamical models with an unknown number of modes for describing complex dynamical phenomena. We presented a Bayesian nonparametric approach and demonstrated both the utility and versatility of the developed HDP-SLDS and HDP-AR-HMM on real applications. Using the same parameter settings, although different model choices, in one case we are able to learn changes in the volatility of the IBOVESPA stock exchange while in another case we learn segmentations of data into \emph{waggle}, \emph{turn-right}, and \emph{turn-left} honey bee dances. We also described a method of applying automatic relevance determination (ARD) as a sparsity-inducing prior, leading to flexible and scalable dynamical models that allow for identification of variable order structure.  We concluded by considering adaptations of the HDP-SLDS to specific forms often examined in the literature such as the Markov switching stochastic volatility model and a standard multiple model target tracking formulation.
	
The batch processing of the Gibbs samplers derived herein may be impractical and offline-training online-tracking infeasible for certain applications. Due both to the nonlinear dynamics and uncertainty in model parameters, exact recursive estimation is infeasible. One could leverage the \emph{conditionally linear} dynamics and use \emph{Rao-Blackwellized particle filtering} (RBPF)~\cite{Doucet:00}. However, one challenge is that such particle filters can suffer from a progressively impoverished particle representation.
	
Overall, the formulation we developed herein represents a flexible, Bayesian nonparametric model for describing complex dynamical phenomena and discovering simple underlying temporal structures. 
%\section*{Acknowledgment}
%This work was supported in part by MURIs funded through AFOSR Grant FA9550-06-1-0324 and ARO Grant W911NF-06-1-0076.
%\ifCLASSOPTIONcaptionsoff 
%\newpage \fi

\appendix
\paragraph{MNIW General Method}
For the experiments of Sec.~\ref{sec:MNIWresults}, we set $M=\BF{0}$ and $K = I_m$. This choice centers the mass of the prior around stable dynamic matrices while allowing for considerable variability. The inverse-Wishart portion is given $n_0=m+2$ degrees of freedom. For the HDP-AR-HMM, the scale matrix $S_0=0.75\bar{\Sigma}$, where $\bar{\Sigma} = \frac{1}{T}\sum (\BF{y}_t-\bar{\BF{y}})(\BF{y}_t-\bar{\BF{y}})^T$. Setting the prior directly from the data can help move the mass of the distribution to reasonable values of the parameter space. For an HDP-SLDS with $\BF{x}_t \in \RR^n$ and $\BF{y}_t \in \RR^d$ and $n=d$, we set $S_0=0.675\bar{\Sigma}$.  We then set the inverse-Wishart prior on the measurement noise, $R$, to have $r_0=d+2$ and $R_0 = 0.075\bar{\Sigma}$. For $n>d$, see~\cite{Fox:PhD}.
\paragraph{Partially Supervised Honey Bee Experiments}
For the partially supervised experiments of Sec.~\ref{sec:BeeResults}, we set $\Sigma_0=0.75 S_0$. Since we are not shifting and scaling the observations, we set $S_0$ to 0.75 times the empirical covariance of the $\emph{first difference}$ observations. We also use $n_0 = 10$, making the distribution tighter than in the unsupervised case. Examining first differences is appropriate since the bee's dynamics are better approximated as a random walk than as i.i.d. observations. Using raw observations in the unsupervised approach creates a larger expected covariance matrix making the prior on the dynamic matrix less informative, which is useful in the absence of other labeled data.
\paragraph{IBOVESPA Stock Index Experiments}
For the HDP-SLDS variant of the MSSV model of Eq.~\eqref{eqn:MSSV}, we rely on the N-IW-N prior described in Sec.~\ref{sec:shared}. For the dynamic parameter $a$ and process noise mean $\symk{\mu}{k}$, we use $\mathcal{N}(0,0.75\bar{\Sigma})$ priors. The IW prior on $\symk{\Sigma}{k}$ was given 3 degrees of freedom and an expected value of $0.75\bar{\Sigma}$. Finally, each component of the mixture-of-Gaussian measurement noise was given an IW prior with 3 degrees of freedom and an expected value of $5*\pi^2$, which matches with the moment-matching technique of Harvey et. al.~\cite{Harvey:94}. For the HDP-AR(r)-HMM's to which we compare in Fig.~\ref{fig:IBOVESPA}, we place a zero-mean normal prior on the dynamic parameter $a$ with covariance set to the expected noise covariance, which in this case is equal to 0.75 times the empirical covariance plus $5*\pi^2$. The mean parameter $\symk{\mu}{k}$ is defined as in the HDP-SLDS.

For the HDP-SLDS comparison using the model of Table~\ref{table:models}, we use a MNIW prior with $M=0$, $K=1$, $n_0=3$, and $S_0=0.75\bar{\Sigma}$. The IW prior on $R$ was given $r_0 = 100$ and an expected covariance of $25$. Our sampler initializes parameters from the prior, and we found it useful to set the prior around large values of $R$ in order to avoid initial samples chattering between dynamical regimes caused by the state sequence having to account for the noise in the observations. After accounting for the residuals of the data in the posterior distribution, we typically learned $R\approx 10$.

\setlength{\bibspacing}{0pt}\linespread{1.1}
\bibliographystyle{IEEEtran} 
{\small \bibliography{IEEEabrv,../../../Bibliography/Bibliography.bib}
}

\appendices	
\section{Dynamic Parameter Posteriors}
In this appendix, we derive the posterior distribution over the
dynamic parameters of a switching VAR($r$) process defined as
follows:
\begin{align}
\BF{y}_t &= \sum_{i=1}^r A_i^{(z_t)}\BF{y}_{t-i} + \BF{e}_t(z_t)
\quad \BF{e}_t(k) \sim \mathcal{N}(\symkB{\mu}{k},\symk{\Sigma}{k}),
\label{eqn:VARk}
\end{align}
where $z_t$ indexes the mode-specific VAR($r$) process at time $t$.
Assume that the mode sequence $\{z_1,\dots,z_T\}$ is known and we
wish to compute the posterior distribution of the $k^{th}$ mode's
VAR($r$) parameters $A_i^{(k)}$ for $i=1,\dots,r$ and
$\symk{\Sigma}{k}$. Let $\{t_1,\dots,t_{n_k}\} = \{t|z_t = k\}$.
Then, we may write
\begin{align}
\begin{bmatrix} \BF{y}_{t_1} & \BF{y}_{t_2} & \dots & \BF{y}_{t_{n_k}}
\end{bmatrix} = \begin{bmatrix} A_1^{(k)} & A_2^{(k)}
& \dots & A_r^{(k)}\end{bmatrix}\begin{bmatrix}
\BF{y}_{t_1-1} & \BF{y}_{t_2-1} & \dots & \BF{y}_{t_{n_k}-1}\\
\BF{y}_{t_1-2} & \BF{y}_{t_2-2} & \dots & \BF{y}_{t_{n_k}-2}\\
\vdots & & & \\
\BF{y}_{t_1-r} & \BF{y}_{t_2-r} & \dots &
\BF{y}_{t_{n_k}-r}\end{bmatrix} +
\begin{bmatrix} \BF{e}_{t_1} & \BF{e}_{t_2} & \dots & \BF{e}_{t_{n_k}}
\end{bmatrix}.
\label{eqn:bigeqn}
\end{align}
We define the following notation for Eq.~\eqref{eqn:bigeqn}:
\begin{align}
\bigk{Y}{k} &= \bigk{A}{k}\bigk{\Ylag{Y}}{k} + \bigk{E}{k},
\end{align}
and let $\bigk{D}{k} = \{\bigk{Y}{k},\bigk{\Ylag{Y}}{k}\}$. In the
following sections, we consider two possible priors on the dynamic
parameter. In Appendix A-~\ref{app:conjugatePrior}, we assume that
$\symkB{\mu}{k}$ is 0 for all $k$ and consider the conjugate
matrix-normal inverse-Wishart (MNIW) prior for
$\{\bigk{A}{k},\symk{\Sigma}{k}\}$. In
Appendix A-~\ref{app:nonConjPrior}, we consider the more general form
of Eq.~\eqref{eqn:VARk} and take independent priors on
$\bigk{A}{k}$, $\symk{\Sigma}{k}$, and $\symkB{\mu}{k}$.
\subsection{Conjugate Prior --- MNIW}
\label{app:conjugatePrior}
To show conjugacy, we place a MNIW prior on the dynamic parameters
$\{\bigk{A}{k},\symk{\Sigma}{k}\}$ and show that the posterior
remains MNIW given a set of data from the model of
Eq.~\eqref{eqn:VARk} (assuming $\symkB{\mu}{k}=0$). The MNIW prior
is given by placing a matrix-normal prior
$\MN{\bigk{A}{k}}{M}{\symk{\Sigma}{k}}{K}$ on $\bigk{A}{k}$ given
$\symk{\Sigma}{k}$:
\begin{align}
p(\bigk{A}{k}\mid \symk{\Sigma}{k}) &=
\frac{|K|^{d/2}}{|2\pi\symk{\Sigma}{k}|^{m/2}}\exp\left(-\frac{1}{2}tr((\BF{A}-M)^T
\Sigma^{-(k)}(\BF{A}-M)K)\right)
\end{align}
and an inverse-Wishart prior $\mbox{IW}(n_0,S_0)$ on
$\symk{\Sigma}{k}$:
\begin{align}
p(\symk{\Sigma}{k}) &=
\frac{|S_0|^{n_0/2}|\symk{\Sigma}{k}|^{-(d+n_0+1)/2}}{2^{n_0d/2}\Gamma_d(n_0/2)}\exp\left(-\frac{1}{2}tr(\Sigma^{-(k)}S_0)\right)
\end{align}
where $\Gamma_d(\cdot)$ is the multivariate gamma function and
$\BF{B}^{-(k)}$ denotes $(\BF{B}^{(k)})^{-1}$ for some matrix
$\BF{B}$.

We first analyze the likelihood of the data, $\bigk{D}{k}$, given
the $k^{th}$ mode's dynamic parameters,
$\{\bigk{A}{k},\symk{\Sigma}{k}\}$. Starting with the fact that each
observation vector, $\BF{y}_t$, is conditionally Gaussian given the
lag observations, $\BF{\Ylag{y}}_t=[\BF{y}^T_{t-1} \dots
\BF{y}^T_{t-r}]^T$, we have
\begin{align}
p(\bigk{D}{k}|\bigk{A}{k},\symk{\Sigma}{k})&=
\frac{1}{|2\pi\symk{\Sigma}{k}|^{n_k/2}}\exp\left(-\frac{1}{2}\sum_i(\BF{y}_{t_i}-\bigk{A}{k}\BF{\Ylag{y}}_{t_i})^T
\Sigma^{-(k)}(\BF{y}_{t_i}-\bigk{A}{k}\BF{\Ylag{y}}_{t_i})\right)\nonumber\\
%&\hspace{-0.5in}=
%\frac{1}{|2\pi\symk{\Sigma}{k}|^{n_k/2}}\exp\left(-\frac{1}{2}tr(\Sigma^{-(k)}(\bigk{Y}{k}-\bigk{A}{k}\bigk{\Ylag{Y}}{k})
%(\bigk{Y}{k}-\bigk{A}{k}\bigk{\Ylag{Y}}{k})^T)\right)\nonumber\\
&\hspace{-0.5in}=
\frac{1}{|2\pi\symk{\Sigma}{k}|^{n_k/2}}\exp\left(-\frac{1}{2}tr((\bigk{Y}{k}-\bigk{A}{k}\bigk{\Ylag{Y}}{k})^T
\Sigma^{-(k)}(\bigk{Y}{k}-\bigk{A}{k}\bigk{\Ylag{Y}}{k})\BF{I})\right)\nonumber\\
&\hspace{-0.5in}=
\MN{\bigk{Y}{k}}{\bigk{A}{k}\bigk{\Ylag{Y}}{k}}{\symk{\Sigma}{k}}{\BF{I}}.
\end{align}

To derive the posterior of the dynamic parameters, it is useful to
first compute
\begin{align}
p(\bigk{D}{k},\bigk{A}{k} \mid \symk{\Sigma}{k}) = p(\bigk{D}{k}
\mid \bigk{A}{k},\symk{\Sigma}{k})p(\bigk{A}{k} \mid
\symk{\Sigma}{k}).
\end{align}
Using the fact that both the likelihood $p(\bigk{D}{k} \mid
\bigk{A}{k},\symk{\Sigma}{k})$ and the prior $\mbox{$p(\bigk{A}{k}
\mid \symk{\Sigma}{k})$}$ are matrix-normally distributed sharing a
common parameter $\symk{\Sigma}{k}$, we have
\begin{align}
\log p(\bigk{D}{k},\bigk{A}{k} \mid \symk{\Sigma}{k}) + C\nonumber\\
&\hspace{-1in}=
-\frac{1}{2}tr((\bigk{Y}{k}-\bigk{A}{k}\bigk{\Ylag{Y}}{k})^T
\Sigma^{-(k)}(\bigk{Y}{k}-\bigk{A}{k}\bigk{\Ylag{Y}}{k}) + (\bigk{A}{k}-M)^T\Sigma^{-(k)}(\bigk{A}{k}-M)K)\nonumber\\
%&\hspace{-1in}=
%-\frac{1}{2}tr(\Sigma^{-(k)}\{(\bigk{Y}{k}-\bigk{A}{k}\bigk{\Ylag{Y}}{k})(\bigk{Y}{k}-\bigk{A}{k}\bigk{\Ylag{Y}}{k})^T\nonumber\\
%&\hspace{0in}+ (\bigk{A}{k}-M)K(\bigk{A}{k}-M)^T\})\nonumber\\
&\hspace{-1in}=
-\frac{1}{2}tr(\Sigma^{-(k)}\{\bigk{A}{k}\bigksub{S}{k}{\Ylag{y}\Ylag{y}}\bigkT{A}{k}
- 2\bigksub{S}{k}{y\Ylag{y}}\bigkT{A}{k} + \bigksub{S}{k}{yy}\})\nonumber\\
&\hspace{-1in}=
-\frac{1}{2}tr(\Sigma^{-(k)}\{(\bigk{A}{k}-\bigksub{S}{k}{y\Ylag{y}}\bigminksub{S}{k}{\Ylag{y}\Ylag{y}})\bigksub{S}{k}{\Ylag{y}\Ylag{y}}(\bigk{A}{k}-\bigksub{S}{k}{y\Ylag{y}}\bigminksub{S}{k}{\Ylag{y}\Ylag{y}})^T
+ \bigksub{S}{k}{y|\Ylag{y}}\}),\label{eqn:useful}
\end{align}
where we have used the definitions:
\begin{align*}
C &= -\log
\frac{1}{|2\pi\bigk{\Sigma}{k}|^{n_k/2}}\frac{|K|^{d/2}}{|2\pi\bigk{\Sigma}{k}|^{rn_k/2}}
\hspace{0.5in} \bigksub{S}{k}{y|\Ylag{y}} = \bigksub{S}{k}{yy} -
\bigksub{S}{k}{y\Ylag{y}}\bigminksub{S}{k}{\Ylag{y}\Ylag{y}}\bigkTsub{S}{k}{y\Ylag{y}},
\end{align*}
\begin{align*}
\bigksub{S}{k}{\Ylag{y}\Ylag{y}} =
\bigk{\Ylag{Y}}{k}\bigkT{\Ylag{Y}}{k} + K \hspace{0.25in}
\bigksub{S}{k}{y\Ylag{y}} = \bigk{Y}{k}\bigkT{\Ylag{Y}}{k} + MK
\hspace{0.25in} \bigksub{S}{k}{yy} = \bigk{Y}{k}\bigkT{Y}{k} +
MKM^T.
\end{align*}

Conditioning on the noise covariance $\symk{\Sigma}{k}$, we see that
the dynamic matrix posterior is given by:
\begin{align}
p(\bigk{A}{k} \mid \bigk{D}{k}, \symk{\Sigma}{k})&\propto
\exp\left(-\frac{1}{2}tr((\bigk{A}{k}-\bigksub{S}{k}{y\Ylag{y}}\bigminksub{S}{k}{\Ylag{y}\Ylag{y}})^T\Sigma^{-(k)}(\bigk{A}{k}-\bigksub{S}{k}{y\Ylag{y}}\bigminksub{S}{k}{\Ylag{y}\Ylag{y}})\bigksub{S}{k}{\Ylag{y}\Ylag{y}})\right)\nonumber\\
&=\MN{\bigk{A}{k}}{\bigksub{S}{k}{y\Ylag{y}}\bigminksub{S}{k}{\Ylag{y}\Ylag{y}}}{\symk{\Sigma}{k}}{\bigksub{S}{k}{\Ylag{y}\Ylag{y}}}.
\end{align}

Marginalizing Eq.~\eqref{eqn:useful} over the dynamic matrix
$\bigk{A}{k}$, we derive
\begin{align}
p(\bigk{D}{k} \mid \symk{\Sigma}{k}) &=
\int_{\bigk{A}{k}}p(\bigk{D}{k},\bigk{A}{k} \mid
\symk{\Sigma}{k})d\bigk{A}{k}\nonumber\\
%&=\int_{\bigk{A}{k}}\frac{|K^{d/2}|}{|2\pi\symk{\Sigma}{k}|^{n_k/2}|2\pi\symk{\Sigma}{k}|^{rn_k/2}}\nonumber\\
%&\hspace{0.25in} \: \exp\left(-\frac{1}{2}tr(\Sigma^{-(k)}(\bigk{A}{k}-\bigksub{S}{k}{y\Ylag{y}}\bigminksub{S}{k}{\Ylag{y}\Ylag{y}})\bigksub{S}{k}{\Ylag{y}\Ylag{y}}(\bigk{A}{k}-\bigksub{S}{k}{y\Ylag{y}}\bigminksub{S}{k}{\Ylag{y}\Ylag{y}})^T)\right)\nonumber\\
%&\hspace{0.25in} \: \exp\left(-\frac{1}{2}tr(\Sigma^{-(k)}\bigksub{S}{k}{y|\Ylag{y}})\right)d\bigk{A}{k}\nonumber\\
&\hspace{-0.75in}= \frac{|K|^{d/2}}{|2\pi\symk{\Sigma}{k}|^{n_k/2}}\exp\left(-\frac{1}{2}tr(\Sigma^{-(k)}\bigksub{S}{k}{y|\Ylag{y}})\right)
\int_{\bigk{A}{k}}\frac{1}{|\bigksub{S}{k}{\Ylag{y}\Ylag{y}}|^{d/2}}\MN{\bigk{A}{k}}{\bigksub{S}{k}{y\Ylag{y}}\bigminksub{S}{k}{\Ylag{y}\Ylag{y}}}{\symk{\Sigma}{k}}{\bigksub{S}{k}{\Ylag{y}\Ylag{y}}}d\bigk{A}{k}\nonumber\\
&\hspace{-0.75in}=\frac{|K|^{d/2}}{|2\pi\symk{\Sigma}{k}|^{n_k/2}|\bigksub{S}{k}{\Ylag{y}\Ylag{y}}|^{d/2}}\exp\left(-\frac{1}{2}tr(\Sigma^{-(k)}\bigksub{S}{k}{y|\Ylag{y}})\right),
\end{align}
which leads us to our final result of the covariance having an
inverse-Wishart marginal posterior distribution:
\begin{align}
p(\Sigma^{(k)} \mid \bigk{D}{k}) &\propto p(\bigk{D}{k}
\mid \symk{\Sigma}{k})p(\symk{\Sigma}{k})\nonumber\\
&\propto
\frac{|K|^{d/2}}{|2\pi\symk{\Sigma}{k}|^{n_k/2}|\bigksub{S}{k}{\Ylag{y}\Ylag{y}}|^{d/2}}\exp\left(-\frac{1}{2}tr(\Sigma^{-(k)}\bigksub{S}{k}{y|\Ylag{y}})\right)|\symk{\Sigma}{k}|^{-(d+n_0+1)/2}\exp\left(-\frac{1}{2}tr(\Sigma^{-(k)}S_0)\right)\nonumber\\
&\propto
|\Sigma^{(k)}|^{-(d+n_k+n_0+1)/2}\exp\left(-\frac{1}{2}tr(\Sigma^{-(k)}(\bigksub{S}{k}{y|\Ylag{y}}+S_0))\right)\nonumber\\
&= \mbox{IW}(n_k + n_0,\bigksub{S}{k}{y|\Ylag{y}}+S_0).
\end{align}

\subsection{Non-Conjugate Independent Priors on $\bigk{A}{k}$,
$\symk{\Sigma}{k}$, and $\symkB{\mu}{k}$} 
\label{app:nonConjPrior}
In this section, we provide the derivations for the posterior
distributions of $\bigk{A}{k}$, $\symk{\Sigma}{k}$, and
$\symkB{\mu}{k}$ when each of these parameters is given an
independent prior.  One example of a non-conjugate prior is our proposed ARD sparsity-inducing prior.
\paragraph{Normal Prior on $\bigk{A}{k}$} 
\label{app:normalPriorA}
Assume we place a Gaussian prior,
$\mathcal{N}(\BF{\mu}_A,\Sigma_A)$, on the vectorization of the
matrix $\bigk{A}{k}$, which we denote by $\vecc{\bigk{A}{k}}$. To
examine the posterior distribution, we first aim to write the data
as a linear function of $\vecc{\bigk{A}{k}}$. We may rewrite
Eq.~\eqref{eqn:VARk} as
\begin{align}
\BF{y}_t &= \bigk{A}{k}\begin{bmatrix} \BF{y}_{t-1}^T & \BF{y}_{t-2}^T & \dots &
\BF{y}_{t-r}^T \end{bmatrix}^T + \BF{e}_t \hspace{0.25in} \forall t|z_t = k\nonumber\\
&\triangleq \bigk{A}{k}\BF{\bar{y}}_t + \BF{e}_t(k).
\end{align}
Recalling that $r$ is the autoregressive order and $d$ the dimension
of the observation vector $\BF{y}_t$, we can equivalently represent
the above as
\begin{align}
\BF{y}_t &= \left[\begin{array}{cccccccccccc} \bar{y}_{t,1} & \bar{y}_{t,2} & \cdots & \bar{y}_{t,d*r} & 0 & 0 & \cdots & 0 & 0 & 0 & \cdots & 0\\
0 & 0 & \cdots & 0 & \bar{y}_{t,1} & \bar{y}_{t,2} & \cdots & \bar{y}_{t,d*r} & 0 & 0 & \cdots & 0\\
\vdots & \vdots & \ddots & \vdots & \vdots & \vdots & \ddots & \vdots & \vdots & \vdots & \ddots & \vdots\\
0 & 0 & \cdots & 0 & 0 & 0 & \cdots & 0 & \bar{y}_{t,1} &
\bar{y}_{t,2} & \cdots
& \bar{y}_{t,d*r}\end{array}\right] \left[\begin{array}{c} a^{(k)}_{1,1}\\
a^{(k)}_{1,2} \\ \vdots \\ a^{(k)}_{1,d*r} \\ a^{(k)}_{2,1} \\ a^{(k)}_{2,2} \\ \vdots \\
a^{(k)}_{d,d*r}\end{array}\right] + \BF{e}_t(k)\nonumber\\
&= \left[\begin{array}{cccc} \bar{y}_{t,1}I_d &
\bar{y}_{t,2}I_d & \cdots & \bar{y}_{t,d*r}I_d\end{array}\right]
\vecc{\bigk{A}{k}} + \BF{e}_t(k) \triangleq\BF{\bar{Y}}_t \vecc{A} +
\BF{e}_t(k).
\end{align}
Here, the columns of $\BF{\bar{y}}_t$ are permutations of those of
the matrix in the first line such that we may write $\BF{y}_t$ as a
function of $\vecc{\bigk{A}{k}}$. Noting that $\BF{e}_t(k) \sim
\mathcal{N}(\symkB{\mu}{k},\symk{\Sigma}{k})$,
\begin{align}
\log p(\bigk{D}{k},\bigk{A}{k}\mid \symk{\Sigma}{k},\symkB{\mu}{k})\nonumber\\
&\hspace{-1in}= C -\frac{1}{2} \sum_{t|z_t = k} (\BF{y}_t
-\symkB{\mu}{k}-
\BF{\bar{Y}}_t\vecc{\bigk{A}{k}})^T\symmink{\Sigma}{k}(\BF{y}_t
-\symkB{\mu}{k}-
\BF{\bar{Y}}_t\vecc{\bigk{A}{k}})\nonumber\\
&\hspace{-0.68in}-\frac{1}{2}(\vecc{\bigk{A}{k}}-\BF{m}_A)^T\Sigma_A^{-1}(\vecc{\bigk{A}{k}}-\BF{m}_A),
\end{align}
%
%&=&C -\frac{1}{2}( \sum_{t|z_t=k}
%(\BF{y}_t-\symkB{\mu}{k})^T\symmink{\Sigma}{k}(\BF{y}_t-\symkB{\mu}{k})
%-
%2\vecc{\bigk{A}{k}}^T\BF{\bar{Y}}_t^T\symmink{\Sigma}{k}(\BF{y}_t-\symkB{\mu}{k})\\
%&&\hspace{0.5in}+
%\vecc{\bigk{A}{k}}^T\BF{\bar{Y}}_t^T\symmink{\Sigma}{k}\BF{\bar{Y}}_t\vecc{\bigk{A}{k}}\\
%&&\hspace{0.5in} +
%\vecc{\bigk{A}{k}}^T\Sigma_A^{-1}\vecc{\bigk{A}{k}} -
%2\vecc{\bigk{A}{k}}\Sigma_A^{-1}\BF{m}_A +
%\BF{m}_A^T\Sigma_A^{-1}\BF{m}_A)\\
%
which can be rewritten as,
\begin{align}
\log p(\bigk{D}{k},\bigk{A}{k}\mid \symk{\Sigma}{k},\symkB{\mu}{k})
&=C -\frac{1}{2}
\vecc{\bigk{A}{k}}^T\left(\Sigma_A^{-1}+\sum_{t|z_t=k}\BF{\bar{Y}}_t^T\symmink{\Sigma}{k}\BF{\bar{Y}}_t\right)\vecc{\bigk{A}{k}}\nonumber\\
&\hspace{0.32in}+ \vecc{\bigk{A}{k}}^T\left(\Sigma_A^{-1}\BF{m}_A +
\sum_{t|z_t=k}\BF{\bar{Y}}_t^T\symmink{\Sigma}{k}(\BF{y}_t-\symkB{\mu}{k})\right)\nonumber\\
&\hspace{0.32in}-\frac{1}{2}\BF{m}_A^T\Sigma_A^{-1}\BF{m}_A
-\frac{1}{2}\sum_{t|z_t=k}
(\BF{y}_t-\symkB{\mu}{k})^T\symmink{\Sigma}{k}(\BF{y}_t-\symkB{\mu}{k})
\end{align}
Conditioning on the data, we arrive at the desired posterior
distribution
\begin{align}
\log p(\bigk{A}{k}\mid \bigk{D}{k},\symk{\Sigma}{k},\symkB{\mu}{k})
&= C -\frac{1}{2}\bigg(
\vecc{\bigk{A}{k}}^T(\Sigma_A^{-1}+\sum_{t|z_t=k}\BF{\bar{Y}}_t^T\symmink{\Sigma}{k}\BF{\bar{Y}}_t)\vecc{\bigk{A}{k}}\nonumber\\
&\hspace{0.32in} - 2\vecc{\bigk{A}{k}}^T(\Sigma_A^{-1}\BF{m}_A +
\sum_{t|z_t=k}\BF{\bar{Y}}_t^T\symmink{\Sigma}{k}(\BF{y}_t-\symkB{\mu}{k}))\bigg)\nonumber\\
&\hspace{-1in}= \mathcal{N}^{-1}\left(\Sigma_A^{-1}\BF{m}_A +
\sum_{t|z_t=k}\BF{\bar{Y}}_t^T\symmink{\Sigma}{k}(\BF{y}_t-\symkB{\mu}{k}),\Sigma_A^{-1}+\sum_{t|z_t=k}\BF{\bar{Y}}_t^T\symmink{\Sigma}{k}\BF{\bar{Y}}_t\right)
\end{align}
\paragraph{Inverse Wishart Prior on $\symk{\Sigma}{k}$}
We place an inverse-Wishart prior, $\mbox{IW}(n_0,S_0)$, on
$\symk{\Sigma}{k}$. Let $n_k = |\{t|z_t=k,t=1,2,\dots,T\}|$.
Conditioned on $\bigk{A}{k}$ and $\symkB{\mu}{k}$, standard conjugacy results imply that the posterior of $\symk{\Sigma}{k}$ is:
\begin{align}
p(\symk{\Sigma}{k}\mid
\bigk{D}{k},\bigk{A}{k},\symkB{\mu}{k})=
\mbox{IW}\left(n_k+n_0,
S+\sum_{t|z_t=k}(\BF{y}_t-\bigk{A}{k}\BF{\bar{y}}_t-\symkB{\mu}{k})(\BF{y}_t-\bigk{A}{k}\BF{\bar{y}}_t-\symkB{\mu}{k})^T\right).
\end{align}
\paragraph{Normal Prior on $\symkB{\mu}{k}$}
\label{app:normalPriorMu}
Finally, we place a Gaussian prior,
$\mathcal{N}(\BF{\mu}_0,\Sigma_0)$, on $\symkB{\mu}{k}$. Conditioned
on $\bigk{A}{k}$ and $\symk{\Sigma}{k}$, the posterior of $\symkB{\mu}{k}$
is:
\begin{align}
p(\symkB{\mu}{k}\mid \bigk{D}{k},\bigk{A}{k},\symk{\Sigma}{k}) =
\mathcal{N}^{-1}\left(\symkB{\mu}{k};\Sigma_0^{-1}\BF{\mu}_0 +
\symmink{\Sigma}{k}\sum_{t|z_t=k}(\BF{y}_t-\bigk{A}{k}\BF{\bar{y}}_t),\Sigma_0^{-1}+
n_k\symmink{\Sigma}{k}\right).
\end{align}
We iterate between sampling $\bigk{A}{k}$, $\symk{\Sigma}{k}$, and
$\symkB{\mu}{k}$ many times before moving on to the next step of the
Gibbs sampler.
\section{Sparsity-Inducing Priors for Inferring Variable Order Models}
\label{app:ARD}
Recall Sec.~\ref{sec:ARD} and the proposed automatic relevance determination (ARD) prior for inferring non-dynamical components of the state vector in the case of the HDP-SLDS or lag components in the HDP-AR-HMM by shrinking components of the model parameters to zero.  However, if we would like to ensure that our choice of $C=[I_d \,\, 0]$ does not interfere with learning a sparse realization if one exists, we must restrict ourselves to considered a constrained class of dynamical phenomenon. For example, imagine a realization of an LDS with
\begin{align*}
	\tilde{A} = 
	\begin{bmatrix}
		0.8 & 0\\
		0.2 & 0 
	\end{bmatrix}
	, \hspace{0.25in} \tilde{C} = 
	\begin{bmatrix}
		1 & 1 
	\end{bmatrix}. 
\end{align*}
Then, the transformation to $C=[1 \,\, 0]$ leads to
\begin{align*}
	A = T^{-1}\tilde{A}T = 
	\begin{bmatrix}
		0.5 & 1\\
		0.15 & 0.3
	\end{bmatrix}, \hspace{0.25in} \mbox{for} \hspace{0.25in} 	T=
		\begin{bmatrix}
			0.5 & 1\\
			0.5 & -1 
		\end{bmatrix}.
\end{align*}
So, for this example, fixing $C=[1 \,\, 0]$ would not lead to learning a sparse dynamical matrix $A$.  Criterion~\ref{crit:nested} provides a set of sufficient, though not necessary, conditions for maintaining the sparsity within each $\bigk{A}{k}$ when transforming to the realization with $C = [I_d \,\, 0]$. That is, given there exists a realization $\mathcal{R}_1$ of our dynamical phenomena that satisfies Criterion~\ref{crit:nested}, the transformation $T$ to an equivalent realization $\mathcal{R}_2$ with $C = [I_d \,\, 0]$ will maintain the sparsity structure seen in $\mathcal{R}_1$, which we aim to infer with the ARD prior. Criterion~\ref{crit:nested}, which states that the observed state vector components are a subset of those relevant to \emph{all} modes, is reasonable for many applications: we often have observations only of components of the state vector that are essential to \emph{all} modes while \emph{some} modes may have additional components that affect the dynamics, but are not directly observed.

To clarify the conditions of Criterion~\ref{crit:nested}, consider a 3-mode SLDS realization $\mathcal{R}$ with
\begin{align}
	\bigk{A}{1} &= 
	\begin{bmatrix}
		\BF{a}_1^{(1)} & \BF{a}_2^{(1)} & \BF{a}_3^{(1)} & 0 & 0 
	\end{bmatrix}
	\quad \bigk{A}{2} = 
	\begin{bmatrix}
		\BF{a}_1^{(2)} & \BF{a}_2^{(2)} & 0 & \BF{a}_4^{(2)} & 0 
	\end{bmatrix}
	\nonumber\\
	&\hspace{0.5in}\bigk{A}{3} =
	\begin{bmatrix}
		\BF{a}_1^{(3)} & \BF{a}_2^{(3)} & \BF{a}_3^{(3)} & 0 & \BF{a}_5^{(3)} 
	\end{bmatrix}
	, 
\end{align}
then the observation matrix must be of the form $C = 
\begin{bmatrix}
	\BF{c}_1 & \BF{c}_2 & 0 & 0 & 0 
\end{bmatrix}
$ 
to satisfy Criterion~\ref{crit:nested}.
\section{HDP-SLDS and HDP-AR-HMM Message Passing}
\label{app:msgPassing}
In this appendix, we explore the computation of the backwards
message passing and forward sampling scheme used for generating
samples of the mode sequence $z_{1:T}$ and state sequence
$\BF{x}_{1:T}$.
\subsection{Mode Sequence Message Passing for Blocked Sampling}
Consider a switching VAR($r$) process. To derive the
forward-backward procedure for jointly sampling the mode sequence
$z_{1:T}$ given observations $\BF{y}_{1:T}$, plus $r$ initial
observations $\BF{y}_{1-r:0}$, we first note that the chain rule and
Markov structure allows us to decompose the joint distribution as
follows:
\begin{multline}
p(z_{1:T}\mid \BF{y}_{1-r:T},\BF{\pi},\BF{\uniqueTheta}) =
p(z_T\mid z_{T-1},\BF{y}_{1-r:T},\BF{\pi},\BF{\uniqueTheta})p(z_{T-1}\mid z_{T-2},\BF{y}_{1-r:T},\BF{\pi},\BF{\uniqueTheta})\\
\cdots p(z_{2}\mid
z_{1},\BF{y}_{1-r:T},\BF{\pi},\BF{\uniqueTheta})p(z_{1}\mid
\BF{y}_{1-r:T},\BF{\pi},\BF{\uniqueTheta}).
\end{multline}
Thus, we may first sample $z_1$ from $p(z_{1}\mid
\BF{y}_{1-r:T},\BF{\pi},\BF{\uniqueTheta})$, then condition on this
value to sample $z_2$ from $p(z_{2}\mid
z_{1},\BF{y}_{1-r:T},\BF{\pi},\BF{\uniqueTheta})$, and so on. The
conditional distribution of $z_1$ is derived as:
\begin{align}
p(z_1\mid \BF{y}_{1-r:T},\BF{\pi},\BF{\uniqueTheta}) &\propto
p(z_1)p(\BF{y}_1\mid
\uniqueTheta_{z_1},\BF{y}_{1-r:0})\sum_{z_{2:T}}\prod_t
p(z_t\mid \pi_{z_{t-1}})p(\BF{y}_t\mid \uniqueTheta_{z_t},\BF{y}_{t-r:t-1})\nonumber\\
&\propto p(z_1)p(\BF{y}_1\mid
\uniqueTheta_{z_1},\BF{y}_{1-r:0})\sum_{z_{2}}
p(z_2\mid \pi_{z_{1}})p(\BF{y}_2\mid \uniqueTheta_{z_2},\BF{y}_{2-r:1})m_{3,2}(z_2)\nonumber\\
&\propto p(z_1)p(\BF{y}_1\mid
\uniqueTheta_{z_1},\BF{y}_{1-r:0})m_{2,1}(z_1),
\end{align}
where $m_{t,t-1}(z_{t-1})$ is the backward message passed from $z_t$
to $z_{t-1}$ and is recursively defined by:
\begin{align}
m_{t,t-1}(z_{t-1}) &\propto \left\{
                               \begin{array}{ll}
                                 \sum_{z_t}p(z_t\mid \pi_{z_{t-1}})p(\BF{y}_t\mid \uniqueTheta_{z_t},\BF{y}_{t-r:t-1})m_{t+1,t}(z_t), & t\leq T; \\
                                 1, & t=T+1.
                               \end{array}
                             \right.
\end{align}
The general conditional distribution of $z_t$ is:
\begin{align}
p(z_t\mid z_{t-1},\BF{y}_{1-r:T},\BF{\pi},\BF{\uniqueTheta})
&\propto p(z_t\mid \pi_{z_{t-1}})p(\BF{y}_t\mid
\uniqueTheta_{z_t},\BF{y}_{t-r:t-1})m_{t+1,t}(z_t).
\end{align}
For the HDP-AR-HMM, these distributions are given by:
\begin{align}
p(z_t=k\mid z_{t-1},\BF{y}_{1-r:T},\BF{\pi},\BF{\uniqueTheta})
&\propto \pi_{z_{t-1}}(k)\mathcal{N}(\BF{y}_t;\sum_{i=1}^r
A_i^{(k)}\BF{y}_{t-i},\Sigma^{(k)})m_{t+1,t}(k)\nonumber\\
m_{t+1,t}(k) &= \sum_{j=1}^L
\pi_{k}(j)\mathcal{N}(\BF{y}_{t+1};\sum_{i=1}^r
A_i^{(j)}\BF{y}_{t-i},\Sigma^{(j)})m_{t+2,t+1}(j)\nonumber\\
m_{T+1,T}(k) &= 1 \quad k=1,\dots,L.
\end{align}
\subsection{State Sequence Message Passing for Blocked Sampling}
\label{app:stateSampling}
A similar sampling scheme is used for generating samples of the
state sequence $\BF{x}_{1:T}$. Although we now have a continuous
state space, the computation of the backwards messages
$m_{t+1,t}(\BF{x}_t)$ is still analytically feasible since we are
working with Gaussian densities. Assume, $m_{t+1,t}(\BF{x}_t)
\propto \mathcal{N}^{-1}(\BF{x}_t;\theta_{t+1,t},\Lambda_{t+1,t})$,
where $\mathcal{N}^{-1}(x;\theta,\Lambda)$ denotes a Gaussian
distribution on $x$ in information form with mean $\mu =
\Lambda^{-1}\theta$ and covariance $\Sigma = \Lambda^{-1}$. Given a
fixed mode sequence $z_{1:T}$, we simply have a time-varying linear
dynamic system. The backwards messages for the HDP-SLDS can
be recursively defined by
\begin{eqnarray}
m_{t,t-1}(\BF{x}_{t-1}) \propto \int_{\mathcal{X}_t}
p(\BF{x}_t|\BF{x}_{t-1},z_t)p(\BF{y}_t|\BF{x}_t)m_{t+1,t}(\BF{x}_t)d\BF{x}_t.
\label{eqn:bwdmsg}
\end{eqnarray}
For this model, the state transition density of
Eq.~\eqref{eqn:bwdmsg} can be expressed as
\begin{align}
p(\BF{x}_t|\BF{x}_{t-1},z_t) &\propto
\exp\left\{-\frac{1}{2}(\BF{x}_t-\symk{A}{z_t}\BF{x}_{t-1}-\symkB{\mu}{z_t})^T\symmink{\Sigma}{z_t}(\BF{x}_t-\symk{A}{z_t}\BF{x}_{t-1}-\symkB{\mu}{z_t})\right\}\\
&\propto \exp\bigg\{-\frac{1}{2}\left[ \begin{array}{c}
\BF{x}_{t-1}\\
\BF{x}_t\end{array} \right]^T \left[ \begin{array}{cc}
\symkT{A}{z_t}\symmink{\Sigma}{z_t}\symk{A}{z_t} & -\symkT{A}{z_t}\symmink{\Sigma}{z_t}\\
-\symmink{\Sigma}{z_t}\symk{A}{z_t} &
\symmink{\Sigma}{z_t}\end{array} \right]\left[
\begin{array}{c}
\BF{x}_{t-1}\\
\BF{x}_t\end{array} \right]\nonumber\\
&\hspace{2in}+ \left[ \begin{array}{c}
\BF{x}_{t-1}\\
\BF{x}_t\end{array} \right]^T\left[ \begin{array}{c}
-\symkT{A}{z_t}\symmink{\Sigma}{z_t}\symkB{\mu}{z_t}\\\
\symmink{\Sigma}{z_t}\symkB{\mu}{z_t}\end{array}
\right]\bigg\}\nonumber.
\end{align}
We can similarly write the likelihood in exponentiated quadratic
form
\begin{align}
p(\BF{y}_t|\BF{x}_t) &\propto
\exp\bigg\{-\frac{1}{2}(\BF{y}_t-C\BF{x}_t)^TR^{-1}(\BF{y}_t-C\BF{x}_t)\bigg\}\\
&\propto \exp\bigg\{-\frac{1}{2}\left[
\begin{array}{c}
\BF{x}_{t-1}\\
\BF{x}_t\end{array} \right]^T \left[ \begin{array}{cc}
0 & 0\\
0 & C^TR^{-1}C\end{array} \right]\left[ \begin{array}{c}
\BF{x}_{t-1}\\
\BF{x}_t\end{array} \right] + \left[ \begin{array}{c}
\BF{x}_{t-1}\\
\BF{x}_t\end{array} \right]^T\left[ \begin{array}{c}
0\\
C^TR^{-1}\BF{y}_t\end{array} \right]\bigg\}\nonumber,
\end{align}
as well as the messages
\begin{align}
m_{t+1,t}(\BF{x}_t) &\propto
\exp\bigg\{-\frac{1}{2}\BF{x}_t^T\Lambda_{t+1,t}\BF{x}_t+\BF{x}_t^T\theta_{t+1,t}\bigg\}\\
&\propto \exp\bigg\{-\frac{1}{2}\left[
\begin{array}{c}
\BF{x}_{t-1}\\
\BF{x}_t\end{array} \right]^T \left[ \begin{array}{cc}
0 & 0\\
0 & \Lambda_{t+1,t}\end{array} \right]\left[ \begin{array}{c}
\BF{x}_{t-1}\\
\BF{x}_t\end{array} \right] + \left[ \begin{array}{c}
\BF{x}_{t-1}\\
\BF{x}_t\end{array} \right]^T\left[ \begin{array}{c}
0\\
\theta_{t+1,t}\end{array} \right]\bigg\}\nonumber.
\end{align}
The product of these quadratics is given by:
\begin{align}
p(\BF{x}_t|\BF{x}_{t-1},z_t)p(\BF{y}_t|\BF{x}_t)m_{t+1,t}(\BF{x}_t) &\propto&\nonumber\\
&\hspace{-2in}\exp\bigg\{-\frac{1}{2}\left[ \begin{array}{c}
\BF{x}_{t-1}\\
\BF{x}_t\end{array} \right]^T \left[ \begin{array}{cc}
\symkT{A}{z_t}\symmink{\Sigma}{z_t}A & -\symkT{A}{z_t}\symmink{\Sigma}{z_t}\\
-\symmink{\Sigma}{z_t}\symk{A}{z_t} &  \symmink{\Sigma}{z_t} +
C^TR^{-1}C + \Lambda_{t+1,t}\end{array} \right]\left[
\begin{array}{c}
\BF{x}_{t-1}\\
\BF{x}_t\end{array} \right]\nonumber\\
&\hspace{-1in}+ \left[ \begin{array}{c}
\BF{x}_{t-1}\\
\BF{x}_t\end{array} \right]^T\left[ \begin{array}{c}
-\symkT{A}{z_t}\symmink{\Sigma}{z_t}\symkB{\mu}{z_t}\\
C^TR^{-1}\BF{y}_t + \symmink{\Sigma}{z_t}\symkB{\mu}{z_t} +
\theta_{t+1,t}\end{array} \right]\bigg\}
\end{align}
Using standard Gaussian marginalization identities we integrate over
$\BF{x}_t$ to get,
\begin{align}
m_{t,t-1}(\BF{x}_{t-1}) \propto
\mathcal{N}^{-1}(\BF{x}_{t-1};\theta_{t,t-1},\Lambda_{t,t-1}),
\end{align}
where,
\begin{equation}
\begin{aligned}
\theta_{t,t-1} &=
-\symkT{A}{z_t}\symmink{\Sigma}{z_t}\symkB{\mu}{z_t} +
\symkT{A}{z_t}\symmink{\Sigma}{z_t}(\symmink{\Sigma}{z_t} +
C^TR^{-1}C + \Lambda_{t+1,t})^{-1}(C^TR^{-1}\BF{y}_t +
\symmink{\Sigma}{z_t}\symkB{\mu}{z_t} +
\theta_{t+1,t})\\
\Lambda_{t,t-1} &= \symkT{A}{z_t}\symmink{\Sigma}{z_t}\symk{A}{z_t}
- \symkT{A}{z_t}\symmink{\Sigma}{z_t}(\symmink{\Sigma}{z_t} +
C^TR^{-1}C +
\Lambda_{t+1,t})^{-1}\symmink{\Sigma}{z_t}\symk{A}{z_t}.
\end{aligned}
\end{equation}
The backwards message passing recursion is initialized with
$m_{T+1,T} \sim \mathcal{N}^{-1}(\BF{x}_{T};0,0)$.  Let,
\begin{equation}
\begin{aligned}
\Lambda^b_{t|t} &= C^TR^{-1}C + \Lambda_{t+1,t}\\
\theta^b_{t|t} &= C^TR^{-1}\BF{y}_t + \theta_{t+1,t}.
\end{aligned}
\end{equation}
Then we can define the following recursion, which we note is
equivalent to a backwards running Kalman filter in information form,
\begin{align}
\Lambda^b_{t-1|t-1} &= C^TR^{-1}C +
\symkT{A}{z_t}\symmink{\Sigma}{z_t}\symk{A}{z_t} -
\symkT{A}{z_t}\symmink{\Sigma}{z_t}(\symmink{\Sigma}{z_t} +
C^TR^{-1}C + \Lambda_{t+1,t})^{-1}\symmink{\Sigma}{z_t}\symk{A}{z_t}\nonumber\\
&= C^TR^{-1}C + \symkT{A}{z_t}\symmink{\Sigma}{z_t}\symk{A}{z_t} -
\symkT{A}{z_t}\symmink{\Sigma}{z_t}(\symmink{\Sigma}{z_t} +
\Lambda^b_{t|t})^{-1}\symmink{\Sigma}{z_t}\symk{A}{z_t}\nonumber\\
\theta^b_{t-1|t-1} &= C^TR^{-1}\BF{y}_{t-1}
-\symkT{A}{z_t}\symmink{\Sigma}{z_t}\symkB{\mu}{z_t}+
\symkT{A}{z_t}\symmink{\Sigma}{z_t}(\symmink{\Sigma}{z_t} +
C^TR^{-1}C + \Lambda_{t+1,t})^{-1}\nonumber\\
&\hspace{3in}\cdot(C^TR^{-1}\BF{y}_t +
\symmink{\Sigma}{z_t}\symkB{\mu}{z_t} +
\theta_{t+1,t})\nonumber\\
&= C^TR^{-1}\BF{y}_{t-1}
-\symkT{A}{z_t}\symmink{\Sigma}{z_t}\symkB{\mu}{z_t}+
\symkT{A}{z_t}\symmink{\Sigma}{z_t}(\symmink{\Sigma}{z_t} +
\Lambda^b_{t|t})^{-1}(\theta^b_{t|t} +
\symmink{\Sigma}{z_t}\symkB{\mu}{z_t})\nonumber
\end{align}
We initialize at time $T$ with
\begin{equation}
\begin{aligned}
\Lambda^b_{T|T} &= C^TR^{-1}C\\
\theta^b_{T|T} &= C^TR^{-1}\BF{y}_T
\end{aligned}
\end{equation}
An equivalent, but more numerically stable recursion is summarized
in Algorithm~\ref{alg:stablefilterBackward}.
\begin{algorithm}[htb]
\vspace*{-1pt} \hspace*{-6pt}
\begin{flushleft}
\begin{enumerate}
  \algtop
\item Initialize filter with
\begin{align*}
\Lambda^b_{T|T} &= C^TR^{-1}C\\
\theta^b_{T|T} &= C^TR^{-1}\BF{y}_T
\end{align*}
\item Working backwards in time, for each $t \in \{T-1,\dots,1\}$:
\begin{enumerate}
\item Compute
\begin{align*}
\tilde{J}_{t+1} &= \Lambda^{b}_{t+1|t+1}(\Lambda^{b}_{t+1|t+1} + \Sigma^{-(z_{t+1})})^{-1}\\
\tilde{L}_{t+1} &= I - \tilde{J}_{t+1}.
\end{align*}
\item Predict
\begin{align*}
\Lambda_{t+1,t} &=
\symkT{A}{z_{t+1}}(\tilde{L}_{t+1}\Lambda^b_{t+1|t+1}\tilde{L}_{t+1}^T
+
\tilde{J}_{t+1}\Sigma^{-(z_{t+1})}\tilde{J}_{t+1}^T)\symk{A}{z_{t+1}}\\
\theta_{t+1,t} &=
\symkT{A}{z_{t+1}}\tilde{L}_{t+1}(\theta^b_{t+1|t+1}-\Lambda_{t+1|t+1}^b\symkB{\mu}{z_{t+1}})
\end{align*}
\item Update
\begin{align*}
\Lambda^b_{t|t} = \Lambda_{t+1,t} +
C^TR^{-1}C\\
\theta^b_{t|t} = \theta_{t+1,t} + C^TR^{-1}\BF{y}_t
\end{align*}
\end{enumerate}
\item Set
\begin{align*}
\Lambda^b_{0|0} = \Lambda_{1,0}\\
\theta^b_{0|0} = \theta_{1,0}
\end{align*}
\end{enumerate}
\end{flushleft}
%\begin{singlespace}
\caption{Numerically stable form of the backwards Kalman information
filter.} \label{alg:stablefilterBackward}
%\end{singlespace}
\end{algorithm}

After computing the messages $m_{t+1,t}(\BF{x}_t)$ backwards in
time, we sample the state sequence $\BF{x}_{1:T}$ working forwards
in time. As with the discrete mode sequence, one can decompose the
posterior distribution of the state sequence as
\begin{align}
p(\BF{x}_{1:T}\mid \BF{y}_{1:T},z_{1:T},\BF{\uniqueTheta}) &=
p(\BF{x}_T\mid \BF{x}_{T-1},\BF{y}_{1:T},z_{1:T},\BF{\uniqueTheta})p(\BF{x}_{T-1}\mid \BF{x}_{T-2},\BF{y}_{1:T},z_{1:T},\BF{\uniqueTheta})\nonumber\\
&\cdots p(\BF{x}_{2}\mid
\BF{x}_{1},\BF{y}_{1:T},z_{1:T},\BF{\uniqueTheta})p(\BF{x}_{1}\mid
\BF{y}_{1:T},z_{1:T},\BF{\uniqueTheta}).
\end{align}
where
\begin{align}
p(\BF{x}_t\mid \BF{x}_{t-1},\BF{y}_{1:T},z_{1:T},\BF{\uniqueTheta})
\propto p(\BF{x}_t\mid
\BF{x}_{t-1},\symk{A}{z_t},\symk{\Sigma}{z_t},\symkB{\mu}{z_t})p(\BF{y}_t\mid
\BF{x}_t, R)m_{t+1,t}(\BF{x}_t).
\end{align}
For the HDP-SLDS, the product of these distributions is equivalent
to
\begin{align}
p(\BF{x}_t\mid \BF{x}_{t-1},\BF{y}_{1:T},z_{1:T},\BF{\uniqueTheta})
   &\propto
\mathcal{N}(\BF{x}_t;\symk{A}{z_t}\BF{x}_{t-1}+\symkB{\mu}{z_t},\symk{\Sigma}{z_t})\mathcal{N}(\BF{y}_t;C\BF{x}_{t},R)m_{t+1,t}(\BF{x}_t)\nonumber\\
&\propto
\mathcal{N}(\BF{x}_t;\symk{A}{z_t}\BF{x}_{t-1}+\symkB{\mu}{z_t},\symk{\Sigma}{z_t})\mathcal{N}^{-1}(\BF{x}_t;\theta^b_{t|t},\Lambda^b_{t|t})\nonumber\\
&\propto
\mathcal{N}^{-1}(\BF{x}_t;\symmink{\Sigma}{z_t}(\symk{A}{z_t}\BF{x}_{t-1}+\symkB{\mu}{z_t})+\theta^b_{t|t},\symmink{\Sigma}{z_t}+\Lambda^b_{t|t}),
\end{align}
which is a simple Gaussian distribution so that the normalization
constant is easily computed. Specifically, for each $t\in
\{1,\dots,T\}$ we sample $\BF{x}_t$ from
\begin{align}
\BF{x}_t
   &\sim
\mathcal{N}(\BF{x}_t;(\symmink{\Sigma}{z_t}+\Lambda^b_{t|t})^{-1}(\symmink{\Sigma}{z_t}\symk{A}{z_t}\BF{x}_{t-1}+\symmink{\Sigma}{z_t}\symkB{\mu}{z_t}+\theta^b_{t|t}),(\symmink{\Sigma}{z_t}+\Lambda^b_{t|t})^{-1}).
\end{align}
\subsection{Mode Sequence Message Passing for Sequential Sampling}
\label{app:sequentialZSampling}
A similar sampling scheme to Carter and Kohn~\cite{Carter:96} is used for
generating samples of the mode sequence $z_{1:T}$ having
marginalized over the state sequence $\BF{x}_{1:T}$. Specifically,
we sample $z_t$ from:
\begin{align}
p(z_t = k\mid z_{\backslash t},
\BF{y}_{1:T},\BF{\pi},\BF{\uniqueTheta}) &\propto p(z_t = k \mid
z_{\backslash t}, \BF{\pi})p(\BF{y}_{1:T}
\mid z_t = k, z_{\backslash t})\nonumber\\
&\propto \pi_{z_{t-1}}(k)\pi_k(z_{t+1})p(\BF{y}_{1:T} \mid z_t = k,
z_{\backslash t}).
\end{align}
We omit the dependency on $\BF{\pi}$ and $\BF{\uniqueTheta}$ for
compactness. To compute the likelihood for each $z_t$, we combine
forward and backward messages along with the local dynamics and
measurements as follows:
\begin{align}
p(\BF{y}_{1:T}\mid z_t=k,z_{\backslash t}) &\propto
\int_{\mathcal{X}_{t-1}}\int_{\mathcal{X}_{t}}m_{t-2,t-1}(\BF{x}_{t-1})p(\BF{y}_{t-1}\mid\BF{x}_{t-1})
p(\BF{x}_{t}|\BF{x}_{t-1},z_{t}=k)\nonumber\\
&\hspace{1in}p(\BF{y}_{t}|\BF{x}_{t})m_{t+1,t}(\BF{x}_{t})d\BF{x}_{t}d\BF{x}_{t-1}\\
&\propto
\int_{\mathcal{X}_{t}}\int_{\mathcal{X}_{t-1}}m_{t-2,t-1}(\BF{x}_{t-1})p(\BF{y}_{t-1}\mid\BF{x}_{t-1})
p(\BF{x}_{t}|\BF{x}_{t-1},z_{t}=k)d\BF{x}_{t-1}\nonumber\\
&\hspace{1in}p(\BF{y}_{t}|\BF{x}_{t})m_{t+1,t}(\BF{x}_{t})d\BF{x}_{t},
\label{eqn:sequentialLikelihood}
\end{align}
where the backwards messages are defined as in
Appendix~\ref{app:stateSampling} and the forward messages by:
\begin{align}
m_{t-1,t}(\BF{x}_{t}) \propto \int_{\mathcal{X}_{t-1}}
p(\BF{x}_t|\BF{x}_{t-1},z_t)p(\BF{y}_{t-1}|\BF{x}_{t-1})m_{t-2,t-1}(\BF{x}_{t-1})d\BF{x}_{t-1}.
\label{eqn:fwdmsg}
\end{align}
To derive the forward message passing recursions, assume that
\begin{align}
m_{t-2,t-1}(\BF{x}_{t-1}) \propto
\mathcal{N}^{-1}(\BF{x}_{t-1};\theta_{t-2,t-1},\Lambda_{t-2,t-1})
\end{align}
and $z_t$ is known.  The terms of the integrand of
Eq.~\eqref{eqn:fwdmsg} can be written as:
\begin{align}
p(\BF{x}_t|\BF{x}_{t-1},z_t) &= \mathcal{N}(\BF{x}_t;
\symk{A}{z_t}\BF{x_{t-1}} + \symkB{\mu}{z_t},\symk{\Sigma}{z_t})\label{eqn:transDist}\\
&\propto \exp\bigg\{-\frac{1}{2}\left[
\begin{array}{c}
\BF{x}_t\\
\BF{x}_{t-1}\end{array} \right]^T \left[ \begin{array}{cc}
\symmink{\Sigma}{z_t} & -\symmink{\Sigma}{z_t}\symk{A}{z_t}\\
-\symkT{A}{z_t}\symmink{\Sigma}{z_t} &
\symkT{A}{z_t}\symmink{\Sigma}{z_t}\symk{A}{z_t}\end{array}
 \right]\left[
\begin{array}{c}
\BF{x}_t\\
\BF{x}_{t-1}\end{array} \right]\nonumber\\
&\hspace{0.5in} + \left[
\begin{array}{c}
\BF{x}_t\\
\BF{x}_{t-1}\end{array} \right]^T\left[ \begin{array}{c}
\symmink{\Sigma}{z_t}\symkB{\mu}{z_t}\\
-\symkT{A}{z_t}\symmink{\Sigma}{z_t}\symkB{\mu}{z_t}\end{array}
\right]\bigg\}\nonumber
\end{align}
\begin{align}
m_{t-2,t-1}(\BF{x}_{t-1})p(\BF{y}_{t-1}|\BF{x}_{t-1}) &\propto
\mathcal{N}(\BF{x}_{t-1};
\Lambda_{t-1|t-1}^{-f}\theta_{t-1|t-1}^f,\Lambda_{t-1|t-1}^{f})\label{eqn:updatedDist}\\
&\hspace{-1.75in}\propto \exp\bigg\{-\frac{1}{2}\left[
\begin{array}{c}
\BF{x}_t\\
\BF{x}_{t-1}\end{array} \right]^T \left[ \begin{array}{cc}
0 & 0\\
0 & \Lambda^f_{t-1|t-1}\end{array} \right]\left[
\begin{array}{c}
\BF{x}_t\\
\BF{x}_{t-1}\end{array} \right] + \left[
\begin{array}{c}
\BF{x}_t\\
\BF{x}_{t-1}\end{array} \right]^T\left[ \begin{array}{c}
0\\
\theta^f_{t-1|t-1}\end{array} \right]\bigg\}\nonumber,
\end{align}
where, similar to the backwards recursions, we have made the
following definitions
\begin{equation}
\begin{aligned}
\theta_{t|t}^f &= \theta_{t-1,t} + C^TR^{-1}\BF{y}_t\\
\Lambda_{t|t}^f &= \Lambda_{t-1,t}+C^TR^{-1}C.
\end{aligned}
\end{equation}
Combining these distributions and integrating over $\BF{x}_{t-1}$,
we have
\begin{align}
m_{t-1,t}(\BF{x}_t) &\propto
\mathcal{N}^{-1}(\BF{x}_{t};\theta_{t-1,t},\Lambda_{t-1,t})
\end{align}
with
\begin{align}
\theta_{t-1,t} &=
\symmink{\Sigma}{z_t}\symkB{\mu}{z_t}+\symmink{\Sigma}{z_t}\symk{A}{z_t}(\symkT{A}{z_t}\symmink{\Sigma}{z_t}\symk{A}{z_t}
+ \Lambda_{t-1|t-1}^f)^{-1}(\theta_{t-1|t-1}^f -
\symkT{A}{z_t}\symmink{\Sigma}{z_t}\symkB{\mu}{z_t})\nonumber\\
\Lambda_{t-1,t} &=
\symmink{\Sigma}{z_t}-\symmink{\Sigma}{z_t}\symk{A}{z_t}(\symkT{A}{z_t}\symmink{\Sigma}{z_t}\symk{A}{z_t}
+
\Lambda_{t-1|t-1}^f)^{-1}\symkT{A}{z_t}\symmink{\Sigma}{z_t}\nonumber,
\end{align}
or equivalently,
\begin{equation}
\begin{aligned}
\theta_{t-1,t} &= \Lambda_{t-1,t}(\symkB{\mu}{z_t} +
\symk{A}{z_t}\Lambda_{t-1|t-1}^{-f}\theta_{t-1|t-1}^f)\\
\Lambda_{t-1,t} &=(\symk{\Sigma}{z_t} +
\symk{A}{z_t}\Lambda_{t-1|t-1}^{-f}\symkT{A}{z_t})^{-1}.
\end{aligned}
\end{equation}
Assuming $\BF{x}_0\sim \mathcal{N}(0,P_0)$, we initialize at time
$t=0$ to
\begin{align}
\theta_{-1,0} &= 0\nonumber\\
\Lambda_{-1,0} &= P_0^{-1}.
\end{align}
An equivalent, but more numerically stable recursion is summarized
in Algorithm~\ref{alg:stablefilterForward}. However, this algorithm
relies on the dynamic matrix $\symk{A}{k}$ being invertible.
\begin{algorithm}[htb]
\vspace*{-1pt} \hspace*{-6pt}
\begin{flushleft}
\begin{enumerate}
  \algtop
\item Initialize filter with
\begin{align*}
\Lambda^b_{0|0} &= P_0\\
\theta^b_{0|0} &= 0
\end{align*}
\item Working forwards in time, for each $t \in \{1,\dots,T\}$:
\begin{enumerate}
\item Compute
\begin{align*}
M_t &= \symminkT{A}{z_{t+1}}\Lambda_{t|t}^{-f}\symmink{A}{z_{t+1}}\\
J_t &= M_t(M_t + \symmink{\Sigma}{z_{t+1}})^{-1}\\
L_t &= I - J_t.
\end{align*}
\item Predict
\begin{align*}
\Lambda_{t-1,t} &= L_{t-1}M_{t-1}L_{t-1}^T +
J_{t-1}\symmink{\Sigma}{z_{t}}J_{t-1}^T\\
\theta_{t-1,t} &= L_{t-1}\symminkT{A}{z_{t}}(\theta_{t-1|t-1}^f +
\theta_{t-1|t-1}^f \symmink{A}{z_t}\symkB{\mu}{z_{t}})
\end{align*}
\item Update
\begin{align*}
\Lambda^f_{t|t} = \Lambda_{t-1,t} +
C^TR^{-1}C\\
\theta^f_{t|t} = \theta_{t-1,t} + C^TR^{-1}\BF{y}_t
\end{align*}
\end{enumerate}
\end{enumerate}
\end{flushleft}
%\begin{singlespace}
\caption{Numerically stable form of the forward Kalman information
filter.} \label{alg:stablefilterForward}
%\end{singlespace}
\end{algorithm}

We now return to the computation of the likelihood of
Eq.~\eqref{eqn:sequentialLikelihood}. We note that the integral over
$\BF{x}_{t-1}$ is equivalent to computing the message
$m_{t-1,t}(\BF{x}_t)$ using $z_t=k$. However, we have to be careful
that any constants that were previously ignored in this message
passing are not a function of $z_t$. For the meantime, let us assume
that there exists such a constant and let us denote this special
message by
\begin{align}
m_{t-1,t}(\BF{x}_t;z_t) \propto
c(z_t)\mathcal{N}^{-1}(\BF{x}_t;\theta_{t-1,t}(z_t),\Lambda_{t-1,t}(z_t))\label{eqn:specialMsg}.
\end{align}
Then, the likelihood can be written as
\begin{align}
p(\BF{y}_{1:T}\mid z_t=k,z_{\backslash t}) &\propto
\int_{\mathcal{X}_t}m_{t-1,t}(\BF{x}_t;z_t=k)p(\BF{y}_{t}|\BF{x}_{t})m_{t+1,t}(\BF{x}_{t})d\BF{x}_t\\
&\propto
\int_{\mathcal{X}_t}c(k)\mathcal{N}^{-1}(\BF{x}_t;\theta_{t-1,t}(k),\Lambda_{t-1,t}(k))\mathcal{N}^{-1}(\BF{x}_t;
\theta_{t|t}^b,\Lambda_{t|t}^b)d\BF{x}_t
\end{align}
Combining the information parameters, and maintaining the term in
the normalizing constant that is a function of $k$, this is
equivalent to
\begin{align}
p(\BF{y}_{1:T}\mid z_t=k,z_{\backslash t}) &\propto
%&\hspace{-2.5in}\propto c(k)
%\int_{\mathcal{X}_t}|\Lambda_{t-1,t}(k)|^{1/2}\exp(-\frac{1}{2}\BF{x}_t^T\Lambda_{t-1,t}(k)\BF{x}_t
%+ \BF{x}_t^T\theta_{t-1,t}(k)
%-\frac{1}{2}\theta_{t-1,t}(k)^T\Lambda_{t-1,t}(k)^{-1}\theta_{t-1,t}(k))\exp(-\frac{1}{2}\BF{x}_t^T\Lambda_{t|t}^b\BF{x}_t
%+ \BF{x}_t^T\theta_{t|t}^b)d\BF{x}_t\\
c(k)
|\Lambda_{t-1,t}(k)|^{1/2}\exp\left(-\frac{1}{2}\theta_{t-1,t}(k)^T\Lambda_{t-1,t}(k)^{-1}\theta_{t-1,t}(k)\right)\nonumber\\
&\hspace{-0.75in}\int_{\mathcal{X}_t}\exp\left(-\frac{1}{2}\BF{x}_t^T(\Lambda_{t-1,t}(k)+\Lambda_{t|t}^b)\BF{x}_t
+ \BF{x}_t^T(\theta_{t-1,t}(k)+\theta_{t|t}^b)\right)d\BF{x}_t
\end{align}
To compute this integral, we write the integrand in terms of a
Gaussian distribution times a constant. The integral is then simply
that constant term:
\begin{align}
p(\BF{y}_{1:T}\mid z_t=k,z_{\backslash t})
 &\propto c(k)
|\Lambda_{t-1,t}(k)|^{1/2}\exp\left(-\frac{1}{2}\theta_{t-1,t}(k)^T\Lambda_{t-1,t}(k)^{-1}\theta_{t-1,t}(k)\right)\nonumber\\
&\hspace{-1in}|\Lambda_{t-1,t}(k)+\Lambda_{t|t}^b|^{-1/2}
\exp\left(\frac{1}{2}(\theta_{t-1,t}(k)+\theta_{t|t}^b)^T(\Lambda_{t-1,t}(k)+\Lambda_{t|t}^b)^{-1}(\theta_{t-1,t}(k)+\theta_{t|t}^b)\right)\nonumber\\
&\hspace{-0.5in}\int_{\mathcal{X}_t}\mathcal{N}^{-1}(\BF{x}_t;\theta_{t-1,t}(k)+\theta_{t|t}^b,\Lambda_{t-1,t}(k)+\Lambda_{t|t}^b)d\BF{x}_t\nonumber\\
&\hspace{-1in}\propto c(k)
\frac{|\Lambda_{t-1,t}(k)|^{1/2}}{|\Lambda_{t-1,t}(k)+\Lambda_{t|t}^b|^{1/2}}\nonumber\\
&\hspace{-0,75in}\exp\bigg(-\frac{1}{2}\theta_{t-1,t}(k)^T\Lambda_{t-1,t}(k)^{-1}\theta_{t-1,t}(k)\nonumber\\
&\hspace{0.5in}+
\frac{1}{2}(\theta_{t-1,t}(k)+\theta_{t|t}^b)^T(\Lambda_{t-1,t}(k)+\Lambda_{t|t}^b)^{-1}(\theta_{t-1,t}(k)+\theta_{t|t}^b)\bigg)\nonumber
\end{align}

Thus,
\begin{multline}
p(z_t = k\mid z_{\backslash t},
\BF{y}_{1:T},\BF{\pi},\BF{\uniqueTheta}) \propto
\pi_{z_{t-1}}(k)\pi_k(z_{t+1})c(k)|\Lambda_t^{(k)}|^{1/2}|\Lambda_t^{(k)}+\Lambda_{t|t}^b|^{-1/2}\\
\exp\left(-\frac{1}{2}\theta_t^{(k)^T}\Lambda_t^{-(k)}\theta_t^{(k)}
+
\frac{1}{2}(\theta_t^{(k)}+\theta_{t|t}^b)^T(\Lambda_t^{(k)}+\Lambda_{t|t}^b)^{-1}(\theta_t^{(k)}+\theta_{t|t}^b)\right)
\end{multline}

We now show that $c(z_t)$ is not a function $z_t$.  The only place
where the previously ignored dependency on $z_t$ arises is from
$p(\BF{x}_t \mid \BF{x}_{t-1},z_t)$. Namely,
\begin{align}
p(\BF{x}_t \mid \BF{x}_{t-1},z_t) &=
\frac{\exp(-\frac{1}{2}\symkT{\BF{\mu}}{z_t}\symmink{\Sigma}{z_t}\symkB{\mu}{z_t})}{|\symk{\Sigma}{z_t}|^{1/2}}
\cdot \mbox{exponential}_1\nonumber\\
&= c_1(z_t)\cdot\mbox{exponential}_1 \label{eqn:exp1}
\end{align}
where $\mbox{exponential}_1$ is the exponentiated quadratic of
Eq.~\eqref{eqn:transDist}. Then, when compute the message
$m_{t-1,t}(\BF{x}_{t};z_t)$ we update the previous message
$m_{t-2,t-1}(\BF{x}_{t-1})$ by incorporating the local likelihood
$p(\BF{y}_{t-1}\mid \BF{x}_{t-1})$ and then propagating the state
estimate with $p(\BF{x}_t \mid \BF{x}_{t-1},z_t)$ and integrating
over $\BF{x}_{t-1}$.  Namely, we combine the distribution of
Eq.~\eqref{eqn:exp1} with the exponentiated quadratic of
Eq.~\eqref{eqn:updatedDist} and integrate over $\BF{x}_{t-1}$:
\begin{align}
m_{t-1,t}(\BF{x}_{t};z_t) &\propto
c_1(z_t)\int_{\mathcal{X}_{t-1}}\mbox{exponential}_1 \cdot
\mbox{exponential}_2d\BF{x}_{t-1}\label{eqn:exp3},
\end{align}
where $\mbox{exponential}_2$ is the exponentiated quadratic of
Eq.~\eqref{eqn:updatedDist}.

Since $m_{t-2,t-1}(\BF{x}_{t-1}) \propto p(\BF{x}_{t-1}\mid
\BF{y}_{1:t-2},z_{1:t-1})$, and the Markov properties of the state
space model dictate
\begin{align}
p(\BF{x}_{t-1}\mid \BF{y}_{1:t-1},z_{1:t-1}) &=
p(\BF{y}_{t-1}|\BF{x}_{t-1})p(\BF{x}_{t-1}\mid
\BF{y}_{1:t-2},z_{1:t-1})\nonumber\\
&\propto p(\BF{y}_{t-1}|\BF{x}_{t-1})m_{t-2,t-1}(\BF{x}_{t-1}),
\end{align}
then
\begin{align*}
p(\BF{x}_{t-1} \mid \BF{y}_{1:t-1},z_{1:t-1}) &=
c_2\cdot\mbox{exponential}_2\label{eqn:exp2}.
\end{align*}
We note that the normalizing constant $c_2$ is not a function of
$z_t$ since we have only considered $z_\tau$ for $\tau<t$.

Once again exploiting the conditional independencies induced by the
Markov structure of our state space model, and plugging in
Eq.~\eqref{eqn:exp1} and Eq.~\eqref{eqn:exp2},
\begin{align}
p(\BF{x}_t,\BF{x}_{t-1} \mid \BF{y}_{1:t-1},z_{1:t}) &=
p(\BF{x}_{t-1} \mid \BF{x}_{t-1},z_t)p(\BF{x}_{t-1} \mid
\BF{y}_{1:t-1},z_{1:t-1})\nonumber\\
&= (c_1(z_t)\cdot\mbox{exponential}_1)(c_2\cdot \mbox{exponential}_2)\nonumber\\
&= c_1(z_t)c_2\cdot \mbox{exponential}_1\cdot\mbox{exponential}_2.
\end{align}

Plugging this results into Eq.~\eqref{eqn:exp3}, we have
\begin{align}
m_{t-1,t}(\BF{x}_{t};z_t) &\propto
c_1(z_t)\int_{\mathcal{X}_{t-1}}\frac{1}{c_1(z_t)c_2}p(\BF{x}_t,\BF{x}_{t-1}
\mid \BF{y}_{1:t-1},z_{1:t})d\BF{x}_{t-1}\nonumber\\
&\propto \frac{1}{c_2}p(\BF{x}_t \mid
\BF{y}_{1:t-1},z_{1:t})\label{eqn:specialMsg2}.
\end{align}
Comparing Eq.~\eqref{eqn:specialMsg2} to Eq.~\eqref{eqn:specialMsg},
and noting that
\begin{align*}
p(\BF{x}_t \mid \BF{y}_{1:t-1},z_{1:t}) =
\mathcal{N}^{-1}(\BF{x}_t;\theta_{t-1,t}(z_t),\Lambda_{t-1,t}(z_t)),
\end{align*}
we see that $c(z_t)=\frac{1}{c_2}$ and is thus not a function of
$z_t$.

Algebraically, we could derive this result as follows.
\begin{align}
m_{t-1,t}(\BF{x}_{t};z_t) &\propto
c_1(z_t)\int_{\mathcal{X}_{t-1}}\mbox{exponential}_1\cdot\mbox{exponential}_2d\BF{x}_{t-1}\\
&= c_1(z_t)c_3(z_t)\int_{\mathcal{X}_{t-1}}\mathcal{N}\left(\left[
\begin{array}{c}
\BF{x}_{t-1}\nonumber\\
\BF{x}_t\end{array} \right];
\BF{\Lambda}(z_t)^{-1}\BF{\theta}(z_t),\BF{\Lambda}(z_t)\right)d\BF{x}_{t-1},
\label{eqn:exp3_2}
\end{align}
where $\BF{\theta}(z_t)$ and $\BF{\Lambda}(z_t)$ are the information
parameters determined by combining the functional forms of
$\mbox{exponential}_1$ and $\mbox{exponential}_2$, and
\begin{align}
c_1(z_t)c_3(z_t) =
\frac{\exp\{-\frac{1}{2}\symkT{\BF{\mu}}{z_{t}}\symmink{\Sigma}{z_t}\symkB{\mu}{z_{t}}\}}{|\symk{\Sigma}{z_t}|^{1/2}}
\frac{\exp\{\frac{1}{2}\BF{\theta}(z_t)^T\BF{\Lambda}(z_t)^{-1}\BF{\theta}(z_t)\}}{|\BF{\Lambda}(z_t)|^{1/2}}.
\end{align}
Computing these terms in parts, and using standard linear algebra
properties of block matrices,
\begin{align}
|\BF{\Lambda}(z_t)| &=
|\symmink{\Sigma}{z_t}||(\symkT{A}{z_t}\symmink{\Sigma}{z_t}\symk{A}{z_t}
+ \Lambda_{t-1|t-1}^f) -
\symkT{A}{z_t}\symmink{\Sigma}{z_t}\symk{A}{z_t}|\nonumber\\
&= |\symmink{\Sigma}{z_t}||\Lambda_{t-1|t-1}^f|\\
\vspace{0.2in}\nonumber\\
\BF{\Lambda}(z_t)^{-1} &= \begin{bmatrix} (\symmink{\Sigma}{z_t} -
\symmink{\Sigma}{z_t}\symk{A}{z_t}\tilde{\BF{\Lambda}}(z_t)^{-1}\symkT{A}{z_t}\symmink{\Sigma}{z_t})^{-1}
&\hspace{-0.2in}
\symk{A}{z_t}\Lambda_{t-1|t-1}^f\nonumber\\
\Lambda_{t-1|t-1}^f\symkT{A}{z_t} &\hspace{-0.2in}
(\tilde{\BF{\Lambda}}(z_t) -
\symkT{A}{z_t}\symmink{\Sigma}{z_t}\symk{A}{z_t})^{-1}
\end{bmatrix}\\
&= \begin{bmatrix} \symk{\Sigma}{z_t} +
\symk{A}{z_t}\Lambda_{t-1|t-1}^{-f}\symkT{A}{z_t} &
\symk{A}{z_t}\Lambda_{t-1|t-1}^f\\
\Lambda_{t-1|t-1}^f\symkT{A}{z_t} & \Lambda_{t-1|t-1}^{-f}
\end{bmatrix},
\end{align}
where $\tilde{\BF{\Lambda}}(z_t) =
(\symkT{A}{z_t}\symmink{\Sigma}{z_t}\symk{A}{z_t} +
\Lambda_{t-1|t-1}^f)$ and we have used the matrix inversion lemma in
obtaining the last equality.  Using this form of
$\BF{\Lambda}(z_t)^{-1}$, we readily obtain
\begin{align}
\BF{\theta}(z_t)^T\BF{\Lambda}(z_t)^{-1}\BF{\theta}(z_t) =
\symkB{\mu}{z_t}\symmink{\Sigma}{z_t}\symkB{\mu}{z_t} +
\theta_{t-1|t-1}^{f^T}\Lambda_{t-1|t-1}^{-f}\theta_{t-1|t-1}^f.
\end{align}
Thus,
\begin{align}
c_1(z_t)c_3(z_t) =
\frac{\exp\{\frac{1}{2}\theta_{t-1|t-1}^{f^T}\Lambda_{t-1|t-1}^{-f}\theta_{t-1|t-1}^f\}}{|\Lambda_{t-1|t-1}^f|^{1/2}},
\end{align}
which does not depend upon the value of $z_t$.
\section{Derivation of Maneuvering Target Tracking Sampler}
\label{app:HDPHMMKF}
In this Appendix we derive the maneuvering target tracking (MTT)
sampler outlined in Sec.~\ref{sec:MTT}. Recall the MTT model of Eq.~\eqref{eqn:SLDS_fixedA}.
As described in Sec.~\ref{sec:MTT}, we are interested in jointly
sampling the control input and dynamical mode $(\BF{u}_t,z_t)$,
marginalizing over the state sequence $\BF{x}_{1:T}$, the transition
distributions $\BF{\pi}$, and the dynamic parameters $\BF{\theta} =
\{\symkB{\mu}{k},\symk{\Sigma}{k}\}$. One can factor the desired
conditional distribution factorizes as,
\begin{align}
p(\BF{u}_t,z_t|z_{\backslash t},\BF{u}_{\backslash
t},\BF{y}_{1:T},\beta,\alpha,\kappa,\lambda)= p(z_t|z_{\backslash
t},\BF{u}_{\backslash
t},\BF{y}_{1:T},\beta,\alpha,\kappa,\lambda)p(\BF{u}_t|z_{1:T},\BF{u}_{\backslash
t},\BF{y}_{1:T},\lambda).
\label{eq:chain2}
\end{align}
The distribution in Eq.\eqref{eq:chain2} is a hybrid distribution:
each discrete value of the dynamical mode indicator variable $z_t$
corresponds to a different continuous distribution on the control
input $\BF{u}_t$. We analyze each of the conditional distributions
of Eq.~\eqref{eq:chain2} by considering the joint distribution on
all of the model parameters, and then marginalizing $\BF{x}_{1:T}$,
$\BF{\pi}$, and $\theta_k$. (Note that marginalization over
$\theta_j$ for $j\neq k$ simply results in a constant.)
\begin{multline}
p(z_t=k|z_{\backslash t},\BF{u}_{\backslash
t},\BF{y}_{1:T},\beta,\alpha,\kappa,\lambda) \propto
\int_{\BF{\pi}}\prod_j p(\pi_j|\beta,\alpha,\kappa) \prod_\tau
p(z_\tau|\pi_{z_{\tau-1}})d\pi\\
\int_{\mathcal{U}_t} \int p(\theta_k\mid \lambda)
\prod_{\tau|z_\tau=k}
p(\BF{u}_\tau|\theta_k)d\theta_k\int_\mathcal{X} \prod_\tau
p(\BF{x}_\tau|\BF{x}_{\tau-1},\BF{u}_{\tau})
p(\BF{y}_\tau|\BF{x}_\tau) d\BF{x}_{1:T}
d\BF{u}_t.\label{eq:marginal_zk}
\end{multline}
Similarly, we can write the conditional density of $\BF{u}_t$ for
each candidate $z_t$ as,
\begin{align}
p(\BF{u}_t|z_t=k,z_{\backslash t},\BF{u}_{\backslash
t},\BF{y}_{1:T},\lambda)
\propto \int p(\theta_k\mid \lambda) \prod_{\tau|z_\tau=k} p(\BF{u}_\tau|\theta_k)d\theta_k
\int_\mathcal{X} \prod_\tau
p(\BF{x}_\tau|\BF{x}_{\tau-1},\BF{u}_{\tau})
p(\BF{y}_\tau|\BF{x}_\tau) d\BF{x}_{1:T} \label{eq:marginal_u2}.
\end{align}
A key step in deriving these conditional distributions is the marginalization of the state sequence $\BF{x}_{1:T}$.  In performing this marginalization, one thing we harness is the fact that conditioning on the control input sequence simplifies the SLDS to an LDS with a deterministic control input $\BF{u}_{1:T}$. Thus, conditioning on $\BF{u}_{1:t-1,t+1:T}$ allows us to marginalize the state sequence in the following manner. We run a forward Kalman filter to pass a message from $t-2$ to $t-1$, which is updated by the local likelihood at $t-1$. A backward filter is also run to pass a message from $t+1$ to $t$, which is updated by the local likelihood at $t$. These updated messages are combined with the local dynamic $p(\BF{x}_t \mid \BF{x}_{t-1},\BF{u}_t,\BF{\theta})$ and then marginalized over $\BF{x}_t$ and $\BF{x}_{t-1}$, resulting in the likelihood of the observation sequence $\BF{y}_{1:T}$ as a function of $\BF{u}_t$, the variable of interest. Because the sampler conditions on control inputs, the filter for this time-invariant system can be efficiently implemented by pre-computing the error covariances and then solely computing local Kalman updates at every time step. Of note is that the computational complexity is linear in the training sequence length, as well as the number of currently instantiated maneuver modes.  In the following sections, we evaluate each of the integrals of Eq.~\eqref{eq:marginal_u2} and Eq.~\eqref{eq:marginal_zk} in turn.
\subsection{Chinese Restaurant Franchise}
The integration over $\BF{\pi}$ appearing in the first line of
Eq.~\eqref{eq:marginal_zk} results in exactly the same predictive
distribution as the sticky HDP-HMM~\cite{Fox:ICML08}.
%Namely,
%%
%\begin{multline}
%p(z_t=k\mid z_{\backslash t},\beta,\alpha,\kappa) \\\propto
%\begin{cases}
%(\alpha\beta_k + n^{-t}_{z_{t-1}k} + \kappa\delta(z_{t-1},k))\\
%\hspace{0.5in}\left(\frac{\alpha\beta_{z_{t+1}} + n^{-t}_{kz_{t+1}}+\kappa\delta(k,z_{t+1}) + \delta(z_{t-1},k)\delta(k,z_{t+1})}{\alpha + n^{-t}_{k \cdot}+\kappa + \delta(z_{t-1},k)}\right) & k \in \{1,\dots,K\}\\
%\frac{\alpha^2\beta_{\tilde{k}}\beta_{z_{t+1}}}{\alpha + \kappa} &
%k=K+1.
%\end{cases}
%\end{multline}
%%
%
\subsection{Normal-Inverse-Wishart Posterior Update}
The marginalization of $\theta_k$, appearing both in
Eq.~\eqref{eq:marginal_zk} and Eq.~\eqref{eq:marginal_u2}, can be
rewritten as follows:
\begin{align}
\int p(\theta_k|\lambda) \prod_{\tau|z_\tau=k}
p(\BF{u}_\tau|\theta_k)d\theta_k &= \int
p(\BF{u}_t|\theta_k)p(\theta_k|\lambda)\prod_{\tau|z_\tau=k,\tau\neq
t}
p(\BF{u}_\tau|\theta_k)d\theta_k\nonumber\\
&\propto \int
p(\BF{u}_t|\theta_k)p(\theta_k|\{\BF{u}_\tau|z_\tau=k,\tau\neq
t\},\lambda)d\theta_k\nonumber\\
&= p(\BF{u}_t|\{\BF{u}_\tau|z_\tau=k,\tau\neq t\},\lambda).
\end{align}
Here, the set $\{\BF{u}_\tau|z_\tau=k,\tau\neq t\}$ denotes all the
observations $\BF{u}_\tau$ other than $\BF{u}_t$ that were drawn
from the Gaussian parameterized by $\theta_k$. When $\theta_k$ has a
normal-inverse-Wishart prior
$\mathcal{NIW}(\kappa,\BF{\vartheta},\nu,\Delta)$, standard conjugacy results imply that the posterior is:
\begin{align}
p(\BF{u}_t|\{\BF{u}_\tau|z_\tau=k,\tau\neq
t\},\kappa,\BF{\vartheta},\nu,\Delta) \simeq \mathcal{N
}\left(\BF{u}_t;\bar{\BF{\vartheta}},\frac{(\bar{\kappa}+1)\bar{\nu}}{\bar{\kappa}(\bar{\nu}-d-1)}\bar{\Delta}\right)
\triangleq \mathcal{N}(\BF{u}_t;\hat{\BF{\mu}}_k,\hat{\Sigma}_k),
\end{align}
where
\begin{equation}
\begin{aligned}
\bar{\kappa} &= \kappa + |\{\BF{u}_s|z_s = k, s\neq t\}|\\
\bar{\nu} &= \nu + |\{\BF{u}_s|z_s = k, s\neq t\}|\\
\bar{\kappa}\bar{\BF{\vartheta}} &= \kappa\BF{\vartheta} +
\sum_{\BF{u}_s \in
\{\BF{u}_s|z_s = k, s\neq t\}}\BF{u}_s\\
\bar{\nu}\bar{\Delta} &= \nu\Delta +\sum_{\BF{u}_s \in
\{\BF{u}_s|z_s = k, s\neq t\}}\BF{u}_s\BF{u}_s^T +
\kappa\BF{\vartheta}\BF{\vartheta}^T -
\bar{\kappa}\bar{\BF{\vartheta}}\bar{\BF{\vartheta}}^T
\end{aligned}
\end{equation}
Here, we are using the moment-matched Gaussian approximation to the
Student-t predictive distribution for $\BF{u}_t$ induced by
marginalizing $\theta_k$.
\subsection{Marginalization by Message Passing}
When considering the control input $\BF{u}_t$ and conditioning on
the values of all $\BF{u}_\tau$, $\tau\neq t$, the marginalization
over all states $\BF{x}_{1:T}$ can be equated to a message passing
scheme that relies on the conditionally linear dynamical system
induced by fixing $\BF{u}_\tau$, $\tau\neq t$. Specifically,
\begin{align}
\int_\mathcal{X} \prod_\tau
&p(\BF{x}_\tau|\BF{x}_{\tau-1},\BF{u}_{\tau})
p(\BF{y}_\tau|\BF{x}_\tau) dx\nonumber\\
&\hspace{-0.05in}\propto\int_{\mathcal{X}_{t-1}}\int_{\mathcal{X}_{t}}m_{t-1,t-2}(\BF{x}_{t-1})
p(\BF{y}_{t-1}|\BF{x}_{t-1})p(\BF{x}_{t}|\BF{x}_{t-1},\BF{u}_t)p(\BF{y}_{t}|\BF{x}_{t})m_{t,t+1}(\BF{x}_{t})dx_{t}dx_{t-1}\nonumber\\
&\hspace{-0.05in}\propto p(\BF{y}_{1:T}|\BF{u}_t;\BF{u}_{\backslash
t}),\label{eq:stateMarg}
\end{align}
where we recall the definitions of the forward messages
$m_{t-1,t}(\BF{x}_t)$ and backward messages $m_{t+1,t}(\BF{x}_t)$
from Appendix~\ref{app:stateSampling}. For our MTT model of
Eq.~\eqref{eqn:SLDS_fixedA}, however, instead of accounting for a
process noise mean $\symkB{\mu}{z_\tau}$ at time $\tau$ in the
filtering equations, we must account for the control input
$\BF{u}_\tau$. Conditioning on $\BF{u}_\tau$, one can equate
$B\BF{u}_\tau$ with a process noise mean, and thus we simply replace
$\symkB{\mu}{z_\tau}$ with $B\BF{u}_\tau$ in the filtering equations
of Appendix~\ref{app:stateSampling}. Similarly, we replace the
process noise covariance term $\symk{\Sigma}{z_\tau}$ with our
process noise covariance $Q$. (Note that although
$\BF{u}_\tau(z_\tau) \sim
\mathcal{N}(\symkB{\mu}{z_\tau},\symk{\Sigma}{z_\tau})$, we
condition on the value $\BF{u}_\tau$ so that the MTT parameters
$\{\symkB{\mu}{z_\tau},\symk{\Sigma}{z_\tau}\}$ do not factor into
the message passing equations.)
\subsection{Combining Messages}
To compute the likelihood of Eq.~\eqref{eq:stateMarg}, we take the
filtered estimates of $\BF{x}_{t-1}$ and $\BF{x}_t$, combine them
with the local dynamics and local likelihood, and marginalize over
$\BF{x}_{t-1}$ and $\BF{x}_t$. To aid in this computation, we
consider the exponentiated quadratic form of each term in the
integrand of Eq.~\eqref{eq:stateMarg}. We then join these terms and
use standard Gaussian integration formulas to arrive at the desired
likelihood. The derivation of this likelihood greatly parallels that
for the sequential mode sequence sampler of
Appendix~\ref{app:sequentialZSampling}.

Recall the forward filter recursions of
Appendix~\ref{app:stateSampling} in terms of information parameters
\begin{align*}
\{\theta_{t-1,t},\Lambda_{t-1,t},\theta^f_{t|t},\Lambda^f_{t|t}\},
\end{align*}
and the backward filter recursions in terms of
\begin{align*}
\{\theta_{t+1,t},\Lambda_{t+1,t},\theta^b_{t|t},\Lambda^b_{t|t}\}.
\end{align*}
Replace $\symkB{\mu}{z_t}$ with $B\BF{u}_t$ and $\symk{\Sigma}{z_t}$
with $Q$ where appropriate. We many then write
$m_{t,t+1}(\BF{x}_{t})$ updated with the likelihood
$p(\BF{y}_{t-1}|\BF{x}_{t-1})$ in exponentiated quadratic form as:
\begin{align*}
\hspace{-0.05in}m_{t-1,t-2}(\BF{x}_{t-1})p(\BF{y}_{t-1}|\BF{x}_{t-1})\\
&\hspace{-0.9in}\propto \exp\bigg\{-\frac{1}{2}\left[
\begin{array}{c}
\BF{x}_{t-1}\\
\BF{x}_{t}\end{array} \right]^T \left[ \begin{array}{cc}
C^TR^{-1}C + \Lambda_{t-1,t-2} & 0\\
0 & 0\end{array} \right]\left[ \begin{array}{c}
\BF{x}_{t-1}\\
\BF{x}_{t}\end{array} \right]\\
&\hspace{0.8in}+ \left[ \begin{array}{c}
\BF{x}_{t-1}\\
\BF{x}_{t}\end{array} \right]^T\left[ \begin{array}{c}
C^TR^{-1}\BF{y}_{t-1} + \theta_{t-1,t-2}\\
0\end{array} \right]\bigg\}.
\end{align*}
The local dynamics can similarly be written as
\begin{align*}
\hspace{-0.05in}p(\BF{x}_{t}|\BF{x}_{t-1},\BF{u}_t)
\propto\exp\bigg\{-\frac{1}{2}\left[
\begin{array}{c}
\BF{u}_t\\
\BF{x}_{t-1}\\
\BF{x}_{t}\end{array} \right]^T \left[ \begin{array}{ccc}
B^TQ^{-1}B & B^TQ^{-1}A & -B^TQ^{-1}\\
A^TQ^{-1}B & A^TQ^{-1}A & -A^TQ^{-1}\\
-Q^{-1}B & -Q^{-1}A & Q^{-1}\end{array} \right]\left[
\begin{array}{c}
\BF{u}_t\\
\BF{x}_{t-1}\\
\BF{x}_{t}\end{array} \right]
+ \left[ \begin{array}{c}
\BF{u}_t\\
\BF{x}_{t-1}\\
\BF{x}_{t}\end{array} \right]^T\left[ \begin{array}{c}
0\\
0\\
0\end{array} \right]\bigg\}.
\end{align*}
Finally, the backward message $m_{t,t+1}(\BF{x}_t)$ updated with the
likelihood $p(\BF{y}_{t}|\BF{x}_{t})$ can be written as
\begin{multline*}
\hspace{-0.05in}p(\BF{y}_{t}|\BF{x}_{t})m_{t,t+1}(\BF{x}_t) \propto
\exp\bigg\{-\frac{1}{2}\left[
\begin{array}{c}
\BF{x}_{t-1}\\
\BF{x}_{t}\end{array} \right]^T \left[ \begin{array}{cc}
0 & 0\\
0 & C^TR^{-1}C + \Lambda_{t,t+1}\end{array} \right]\left[
\begin{array}{c}
\BF{x}_{t-1}\\
\BF{x}_{t}\end{array} \right]\\
+ \left[ \begin{array}{c}
\BF{x}_{t-1}\\
\BF{x}_{t}\end{array} \right]^T\left[ \begin{array}{c}
0\\
C^TR^{-1}\BF{y}_{t} + \theta_{t,t+1}\end{array} \right]\bigg\}.
\end{multline*}
Using the definitions
\begin{align*}
\Lambda^b_{t|t} &= C^TR^{-1}C + \Lambda_{t+1,t}\\
\theta^b_{t|t} &= C^TR^{-1}\BF{y}_t + \theta_{t+1,t}\\
\Lambda_{t|t}^f &= C^TR^{-1}C + \Lambda_{t-1,t}\\
\theta_{t|t}^f &= C^TR^{-1}\BF{y}_t + \theta_{t-1,t},
\end{align*}
we may express the entire integrand as
\begin{align*}
\hspace{-0.1in}m_{t-1,t-2}(\BF{x}_{t-1})p(\BF{y}_{t-1}|\BF{x}_{t-1})p(\BF{x}_{t}|\BF{x}_{t-1},\BF{u}_t)p(\BF{y}_{t}|\BF{x}_{t})m_{t,t+1}(\BF{x}_{t})
&\propto\\
&\hspace{-3.75in} \exp\bigg\{-\frac{1}{2}\left[
\begin{array}{c}
\BF{u}_t\\
\BF{x}_{t-1}\\
\BF{x}_{t}\end{array} \right]^T \left[ \begin{array}{ccc}
B^TQ^{-1}B & B^TQ^{-1}A & -B^TQ^{-1}\\
A^TQ^{-1}B & A^TQ^{-1}A + \Lambda^f_{t-1|t-1}  & -A^TQ^{-1}\\
-Q^{-1}B & -Q^{-1}A & Q^{-1} + \Lambda^b_{t|t}\end{array}
\right]\left[
\begin{array}{c}
\BF{u}_t\\
\BF{x}_{t-1}\\
\BF{x}_{t}\end{array} \right]\\
&\hspace{-0.75in} + \left[ \begin{array}{c}
\BF{u}_t\\
\BF{x}_{t-1}\\
\BF{x}_{t}\end{array} \right]^T\left[ \begin{array}{c}
0\\
\theta^f_{t-1|t-1}\\
\theta^b_{t|t}\end{array} \right]\bigg\}
\end{align*}

Integrating over $\BF{x}_t$, we obtain an expression proportional to
\begin{align*}
\mathcal{N}^{-1}\left(\begin{bmatrix}\BF{u}_t^T\\
\BF{x}_{t-1}\end{bmatrix}; \theta\left(\begin{bmatrix}
\BF{u}_t\\
\BF{x}_{t-1}\end{bmatrix}\right), \Lambda\left(\begin{bmatrix}
\BF{u}_t\\
\BF{x}_{t-1}\end{bmatrix}\right)\right),
\end{align*}
with
\begin{align*}
\Lambda\left(\begin{bmatrix}
\BF{u}_t\\
\BF{x}_{t-1}\end{bmatrix}\right) &= \left[ \begin{array}{cc}
B^TQ^{-1}B & B^TQ^{-1}A\\
A^TQ^{-1}B & A^TQ^{-1}A + \Lambda^f_{t-1|t-1}\end{array} \right] - \left[ \begin{array}{c}
B^TQ^{-1}\\
A^TQ^{-1}\end{array} \right] (Q^{-1} + \Lambda^b_{t|t})^{-1}\left[
\begin{array}{cc} Q^{-1}B & Q^{-1}A\end{array}
\right]\\
&= \left[ \begin{array}{cc}
B^T\Sigma_{t}^{-1}B & B^T\Sigma_{t}^{-1}A\\
A^T\Sigma_{t}^{-1}B & A^T\Sigma_{t}^{-1}A
\end{array} \right]\\
\theta\left(\begin{bmatrix}
\BF{u}_t\\
\BF{x}_{t-1}\end{bmatrix}\right) &= \left[ \begin{array}{c}
0\\
\theta^f_{t-1|t-1}\end{array} \right] + \left[
\begin{array}{c}
B^TQ^{-1}\\
A^TQ^{-1}\end{array} \right](Q^{-1} + \Lambda^b_{t|t})^{-1}\theta^b_{t|t}= \left[ \begin{array}{c}
B^TQ^{-1}K_{t}^{-1}\theta^b_{t|t}\\
\theta^f_{t-1|t-1} + A^TQ^{-1}K_{t}^{-1}\theta^b_{t|t}\end{array}
\right].
\end{align*}
Here, we have defined
\begin{align*}
\Sigma_{t} = Q^{-1} + Q^{-1}(Q^{-1} + \Lambda^b_{t|t})^{-1}Q^{-1} =
Q^{-1} + Q^{-1}K_{t}^{-1}Q^{-1}.
\end{align*}

Finally, integrating over $\BF{x}_{t-1}$ yields an expression
proportional to
\begin{align*}
\mathcal{N}^{-1}(\BF{u}_t^T; \theta(\BF{u}_t), \Lambda( \BF{u}_t)),
\end{align*}
with
\begin{align*}
\Lambda(\BF{u}_t) &= B^T\Sigma_{t}^{-1}B - B^T\Sigma_{t}^{-1}A(A^T\Sigma_{t}^{-1}A + \Lambda^f_{t-1|t-1})^{-1}A^T\Sigma_{t}^{-1}B\\
\theta(\BF{u}_t) &= B^TQ^{-1}K_{t}^{-1}\theta^b_{t|t} - B^T\Sigma_{t}^{-1}A(A^T\Sigma_{t}^{-1}A +
\Lambda^f_{t-1|t-1})^{-1}(\theta^f_{t-1|t-1} +
A^TQ^{-1}K_{t}^{-1}\theta^b_{t|t}).
\end{align*}
\subsection{Joining Distributions that Depend on $\BF{u}_t$}
We have derived two terms which depend on $\BF{u}_t$: a prior and a
likelihood. Normally, one would consider $p(\BF{u}_t|\theta_k)$ the
prior on $\BF{u}_t$. However, through marginalization of this
parameter, we induced dependencies between the control inputs
$\BF{u}_\tau$ and all the $\BF{u}_\tau$ that were drawn from a
distribution parameterized by $\theta_k$ inform us of the
distribution over $\BF{u}_t$. Therefore, we treat
$p(\BF{u}_t|\{\BF{u}_\tau|z_\tau = k, \tau \neq t\})$ as a prior
distribution on $\BF{u}_t$. The likelihood function
$p(\BF{y}_{1:T}|\BF{u}_t;\BF{u}_{\backslash t})$ describes the
likelihood of an observation sequence $\BF{y}_{1:T}$ given the input
sequence $\BF{u}_{1:T}$, containing the random variable is
$\BF{u}_t$.

We multiply the prior distribution by the likelihood function to get
the following quadratic expression:
\begin{align}
p(\BF{u}_t|\{\BF{u}_\tau|z_\tau = k, \tau \neq t\})p(\BF{y}_{1:T}|\BF{u}_t;\BF{u}_{\backslash t})\nonumber\\
&\hspace{-2.3in}\propto \frac{1}{(2\pi)^{N/2}|\hat{\Sigma}_k|^{1/2}}
\exp\bigg\{-\frac{1}{2}(\BF{u}_t-\hat{\BF{\mu}}_k)^T\hat{\Sigma}_k^{-1}(\BF{u}_t-\hat{\BF{\mu}}_k)\nonumber\\
&\hspace{-0.3in} -\frac{1}{2}(\BF{u}_t -
\Lambda(\BF{u}_t)^{-1}\theta(\BF{u}_t))^T\Lambda(\BF{u}_t)(\BF{u}_t
-
\Lambda_ty^{-1}\theta(\BF{u}_t))\bigg\}\nonumber\\
&\hspace{-2.3in}=\frac{1}{(2\pi)^{N/2}|\hat{\Sigma}_k|^{1/2}}
\exp\bigg\{-\frac{1}{2}\bigg[\BF{u}_t^T(\hat{\Sigma}_k^{-1}+\Lambda(\BF{u}_t))\BF{u}_t
- 2\BF{u}_t^T(\hat{\Sigma}_k^{-1}\hat{\BF{\mu}}_k\nonumber\\
&\hspace{0.03in} + \theta(\BF{u}_t)) +
\hat{\BF{\mu}}_k^T\hat{\Sigma}_k^{-1}\hat{\BF{\mu}}_k +
\theta(\BF{u}_t)^T\Lambda(\BF{u}_t)^{-1}\theta(\BF{u}_t)\bigg]\bigg\}\nonumber\\
&\hspace{-2.3in}=\frac{(2\pi)^{N/2}|(\hat{\Sigma}_k^{-1} +
\Lambda(\BF{u}_t))^{-1}|^{1/2}}{(2\pi)^{N/2}|\hat{\Sigma}_k|^{1/2}}
\exp\bigg\{-\frac{1}{2}\bigg[\hat{\BF{\mu}}_k^T\hat{\Sigma}_k^{-1}\hat{\BF{\mu}}_k+
\theta(\BF{u}_t)^T\Lambda(\BF{u}_t)^{-1}\theta(\BF{u}_t)\nonumber\\
&\hspace{-0.58in} - (\hat{\Sigma}_k^{-1}\hat{\BF{\mu}}_k +
\theta(\BF{u}_t))^T(\hat{\Sigma}_k^{-1} +
\Lambda(\BF{u}_t))^{-1}(\hat{\Sigma}_k^{-1}\hat{\BF{\mu}}_k +
\theta(\BF{u}_t))\bigg]\bigg\}\nonumber\\
&\hspace{-1.8in}\cdot\mathcal{N}(\BF{u}_t;(\hat{\Sigma}_k^{-1} +
\Lambda(\BF{u}_t))^{-1}(\hat{\Sigma}_k^{-1}\hat{\BF{\mu}}_k +
\theta(\BF{u}_t)),(\hat{\Sigma}_k^{-1} +
\Lambda(\BF{u}_t))^{-1})\nonumber\\
&\hspace{-2.3in}\triangleq
C_k\cdot\mathcal{N}(\BF{u}_t;(\hat{\Sigma}_k^{-1} +
\Lambda(\BF{u}_t))^{-1}(\hat{\Sigma}_k^{-1}\hat{\BF{\mu}}_k +
\theta(\BF{u}_t)),(\hat{\Sigma}_k^{-1} + \Lambda(\BF{u}_t))^{-1}),
\end{align}
where we note that the defined constant $C_k$ is a function of
$z_t=k$, but not of $\BF{u}_t$.
\subsection{Resulting $(\BF{u}_t,z_t)$ Sampling Distributions}
We write Eq.~\eqref{eq:marginal_zk} and Eq.~\eqref{eq:marginal_u2}
in terms of the derived distributions:
\begin{multline}
p(z_t=k|z_{\backslash t},\BF{u}_{\backslash
t},\BF{y}_{1:T},\beta,\alpha,\kappa,\lambda) \propto
p(z_t=k\mid z_{\backslash t},\beta,\alpha,\kappa)\\
\int_{\mathcal{U}_t} p(\BF{u}_t|\{\BF{u}_\tau|z_\tau = k, \tau \neq
t\})p(\BF{y}_{1:T}|\BF{u}_t;\BF{u}_{\backslash t})d\BF{u}_t,
\end{multline}
\vspace{-0.3in}
\begin{align}
p(\BF{u}_t|z_t=k,z_{\backslash t},\BF{u}_{\backslash
t},\BF{y}_{1:T},\lambda) \propto p(\BF{u}_t|\{\BF{u}_\tau|z_\tau =
k, \tau \neq t\})p(\BF{y}_{1:T}|\BF{u}_t;\BF{u}_{\backslash t}).
\end{align}
Thus, the distribution over $z_t$, marginalizing $\BF{u}_t$, is
given by
\begin{align}
p(z_t=k|z_{\backslash t},\BF{u}_{\backslash
t},\BF{y}_{1:T},\beta,\alpha,\kappa,\lambda)\nonumber\\
&\hspace{-1.75in}\propto
p(z_t=k\mid z_{\backslash t},\beta,\alpha,\kappa)\int_{\mathcal{U}_t}
C_k\cdot\mathcal{N}(\BF{u}_t;(\hat{\Sigma}_k^{-1} +
\Lambda(\BF{u}_t))^{-1}(\hat{\Sigma}_k^{-1}\hat{\BF{\mu}}_k +
\theta(\BF{u}_t)),(\hat{\Sigma}_k^{-1} +
\Lambda(\BF{u}_t))^{-1})d\BF{u}_t\nonumber\\
&\hspace{-1.75in}\propto C_k \cdot p(z_t=k\mid z_{\backslash
t},\beta,\alpha,\kappa).
\end{align}
and the distribution over $\BF{u}_t$ (for $z_t=k$ fixed) is
\begin{align}
p(\BF{u}_t|z_t=k,z_{\backslash t},\BF{u}_{\backslash
t},\BF{y}_{1:T},\lambda) =
\mathcal{N}(\BF{u}_t;(\hat{\Sigma}_k^{-1} +
\Lambda(\BF{u}_t))^{-1}(\hat{\Sigma}_k^{-1}\hat{\BF{\mu}}_k +
\theta(\BF{u}_t)),(\hat{\Sigma}_k^{-1} + \Lambda(\BF{u}_t))^{-1}).
\end{align}
%

%% biography section
%\begin{IEEEbiography}
%	{Emily Fox} Biography text here. 
%\end{IEEEbiography}
%\begin{IEEEbiography}
%	{Erik Sudderth} Biography text here. 
%\end{IEEEbiography}
%\begin{IEEEbiography}
%	{Michael Jordan} Biography text here. 
%\end{IEEEbiography}
%\begin{IEEEbiography}
%	{Alan Willsky} Biography text here. 
%\end{IEEEbiography}

\end{document}